\documentclass[preprint2]{aastex}

\begin{document}
\title{A {\em Chandra}/ACIS Study of 30~Doradus I.  Superbubbles and Supernova Remnants}

\author{Leisa K. Townsley\altaffilmark{1}, Patrick S.
Broos\altaffilmark{1}, Eric D. Feigelson\altaffilmark{1}, Bernhard R. Brandl\altaffilmark{2}, You-Hua Chu\altaffilmark{3}, Gordon P. Garmire\altaffilmark{1}, George G. Pavlov\altaffilmark{1}}

\altaffiltext{1}{Department of Astronomy \& Astrophysics, 525 Davey
Laboratory, Pennsylvania State University, University Park, PA 16802}
\altaffiltext{2}{Sterrewacht Leiden, PO Box 9513, 2300 RA Leiden, 
The Netherlands}
\altaffiltext{3}{Astronomy Department, University of Illinois at
Urbana-Champaign, 1002 West Green Street, Urbana, IL 61801}

\begin{abstract}

We present an X-ray tour of diffuse emission in the 30~Doradus
star-forming complex in the Large Magellanic Cloud using
high-spatial-resolution X-ray images and spatially-resolved spectra
obtained with the Advanced CCD Imaging Spectrometer aboard the {\em
Chandra X-ray Observatory}.  The dominant X-ray feature of the
30~Doradus nebula is the intricate network of diffuse emission
generated by interacting stellar winds and supernovae working together
to create vast superbubbles filled with hot plasma.  We construct maps
of the region showing variations in plasma temperature (T = 3--9
million degrees), absorption ($N_H$ = 1--6$\times 10^{21}$~cm$^{-2}$),
and absorption-corrected X-ray surface brightness ($S_X$ =
3--126$\times 10^{31}$~ergs~s$^{-1}$~pc$^{-2}$).  Enhanced images
reveal the pulsar wind nebula in the composite supernova remnant N157B
and the {\em Chandra} data show spectral evolution from non-thermal
synchrotron emission in the N157B core to a thermal plasma in its outer
regions.  In a companion paper we show that R136, the central massive
star cluster, is resolved at the arcsecond level into almost 100 X-ray
sources.  Through X-ray studies of 30~Doradus the complete life cycle
of such a massive stellar cluster can be revealed.

\end{abstract}


%==========================================================================
%=============================================================================
\section{INTRODUCTION \label{sec:intro}}

Images of spiral and irregular galaxies have been defined historically
by regions of massive star formation, signposts demarcating important
features such as spiral arms, bars, and starbursts.  They remind us
that galaxies really are evolving, with continuous injection of energy
and processed material into the galaxy.  Massive star-forming regions
(MSFRs) present us with a microcosm of starburst astrophysics, where
stellar winds from O and Wolf-Rayet (WR) stars compete with supernovae
(SNe) to carve up the neutral medium from which they formed, igniting a
new generation of stars.  With X-ray observations, we see the sources
that shape the larger view of a galaxy, injecting energy and processed
material into the disk and halo ISM through winds, ionization fronts,
SNe, superbubbles, and chimneys.  About 80\% of all SNe occur in
superbubbles, so cosmic rays generated by SNe also may be accelerated
in superbubbles \citep[e.g.][]{Higdon98, Parizot04}.  X-ray
observations probe different energetic components than traditional
visual and infrared (IR) studies, penetrating the obscuring material of
the natal cloud while minimizing confusion from foreground and
background objects.

X-ray studies also detect the presence of past SNe through the shocks
in their extended remnants.  Some superbubbles are well-known bright
X-ray sources, where multiple SNe from past OB stars produce soft
X-rays filling bubbles 50--100~pc in size with $L_x \sim
10^{35-36}$~ergs~s$^{-1}$ \citep{Chu90}, while others (not recently
brightened by such ``cavity'' SNe) are X-ray faint \citep{Chu95}.
MSFRs thus exhibit a complicated mixture of point and extended
structures that are easily confused by low-resolution X-ray
telescopes.  30~Doradus (30~Dor) is a classic example of this.  The
{\em Chandra X-ray Observatory} \citep[Figure~\ref{fig:csmoothimage}
and][hereafter Paper~II]{Townsley05b} and {\em XMM-Newton}
\citep{Dennerl01} share the highest-quality views yet achieved of this
extraordinary and complicated X-ray field.

30~Dor is a field of superlatives.  Located in the Large Magellanic
Cloud (LMC), it is the most luminous Giant Extragalactic H{\sc II}
Region (GEHR) and ``starburst cluster'' in the Local Group.  It hosts
the massive compact star cluster R136 \citep[][called ``RMC~136'' in
SIMBAD]{Feast60}, a testbed for understanding recent and ongoing star
formation in the 30~Dor complex \citep{Walborn02}.  Although R136, with
a mass of $\sim 6 \times 10^4$~M$_{\odot}$ \citep{Brandl05}, does not
quite qualify as a ``super star cluster'' \citep{Meurer95}, it has more
than 50 times the ionizing radiation of the Orion Nebula.  Nearby, one
finds the large supernova remnant (SNR) N157B (30~Dor~B), embedded in
an H{\sc II} region generated by the OB association LH~99 \citep{Chu92}
and containing an X-ray pulsar \citep{Marshall98}.  {\em ROSAT} images
showed two fainter sources associated with WR stars near the central
star cluster, at least five plasma-filled superbubbles, and the N157B
SNR \citep{Wang95, Wang99}.  30~Dor's proximity and complexity provide
us with a unique and important microscope into the starburst phenomenon
in galaxies.

Figure~\ref{fig:csmoothimage} illustrates the spatial and spectral
complexity of the X-ray emission from 30~Dor with a smoothed image of
our {\em Chandra} observation obtained with the Advanced CCD Imaging
Spectrometer (ACIS).  The aimpoint of the $17\arcmin \times 17\arcmin$
ACIS Imaging Array (ACIS-I) was the R136 cluster.  At $D=50$~kpc, the
distance we assume throughout this paper and Paper~II, $1^{\prime} \sim
14.5$~pc, so ACIS-I covers $\sim 240$~pc $\times$ 240~pc.  Two off-axis
CCDs from the ACIS Spectroscopy Array (ACIS-S) were operated for this
observation, adding a $\sim 120$~pc $\times$ 240~pc image of nearby
interesting structures (see Figure~\ref{fig:introimage}), albeit with
poor spatial resolution.  A more finely binned image of the ACIS-I
data, highlighting the X-ray point sources and showing a J2000
coordinate grid, can be found in Figure~1 of Paper~II.  

30~Dor lies at the confluence of two LMC supergiant shells and just the
main nebula is $\sim 250$~pc in diameter.  Since the LMC is viewed
nearly face-on, confusion between 30~Dor and other disk structures is
minimized.  SNe pervade the region but most go undetected due to age
and environment \citep{Chu90}.  Nearby are two pulsars (B0540$-$69.3 in
the SNR N158A and J0537$-$6910 in the SNR N157B) and SN1987A
(Figure~\ref{fig:introimage}).  Wide-field ground-based H$\alpha$
images \citep[e.g.][]{Meaburn84,Chu90,Wang99}, enhanced by recent {\em
HST} \citep{Walborn02} and {\em Spitzer}
data\footnote{\url{http://www.spitzer.caltech.edu/Media/releases/ssc2004-01/release.shtml}},
show that the combined actions of stellar winds and SNe from several
generations of high-mass stars in 30~Dor have carved its ISM into an
amazing display of arcs, shells, pillars, voids, and bubbles, ranging
over spatial scales of 1--100~pc \citep{Brandl05}. 

This field provides a unique bridge between Galactic giant H{\sc II}
regions (e.g.\ W49A, W51A, NGC~3603) and GEHRs \citep[e.g.\ NGC~5471
and other regions in M101,][]{Chen05}.  30~Dor is the result of
multiple epochs of high-mass star formation in a vast molecular cloud
complex chemically enriched by many SNe; it has no Galactic counterpart
in terms of mass or complexity of its star formation history and is not
even equalled by other GEHRs in the Local Group.  It provides us with a
unique view of the most fundamental building block of the starburst
phenomenon in galaxies; currently we are not able to resolve comparable
star-forming complexes in starburst galaxies, only those many times
larger and more powerful \citep[e.g.\ NGC~253,
NGC~7714/7715;][]{Strickland02,Smith05}.

In \S\ref{sec:background} we provide a brief summary of X-ray
observations of the 30~Dor complex.  Our {\em Chandra} observations and
data analysis are outlined in \S\ref{sec:observation}.  An overview of
the X-ray morphology and global spectral properties of the complex is
presented in \S\ref{sec:global}.  The rest of the paper provides more
in-depth studies of specific diffuse X-ray components of the 30~Dor
complex:  the superbubbles centered on R136 (\S\ref{sec:superbubbles}),
the SNR N157B (\S\ref{sec:N157B}), and other diffuse structures, with a
comparison to the H$\alpha$ kinematic study of 30~Dor by \citet{Chu94}
(\S\ref{sec:otherdiffuse}).  We conclude with a comparison of the X-ray
data to recent H$\alpha$ and IR data and a summary of our results.  The
X-ray point source population of 30~Dor revealed by this dataset is
described in Paper~II.

%==========================================================================
%==========================================================================
\section{PAST X-RAY OBSERVATIONS
\label{sec:background}}

\subsection{The Large Magellanic Cloud and 30~Doradus}

X-rays were first detected from the LMC in 1968 using rocket-launched
proportional counters \citep{Mark69}.  An early survey of the LMC with
the {\em Einstein Observatory} \citep{Long81} found 75 discrete sources
associated with the galaxy, including 25 SNRs and about 25 other
extended sources.  A re-analysis of the {\em Einstein} LMC survey
resulted in a point source catalog of 105 sources, ``roughly half''
associated with the LMC \citep{Wang91c}; the other sources are
foreground stars or background active galactic nuclei (AGN).  These
authors noted source confusion associated with the 30~Dor complex.

Detailed studies of 30~Dor and other LMC superbubbles were performed
with {\em Einstein} \citep[e.g.][]{Chu90,Wang91a,Wang91b}; they
concluded that the bright, diffuse X-ray emission filling the 30~Dor
superbubbles was probably created by off-center cavity SNe inside the
superbubbles shocking the superbubble shells.  Extending the work of
\citet{Meaburn84}, \citet{Chu94} performed a detailed echelle study of
the kinematics of 30~Dor and compared their findings to the {\em
Einstein} data, concluding that 30~Dor is dominated by a hierarchy of
expanding structures with spatial scales ranging over 1--100~pc and
expansion velocities of 20--200~km~s$^{-1}$.  The diffuse X-ray sources
are coincident with these expanding shells, with the brightest X-ray
emission coincident with high-velocity features.  \citet{Chu94} argued
that both the high-velocity features and the X-ray emission are related
to high-velocity shocks created by fast stellar winds and SNe.

Two catalogs of LMC point sources emerged from the {\em ROSAT}
Observatory; a few sources (often SNRs or SNR candidates) are
associated with the 30~Dor complex \citep{Haberl99,Sasaki00}.  In
mosaicked {\em ROSAT} maps of the hot ISM in the LMC, 30~Dor is an
outstanding feature---hotter, brighter, and at higher pressure than
most other parts of the galaxy \citep{Snowden94,Points01,Sasaki02}.
{\em ROSAT} studies of individual LMC superbubbles showed that several
are X-ray faint \citep{Chu95}, consistent with earlier conclusions that
superbubbles are only brightened in X-rays when SNe hit the shell
walls, while others exhibit ``breakout regions'' and may be venting hot
gas into the LMC's ISM \citep{Dunne01}.

A {\em ROSAT}/PSPC study of 30~Dor reported 22 point sources within the
30~Dor nebula, including sources associated with R136 and N157B
\citep{Norci95}.  Spectral fits to the diffuse emission yielded $N_H
\sim 8 \times 10^{21}$~cm$^{-2}$ and $kT \sim 0.4$~keV, with a
luminosity (0.1--2.4~keV) of $L_X = $3--6$ \times
10^{37}$~ergs~s$^{-1}$ depending on choice of background.  Since the
Galactic absorbing column toward 30~Dor is $7 \times
10^{20}$~cm$^{-2}$, most of the absorption found in their spectral fits
is local to 30~Dor.  They confirm the {\em Einstein} results
\citep{Wang91a} that the diffuse X-ray emission fills the voids in the
$H\alpha$ emission and further conclude that stars in 30~Dor contribute
$\leq 2$\% of its total X-ray luminosity.  While Norci and {\" O}gelman
note that the central cluster R136 is too young to have produced X-ray
binaries that could contribute to the X-ray emission in the region, a
{\em ROSAT}/HRI study suggested that the two point sources found by HRI
were associated with the WR stars R140a2 (a WN6 binary) and Melnick 34
(WN4.5) and could be WR/black hole binaries \citep{Wang95}.  This
conjecture was supported by {\em ASCA} spectral results that indicated
hard X-ray emission from the central regions of 30~Dor.  Further {\em
ROSAT}/HRI and {\em ASCA} observations led \citet{Wang99} to conclude
that the diffuse emission came from 2--9~MK thermal plasmas and that
the primary mechanism to explain the temperature and X-ray luminosity
of this emission is mass-loading via mixing of the hot wind-generated
gas with H{\sc II} gas from the large number of ionization fronts in
the region \citep{Scowen98}.

The first light image from the {\em XMM-Newton Observatory} was
obtained from a 106~ks pointing just west of the main 30~Dor nebula
\citep{Dennerl01}.  This observation yielded broad-band (0.1--10~keV)
spectra for several important objects in the neighborhood of 30~Dor:
N157B, the Honeycomb Nebula \citep[a SNR,][]{Chu95b}, and faint diffuse
emission northwest of 30~Dor \citep{Dennerl01}, as well as several
likely AGN seen through the LMC disk \citep{Haberl01}.  A 38~ks {\em
XMM} observation of N157B (Observation 0113020201, PI Aschenbach)
images even more of the main 30~Dor nebula; although it is beyond the
scope of this effort to include that observation in this paper, future
efforts to obtain spectra of various features in 30~Dor from this
dataset would be worthwhile.

%==========================================================================
\subsection{N157B and PSR~J0537$-$6910}
\label{sec:n157bbkgd}

The N157B SNR has itself been the subject of many X-ray studies.  A
known radio source, it was first detected in X-rays by {\em Einstein}
\citep{Long79}.  A later {\em Einstein} study \citep{Clark82}
demonstrated that its spectrum showed no strong emission lines and was
non-thermal, well-fit by a power law with slope $\Gamma = 2.9$, $N_H =
1.2 \times 10^{21}$~cm$^{-2}$, and luminosity (0.5--4.5~keV) of $L_X =
2.3 \times 10^{36}$~ergs~s$^{-1}$; they suggested that a search for a
pulsar in N157B should commence.  \citet{Chu92} determined the boundary
of the SNR based on kinematic information and suggested that N157B is
another example of an off-center SNR expanding into a pre-existing
superbubble and that the SNR may be interacting with a dark cloud on
its southern edge; this dark cloud may shadow the soft X-ray emission
from the SNR.

A detailed {\em ROSAT} and {\em ASCA} study of N157B was presented by
\citet{Wang98a}.  Their spectrum is dominated by a non-thermal
component with power law slope $\Gamma \sim 2.5$, although there are
hints of emission lines below 2~keV that they attributed to a thermal
plasma from the SNR with $kT = $0.4--0.7~keV.  {\em ROSAT}/HRI revealed
a comet-shaped central nebula that dominates the X-ray emission,
explained as a synchrotron nebula from a pulsar moving through the ISM
at a rate of $\sim 1000$~km~s$^{-1}$, although no pulsed emission was
apparent in these data. 

The long-awaited detection of PSR~J0537$-$6910 in N157B, with a period
of 16~msec and a power law photon index of $\Gamma = 2.6$, came
serendipitously from an {\em RXTE} observation of SN1987A
\citep{Marshall98} and was soon confirmed by {\em BeppoSAX}
\citep{Cusumano98}.  A slightly later {\em ROSAT}/HRI detection
provided an arcsecond position for the pulsar \citep{Wang98b}.  This is
the shortest period known for a pulsar associated with a SNR.  This
pulsar remains undetected in the radio \citep{Crawford05} and no
unambiguous optical counterpart has been found, even with deep {\em
HST}/ACS imaging \citep{Mignani05}.

The thermal emission component in N157B's spectrum is confirmed by {\em
XMM} \citep{Dennerl01}, also confirming N157B as a composite SNR,
containing a Crab-like synchrotron core with $\Gamma = 2.83$
\citep{Dennerl01} surrounded by a thermal shell.  {\em Chandra} has
observed N157B directly with both the HRC \citep{Wang01} and ACIS-S
using subarray readout (ObsID 2783, PI Wang).  With the HRC data,
\citet{Wang01} were able to pinpoint the position of the pulsar to
$<1\arcsec$ and to resolve a compact pulsar wind nebula at the ``head''
of the comet-shaped nebula seen by {\em ROSAT} \citep{Wang98a}.  This
$\sim 2\arcsec \times 7\arcsec$ feature is centered on the pulsar and
oriented perpendicular to the larger cometary nebula.  \citet{Wang01}
explain this feature as a toroidal pulsar wind nebula confined by the
bow shock from the supersonic motion of the pulsar and depositing its
energy in the trailing cometary nebula.

%==========================================================================
\subsection{Past Work on This {\em Chandra} Observation}
	
Our short {\em Chandra}/ACIS-I observation of 30~Dor (described in
detail below) was one of the first observations made by {\em Chandra}
and was part of the ACIS Team's Guaranteed Time Observation program.  We
have reported preliminary results from this observation at several
meetings \citep{Townsley99,Townsley00b,Townsley02} and have used it to
estimate the position of PSR~J0537$-$6910 in the N157B SNR
\citep{Mignani05}.  With this study we finally are able to provide a
detailed analysis of this observation, after several years' worth of
code development and special analysis techniques were implemented.  Our
custom tools are also useful for analysis of other {\em Chandra}
fields, so we have made them publicly available (see
\S\ref{sec:observation}).

This dataset has also been studied by other groups.  Recently,
\citet{Lazendic03} proposed two new SNR candidates in the 30~Dor nebula
based on radio continuum maps obtained with the {\em Australia
Telescope Compact Array}, although the {\em Chandra} data showed no
X-ray enhancements associated with these regions.  \citet{Chu04}
examined this claim and concluded, based in part on the absence of
associated X-ray emission, that these small radio sources are not SNRs;
rather one is likely a young star-forming region and the other a
molecular cloud.

\citet{PPL02} found 20 mostly pointlike sources in the inner $11\arcmin
\times 11\arcmin$ of the ACIS-I field of view and studied 11 of those
in the central region, within $100\arcsec$ from the core of R136.  They
matched these sources with known early-type (WR and O) stars, some
known to be binaries, and concluded that they are bright in X-rays
because they are all probably colliding-wind binaries.
\citet{Stevens03} included a spectral analysis of unresolved X-ray
emission around R136 in their study of the properties of super star
clusters.  Some unspecified number of point sources was removed from
the inner $16\arcsec$ of R136, then the remaining X-ray emission was
fit with a thermal plasma model, resulting in fit parameters of $N_H =
4.2 \times 10^{21}$~cm$^{-2}$, $kT = 2.1$~keV, and absorption-corrected
luminosity $L_X = 5.5 \times 10^{34}$~ergs~s$^{-1}$.  In Paper II we
show that more X-ray point sources in R136 are discernable in this
dataset than these two papers considered; their results would change
somewhat if these faint sources were included in their analyses.

Since our observation included the two off-axis ACIS-S chips S3 and S4,
we were able to image an $8\arcmin.5 \times 17\arcmin$ region to the
southwest of 30~Dor, featuring SN1987A, the large superbubble 30~Dor~C,
and the Honeycomb Nebula (see Figures~\ref{fig:csmoothimage} and
\ref{fig:introimage}).  As mentioned above, the {\em XMM-Newton}
first-light image was centered on this field and gives much better
spatial and spectral information on these structures than this short
ACIS exposure \citep{Dennerl01}.  The X-ray brightening of SN1987A has
been monitored by {\em Chandra} since early in the mission; see
\citet{Park05} and references therein.  30~Dor~C, an unusual
non-thermal X-ray emitter, has been studied with {\em Chandra}
\citep{Bamba04} and {\em XMM} \citep{Smith04}.  Given these better
datasets and existing studies, we will not include detailed analysis of
these sources in this paper.

The results we present below represent only one step in the
high-resolution characterization of X-rays from the 30~Dor region.  A
100~ks {\em Chandra} ACIS-I observation of 30~Dor is scheduled for
later this year; it should give a $\sim 5$-fold improvement in
sensitivity and signal over the results reported here.

%==========================================================================
%==========================================================================
\section{{\em CHANDRA} OBSERVATIONS AND ANALYSIS
\label{sec:observation}}

\subsection{Instrumental Set-up}

We observed 30~Dor and its environs for $\sim 26$~ks with ACIS-I on
1999 September 21--22.  R136 was placed at the aimpoint.  The focal
plane temperature at this early point in the mission was $-110$C to
facilitate water outgassing; it was later reduced to $-120$C to improve
detector performance.  This higher focal plane temperature causes the
effects of radiation damage to the frontside-illuminated CCDs in the
ACIS-I array to be particularly pronounced \citep{Townsley02a}.  For
example, the spectral resolution at 1.5~keV varies across the device
from $\sim 75$~eV FWHM at low CCD row numbers to $\sim 170$~eV at high
row numbers.  Unfortunately, the aimpoint of the ACIS-I array falls
near the top of the I3 CCD, so spectral resolution for the sources in
R136 is particularly bad.  A summary of the observations is given in
Table~\ref{tbl:obslog}.  

The first $\sim 1$~ks of the observation (Observation Identification or
``ObsID'' 22) was made in the usual ``timed event, faint'' mode, with
$3 \times 3$ pixel events, 3.2~s CCD frames, and the Spectroscopy Array
(ACIS-S) CCDs S2 and S3 operating in addition to the ACIS-I devices.
The remainder of the observation (ObsID 62520) was taken in ``timed
event, very faint'' mode ($5 \times 5$ pixel events) using the
``alternating'' readout mode with a 10:1 cycle; every ten CCD frames
were exposed for 3.3~s, then every eleventh frame had a short 0.3~s
exposure.  The data processing pipeline divides these short and long
frames from ObsID 62520 into separate event lists.  In order to image
interesting structures near the 30~Dor field, ACIS-S CCDs S3 and S4
were operational for this observation.

We chose to use the ACIS camera's alternating-exposure mode because the
30~Dor field contains PSR~J0537$-$6910, a bright 16~ms pulsar
\citep{Marshall98}.  Although it is imaged $\sim 7\arcmin$ off-axis, we
knew that it would still suffer from photon pile-up in regular-length
CCD frames, so we included the short frames to try to mitigate pile-up
on this source.  In hindsight, this was not an ideal approach, as there
are too few photons in the short frames to provide much information on
PSR~J0537$-$6910 and the use of alternating mode this early in the
mission caused problems for the data processing pipelines, resulting in
absent or incorrect Good Time Interval (GTI) tables, incorrect keyword
values, and other complications.

For this analysis, we used Version 3 of the pipeline processing
(created 26 July 2001), which has correct GTI tables.  Even this third
version of the data still contains some errors though; for example, the
EXPOSURE and LIVETIME keywords in the 0.3~s event list of ObsID 62520
are incorrect.  After recalculating these quantities for each CCD and
correcting the header keywords in the event list, a total of $\sim
200$~s of data taken with 0.3~s CCD frames is obtained.

%==========================================================================
\subsection{Data Analysis}

Starting with the Level 1 event lists, the data were reduced using our
standard methods, as described in \citet{Townsley03} and
\citet{Getman05}, using {\it CIAO} 3.0.2.  {\em Chandra} source
positions were determined from the source-finding routine {\it
wavdetect} \citep{Freeman02} run on a $2 \times 2$-pixel binned image
of the hard band (2--7~keV) data from the ``long'' observations in
ObsID 62520; data below 2~keV were excluded to minimize the
contamination of source positions by the soft diffuse emission that
pervades the field.  ObsID 62520 was registered to the astrometric
reference frame of 2MASS using 4 matches.  Offsets were $+0.95\arcsec$
in RA, $+1.0\arcsec$ in Dec.  

The {\em Clean55}
algorithm\footnote{\url{http://hea-www.harvard.edu/\-~alexey/\-vf\_bg/vfbg.html}}
developed by Alexey Vikhlinin was used to flag events in ObsID 62520;
these flagged events were removed to suppress background for
source-finding and image generation, but this flag can also erroneously
remove events from the cores of point sources so it was not applied in
spectral analysis.  Due to a strong background flare caused by
particles generated by solar activity, ObsID 62520 has some telemetry
saturation that only affects the S3 chip, from frame $\sim 7248$ ($\sim
24300$~sec) to the end of the observation.  Background flares affected
CCD S3 both early and late in the exposure and those times were
filtered from the S3 data; other CCDs were unaffected by these flares.
The {\it CIAO} ``destreaking'' algorithm by John
Houck\footnote{\url{http://asc.harvard.edu/ciao/threads/destreak/}} was
applied to CCD8 to suppress noise on that device.
  
We did not adjust the astrometry of ObsID 22 due to only two 2MASS
matches and inconsistent, small offsets with those matches.  Although
we would normally not remove the event position randomization added in
the standard data processing pipeline for such a short observation, it
was removed for ObsID 22 in order to combine these data with ObsID
62520.  S3 was experiencing background flares for the duration of this
short observation.  Given that and the short amount of total
integration time for the area covered by CCD S2, only the ACIS-I array
component of this observation was merged with ObsID 62520.

For ObsID 62520, individual CCD columns were examined in chip
coordinates using our {\it Event Browser} tool\footnote{Code available
at \url{http://www.astro.psu.edu/xray/docs/TARA/}} \citep{Broos00} to
search for additional hot pixels and columns that were not part of the
pre-defined bad pixel file and to establish the energies of those noise
events.  At $-110$C, there are many more hot columns (exhibiting extra
noise below 1~keV) than at $-120$C (the focal plane temperature used
from January 2000 to the present), especially on CCD S4.  Those events
were removed from both ObsID 62520 and ObsID 22 using CIAO 2.3 because
the necessary ``exclude'' syntax (to remove only certain energies from
specified spatial coordinates) was not working correctly in CIAO
3.0.2.  Although the effects of these deleted events cannot be included
in the exposure map (because only events below certain energies, not
all events from a given column, were removed), we chose to suffer this
slight calibration error rather than to retain the noisy events because
they become especially pronounced and distracting in smoothed soft-band
images.

Both datasets were corrected for charge transfer inefficiency (CTI)
using the Penn State CTI corrector, which was originally developed
specifically to address the CTI problem for this
dataset\footnote{Details, source code, and calibration products for CTI
correction are available at
\url{http://www.astro.psu.edu/users/townsley/cti}.} \citep{Townsley00,
Townsley02a}, due to our interest in spectral analysis of the many
pointlike and diffuse structures in this field.  It remains the only
CTI corrector available for the backside-illuminated CCD S3 and for all
ACIS data taken at a focal plane temperature of $-110$C (where CTI is
worse than at $-120$C for frontside-illuminated CCDs, as mentioned
above).  Custom response matrices (RMFs) and quantum efficiency
uniformity (QEU) files were generated to facilitate spectral fitting
for $-110$C data \citep{Townsley02b}.  We also refined the event
positions in both datasets using the subpixel positioning algorithm
developed by Koji Mori\footnote{Code available at
\url{http://www.astro.psu.edu/users/mori/chandra/subpixel\_resolution.html}}
\citep{Mori01, Tsunemi01}.

The reduced event lists were then merged and the exposure maps added.
A {\em status=0} filter was applied to create an event list appropriate
for source searching and image generation.  The data were once again
examined in {\it Event Browser}, showing that the best energy range to
consider for imaging studies is 350--7300~eV; the lower limit retains
events below 500 eV that are clearly from the diffuse superbubble
structures in the field while limiting low-energy noise, while the
upper limit is set just below the instrumental Ni~K$\alpha$ line, which
marks the beginning of enhanced high-energy background.

Next we conducted a multistage process for identifying and extracting
photons of unresolved sources that is described in detail in Paper II.
It involves source identification using a wavelet transform and a
maximum likelihood image restoration algorithm.  Point source events
for the final list of 180 sources were then extracted using PSB's
IDL-based code {\it ACIS Extract} \citep{Broos02} developed to treat
complicated fields with multiple point-like and diffuse
components\footnote{The {\it ACIS Extract} code is available at
\url{http://www.astro.psu.edu/xray/docs/TARA/ae\_users\_guide.html}.}.
For analyzing the diffuse components in the 30~Dor field, the only
point source properties required are their locations and extraction
regions; these are provided in Tables 1 and 2 of Paper II.  These are
used to construct a point source mask and a source-free dataset
suitable for analysis of diffuse emission.

We used a mask produced by the {\it ACIS Extract} tool {\it
ae\_optimal\_masking} which identifies regions where the expected
surface brightness from the calibrated point sources is larger than one
half the smoothed observed local background.  These point source masks
are illustrated in Figure~\ref{fig:mask_emap}, which shows the I-array
exposure map for ObsID 62520.  Chip gaps and bad columns are seen as
areas of reduced exposure, while the small white regions distributed
across the field and concentrated in the center show areas of zero
exposure representing areas that were masked to exclude point sources.
This figure also illustrates that the {\em Chandra} PSF is a strong
function of off-axis angle; most of the X-ray point sources in the field
are concentrated in R136 (field center), yet the sharp PSF on-axis leads
to relatively little area removed for diffuse emission studies there.

Extraction regions for a variety of diffuse structures were then
defined by hand, based on the apparent surface brightness in a
smoothed, sources-removed image.  Spectra, backgrounds, ARFs, and RMFs
for all point sources and diffuse structures were constructed by {\it
ACIS Extract} and used for both automated and interactive spectral
fitting using Keith Arnaud's {\it XSPEC} fitting package
\citep{Arnaud96} and {\it XSPEC} scripts provided by Konstantin Getman.

%==========================================================================
%==========================================================================
\section{GLOBAL PROPERTIES OF THE X-RAY EMISSION
\label{sec:global}}

We start by describing the large-scale features revealed by this ACIS
observation.  Figure~\ref{fig:csmoothimage} shows the soft-band image
of 30~Dor created with the adaptive-kernel smoothing tool {\it csmooth}
in {\it CIAO} (a translation of the {\it asmooth} code by Harald
Ebeling).  This smoothed image emphasizes the soft diffuse structures,
mapping red to the spectral range 500--700~eV, green to 700--1120~eV,
and blue to 1120--2320~eV.  The reduced, full-field, full-band binned
data are shown in Figure~\ref{fig:introimage}.  The event data were
binned into $8\arcsec \times 8\arcsec$ pixels to create this image.
Some well-known objects are labeled, as are the ACIS CCDs used in this
observation.  Areas of reduced exposure between the CCDs are clearly
visible, as is the extensive diffuse X-ray emission present in this
field.  Several famous structures are imaged far off-axis on the S3 and
S4 CCDs.

%==========================================================================
\subsection{X-ray Morphology
\label{sec:globalmorph}}

In these images we see a bright concentration of hard X-ray events
associated with the R136 star cluster at the center of the ACIS-I array
and fainter emission from several other well-studied stellar objects,
mostly WR stars (see Figure 1 in Paper II).  The bright SNR N157B is a
distinct, extended feature to the southwest of the main 30~Dor
complex.  It is the hardest source in the field due to non-thermal
emission from its central pulsar and synchrotron nebula, but the larger
SNR softens with distance from the pulsar.  A number of
widely-distributed compact X-ray sources are scattered across the
field; some were seen in previous X-ray studies
(\S\ref{sec:background}).  The hard sources are probably mostly
background AGN (see Paper II).

By far the dominant features, however, are the large diffuse structures
associated with the superbubbles produced by existing and past OB
associations and their supernova events.  These structures were
well-known from {\em Einstein} and {\em ROSAT} studies, as outlined in
\S\ref{sec:background} \citep[see especially the HRI data
in][]{Wang99}, but  higher on-axis spatial resolution of {\em Chandra}
and the ACIS camera's intrinsic spectral capabilities reveal a new
level of complexity.  The center of the field is complicated by
crossing ACIS-I chip gaps but clearly contains some of the hardest
emission in the field, due to the R136 cluster, R140, and R145.  Faint,
softer diffuse emission appears to pervade the central region.  Just
northwest of R140 (due west of R139) is a bright, medium-soft clump
$\sim 1\arcmin$ in size.

Shell 1 in Figure~\ref{fig:introimage} contains several point sources
but is dominated by a bright, medium-energy clump of diffuse emission
roughly $1\arcmin.5 \times 2\arcmin$ in size.  An ACIS-I chip gap
bifurcates this feature, so the two-lobed structure it shows in
Figure~\ref{fig:csmoothimage} is largely an artifact \citep[see Figure
1 in][]{Wang99}.  Shell 2 is shown as a larger structure in
Figure~\ref{fig:introimage} than originally defined
\citep{Meaburn84,Wang91a} because the kinematic study of \citet{Chu94}
showed that the expanding shell is larger than the morphological Shell
2 identified in an H$\alpha$ image \citep{Meaburn84}.  It is
edge-brightened in X-rays, with a large central void and a sharp
boundary along its southeast edge.  There are small clumps of emission
around its periphery, many of which are consistent with point sources.
Its faint emission is harder than the other shell structures in the
30~Dor nebula.  At its upper center is an interesting comet-shaped
feature showing substantial spectral complexity, with a soft point
source at its head.

Shell 3 is also largely edge-brightened, quite faint and soft around
its periphery but with a harder, nearly-complete, clumpy central ring
about $2\arcmin.5$ in diameter (hereafter called the ``West Ring'').
This ring structure also shows spectral complexity, appearing harder on
its western half.  A faint, small clump is nearly centered on this
ring, but it is too large and diffuse to be a point source.  The $\sim
20$~Myr old stellar cluster Hodge~301 \citep{Grebel00} sits at the
northeast edge of this ring, but we do not resolve any X-ray point
sources in it.  Shell 3 encompasses two dark voids, one on either side
of its central ring.

Shells 4 and 5 are both distinctly soft and bright, and differ from the
other shells in that they appear center-filled rather than
edge-brightened.  Shell 5, unfortunately, has a chip gap running all
the way through its long dimension, so some of the structure we see may
be affected by that.  Both it and Shell 4 are spectrally uniform and
similar to each other.  Shell 5 is quite clumpy, with a bright lower
clump about $1\arcmin \times 2\arcmin$ in size containing the
WR star R144 and a string of smaller, fainter clumps populating
its interior.

The off-axis ACIS-S CCDs appear to be pervaded by soft, faint diffuse
emission.  (Small point-like features around the edges of this field
seen in Figure~\ref{fig:csmoothimage} are smoothing artifacts.)  The
most striking feature here is the superbubble 30~Dor~C, seen as a
nearly complete ring of clumpy emission spanning S3 and S4; its western
half is very hard in X-rays due to strong non-thermal emission (see
\S\ref{sec:background}).  SN1987A is the bright point source on S4; its
light is spread across a region more than an arcminute in size due to
{\em Chandra's} broad PSF at this large off-axis angle ($>20\arcmin$).
The bright, soft feature at the lower center of S4 is the Honeycomb
SNR; superposed on it appears to be a large ring of soft, diffuse
emission that arcs to the north and east of the Honeycomb.  Such faint,
far off-axis features are more clearly seen in the {\em XMM}
first-light image \citep{Dennerl01}.

%==========================================================================
\subsection{Other Views
\label{sec:diffsrcs}}

\begin{figure}
\centering
  \includegraphics[width=1.0\textwidth]{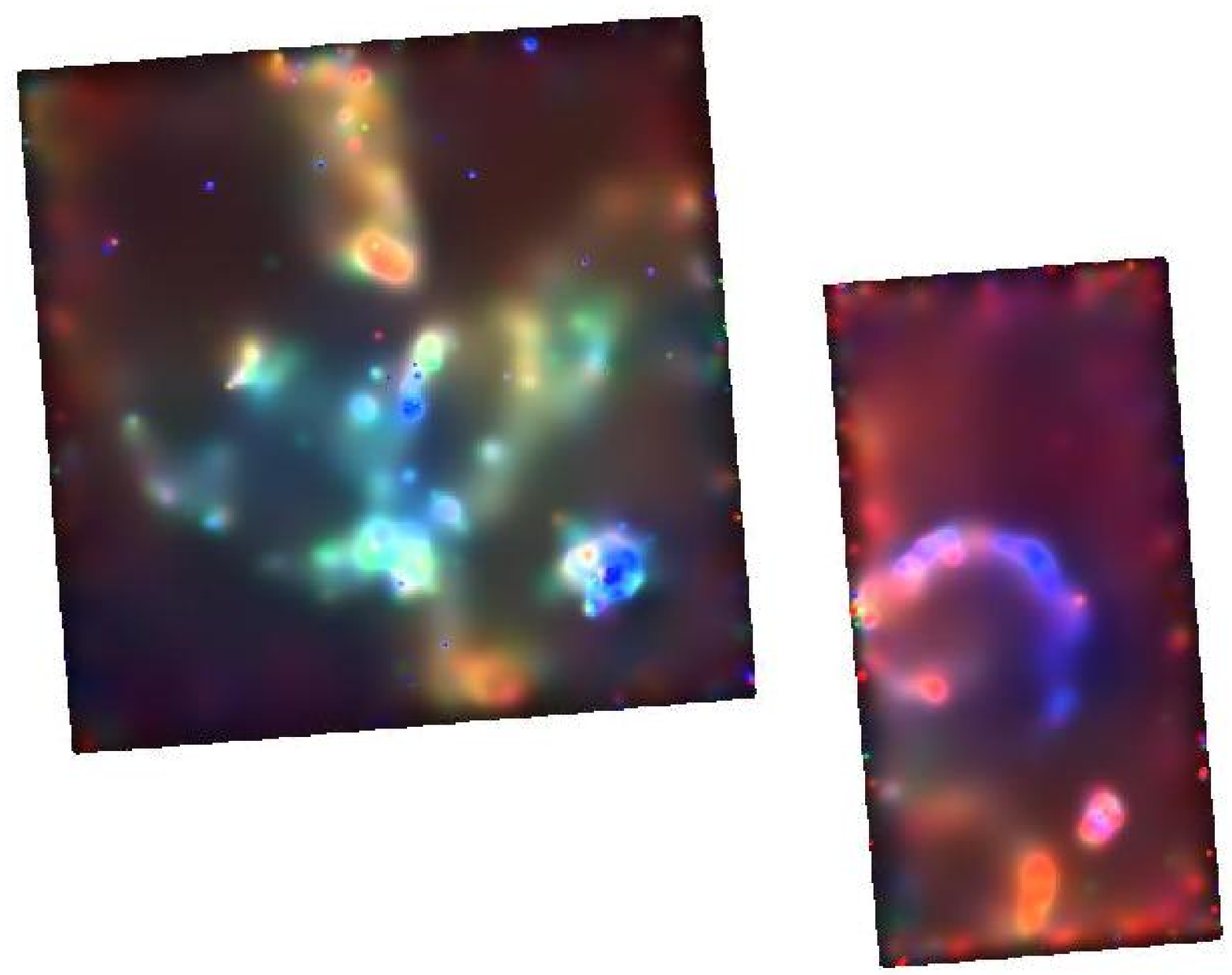}
\caption{ An adaptively smoothed image of our {\em Chandra}/ACIS
observation of 30~Doradus, centered on the massive stellar cluster R136
and partially covering a $\sim 24\arcmin \times 30\arcmin$ region
(roughly 350~pc $\times$ 430~pc at an assumed distance of $D =
50$~kpc).  This rendering emphasizes soft diffuse structures; red =
500--700~eV, green = 700--1120~eV, blue = 1120--2320~eV. 
\label{fig:csmoothimage}} 
\end{figure}

\begin{figure}
\centering
  \includegraphics[width=1.0\textwidth]{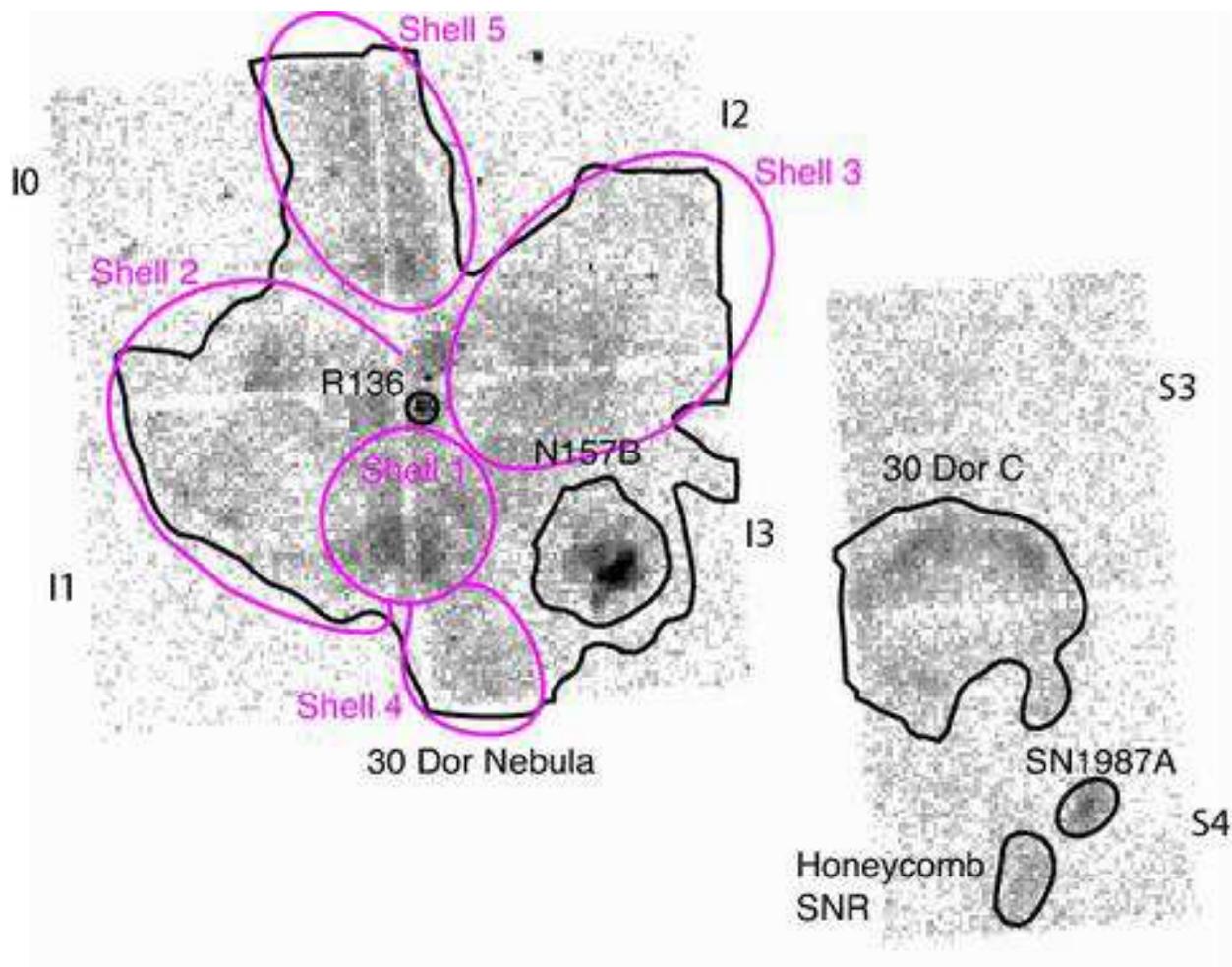}
\caption{ 30~Doradus in X-rays, from the 21,870-sec GTO1 {\em
Chandra}/ACIS observation (binned to $16 \times 16$ ACIS pixels or
$8\arcsec \times 8\arcsec$), with the ACIS-I array centered on the
massive stellar cluster R136.  Other well-known structures in the field
are labeled; the ``shells'' are the superbubbles sketched in
\citet{Wang91a} and \citet{Chu94} based on an H$\alpha$ image from
\citet{Meaburn84}.  The names of the six ACIS CCDs used in this
observation are shown; each CCD covers $\sim 8\arcmin.5 \times
8\arcmin.5$, or $\sim 120$~pc $\times$ 120~pc at $D = 50$~kpc.
\label{fig:introimage}} 
\end{figure}

\begin{figure}
\centering
  \includegraphics[width=1.0\textwidth]{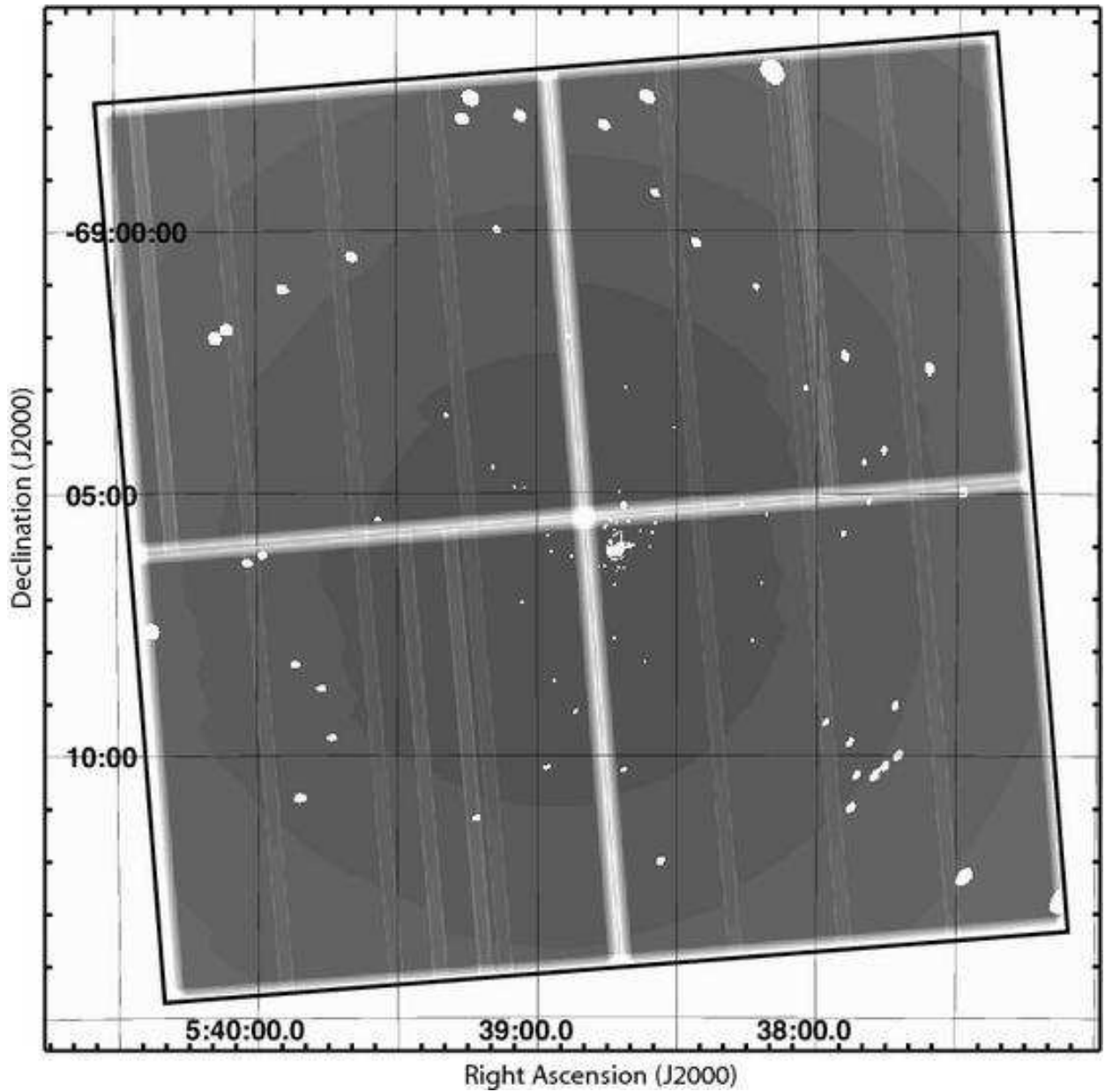}
\caption{ The exposure map for ObsID~62520 with point source masks applied,
displayed with log scaling (darker areas have higher exposure).
Gaps between the four ACIS-I CCDs and bad columns are seen as areas of
reduced exposure, while areas of zero exposure due to point source masks
are seen as white patches scattered across the field. 
\label{fig:mask_emap}} 
\end{figure}

\begin{figure}
\centering
  \includegraphics[width=0.9\textwidth]{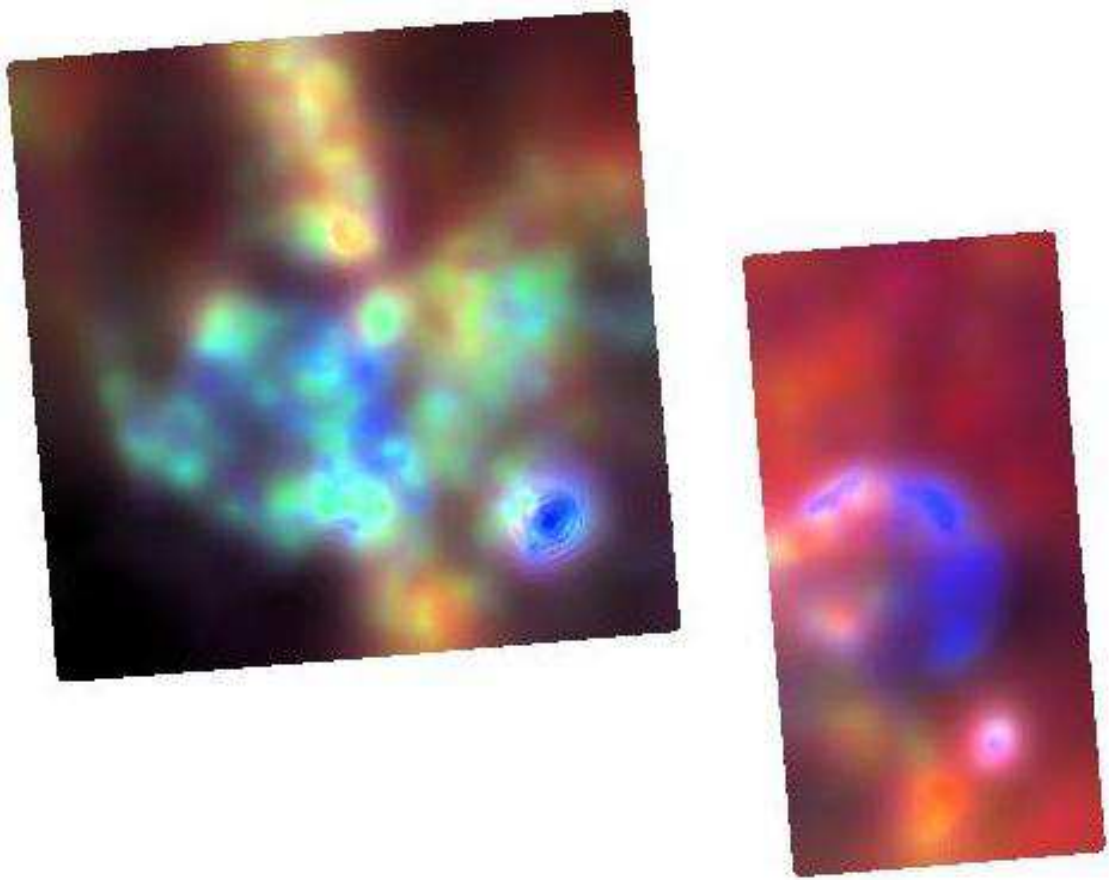}
\caption{ 30~Doradus diffuse structures: red = 350--700~eV, green =
700--1100~eV, blue = 1100--2200~eV.  This adaptively-smoothed image was
generated after 180 point sources were removed from the data and uses a
different smoothing tool than Figure~\ref{fig:csmoothimage}, for comparison.  
\label{fig:diffuseintro}} 
\end{figure}

Given the great variety of diffuse X-ray structures apparent in this
observation, it is worthwhile to find new ways to highlight them.
Figure~\ref{fig:diffuseintro} shows a color composite of three
adaptively-smoothed images \citep[from PSB's adaptive kernel smoothing
tool {\em adaptive\_density\_2d},][]{Townsley03} where the 180 point
sources in the field have been removed and the resulting ``holes''
smoothed over.  The images were made in three soft bands, 350--700~eV,
700--1100~eV, and 1100--2200~eV, similar to the bands used to create
Figure~\ref{fig:csmoothimage}.  They were combined to show the spectral
variety that must be present in the diffuse structures, even on spatial
scales too small for us to extract and fit spectra to demonstrate this
variety; the intensities were set to emphasize faint diffuse features.

This smoothing technique has the advantage (over that used in
Figure~\ref{fig:csmoothimage}) that it does not create artifacts around
the edges of the field, although the overall appearance is more
blurry.  We think that the best impression of the field comes from
actively comparing these two renderings.  For example, the West Ring in
Shell 3 is much more clearly portrayed by Figure~\ref{fig:csmoothimage},
but there are two small, faint loops ($< 2\arcmin$ in diameter) on its
north and east edges that are only visible in
Figure~\ref{fig:diffuseintro}.  We do not have enough data to know
whether or not these features are real.

With the point sources gone from Figure~\ref{fig:diffuseintro}, we are
free to concentrate on the diffuse structures.  Here the outer loop in
Shell 3 is more pronounced, as are other faint diffuse features,
especially on the S-array CCDs.  Another intriguing small loop not
apparent in Figure~\ref{fig:csmoothimage} is seen in the eastern half
of Shell 2, visible as a green arc $\sim 1.5\arcmin$ across just above
a bright green knot in the sharp lower edge of Shell 2.  Again, only
more data will allow us to understand this feature in detail.

%==========================================================================
\subsection{Spectra
\label{sec:globalspectra}}

\begin{figure}
\centering
\includegraphics[width=0.4\textwidth]{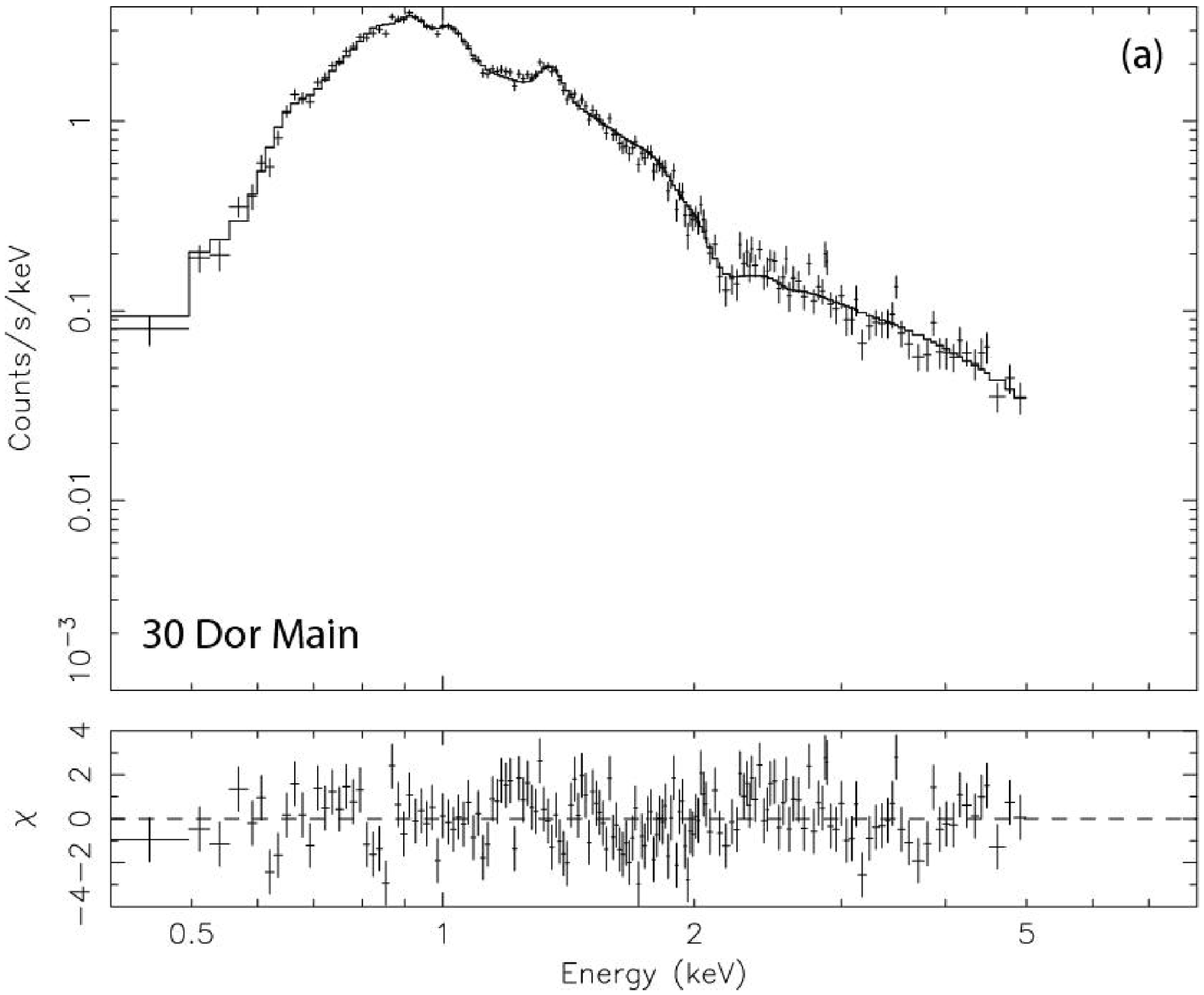}
\hspace*{0.15in}
\includegraphics[width=0.4\textwidth]{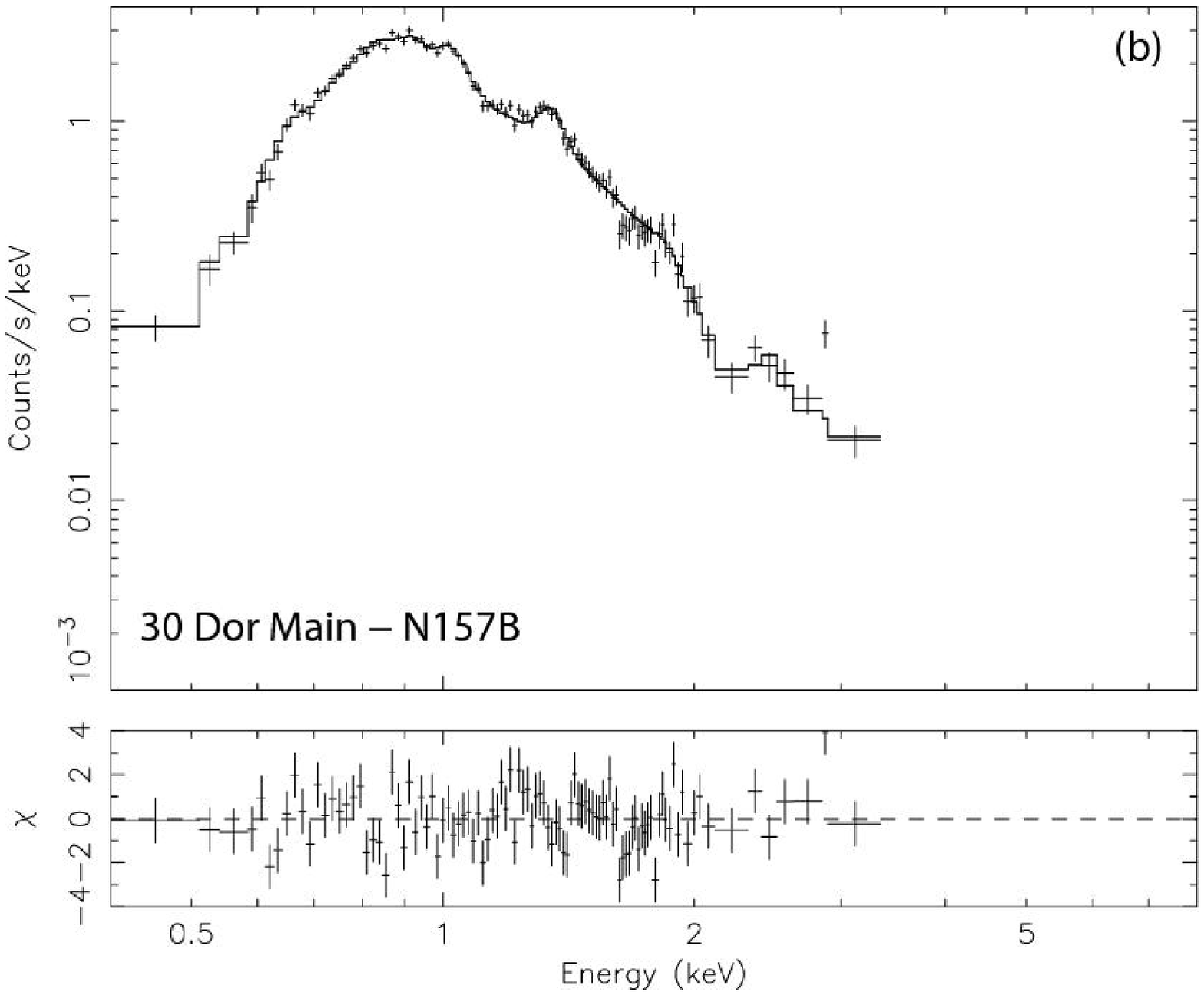}\\
\bigskip
\includegraphics[width=0.4\textwidth]{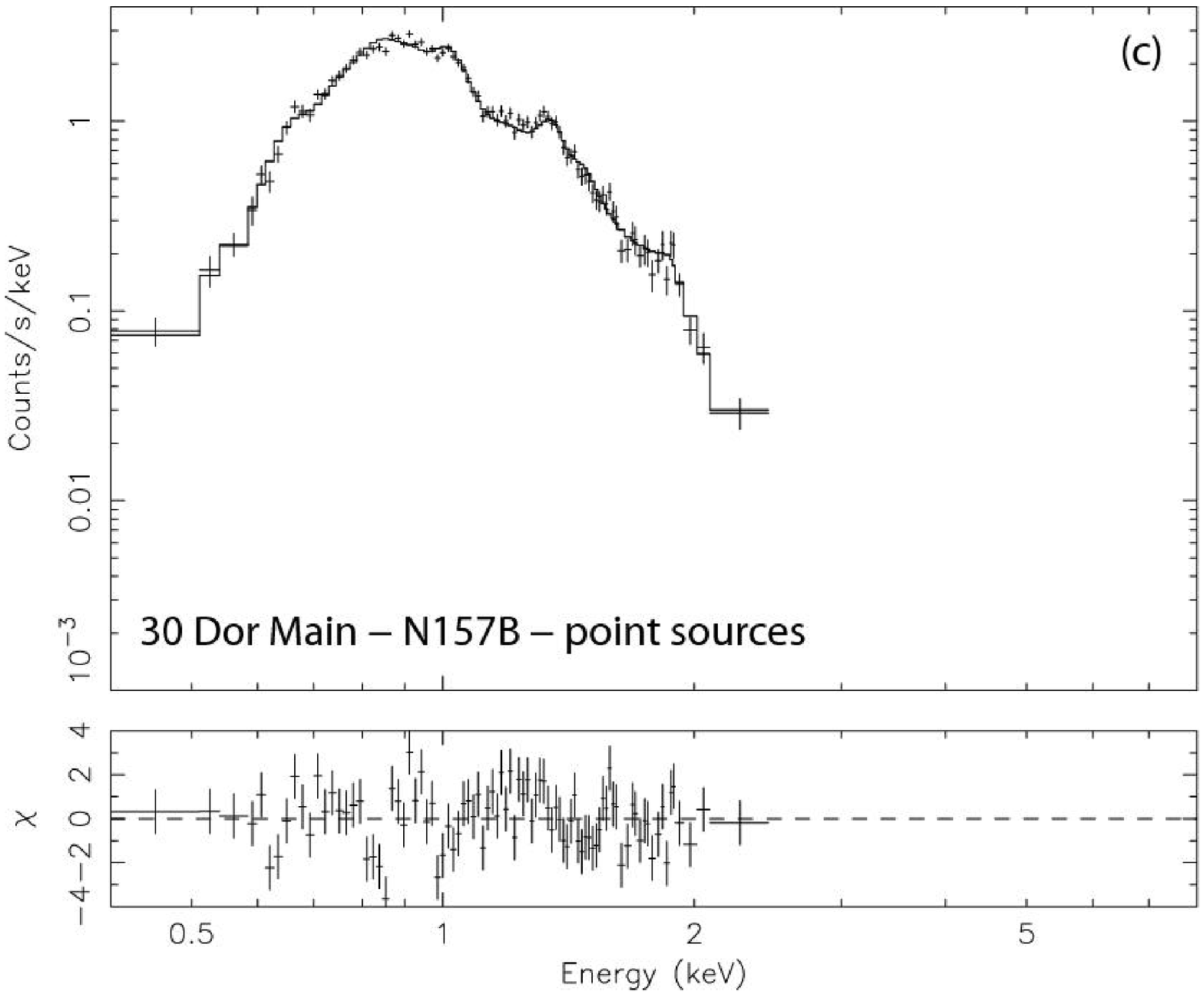}
\hspace*{0.15in}
\includegraphics[width=0.4\textwidth]{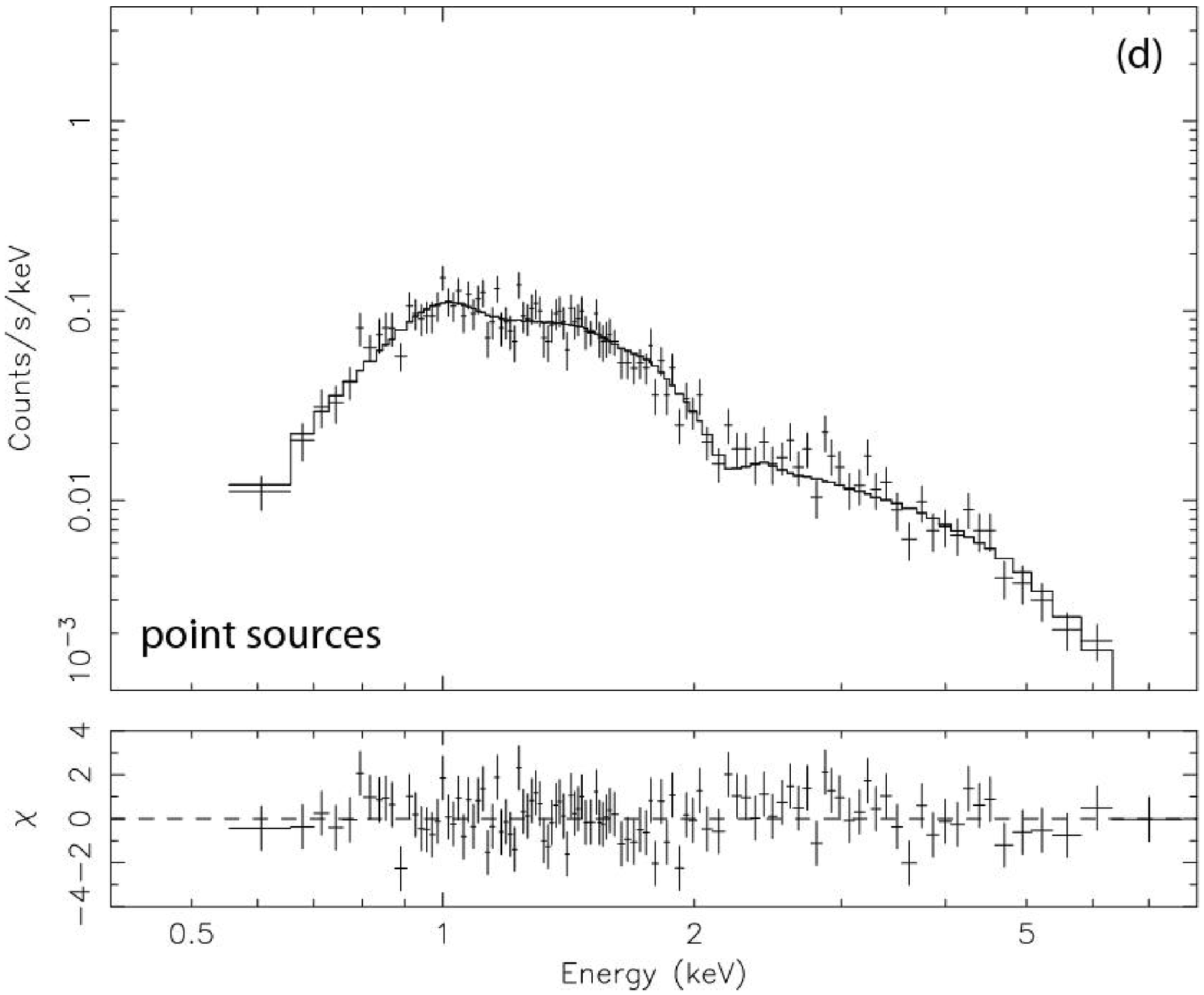}
\caption{Global spectra of the X-ray emission in 30~Dor.  For these and
all later spectral fits, the upper panels show the instrumental
spectra and best-fit models from {\it XSPEC} while the lower panels
show the residuals.  The fit parameters are given in Table~\ref{tbl:globalspectable}.
(a) The composite spectrum of the entire 30~Dor main nebula on the ACIS-I
array (as outlined in Figure~\ref{fig:introimage}), including all point
sources and the SNR N157B.
(b) The spectrum of the same region, including point sources but excluding
N157B (region shown in Figure~\ref{fig:introimage}) and the point sources
associated with it.
(c) The spectrum of the superbubbles alone; this uses the same extraction
region as (b) but the point sources it encloses have been removed.
(d) The composite spectrum of the point sources that were removed in (c).
\label{fig:globalspectra}} 
\end{figure}

For comparison with more distant GEHRs, Figure~\ref{fig:globalspectra}
shows spectra of large-scale regions of X-ray emission on the ACIS-I
array and {\it XSPEC} fits.  Fit results are given in
Table~\ref{tbl:globalspectable}.  Column 1 lists the various composite
regions on which spectral fits were performed; rows 1--4 correspond to
the fits shown in Figure~\ref{fig:globalspectra}.  Column 2 gives the
number of counts used in the fit.  Columns 3--8 give the spectral fit
parameters, while columns 9--12 show the elemental abundances relative
to solar values from the $\it vapec$ thermal plasma model
\citep{Smith01} for elements that required $Z > 0.3Z_{\odot}$.  Columns
13--17 give the X-ray luminosities based on the spectral fits for a
variety of wavebands, with and without correction for absorption.

Column 8 shows that many of these fits are not formally acceptable (the
reduced $\chi^2 > 1.1$).  This is due mainly to the large number of
degrees of freedom, which is in turn due to the large number of events
in these spectra; while in principle more events should allow the
reduced $\chi^2$ to approach unity as signal-to-noise improves, in
practice systematic errors in the ARF and RMF become more important and
the reduced $\chi^2$ remains high.  The fits in
Figure~\ref{fig:globalspectra} show that the spectral models used in
Table~\ref{tbl:globalspectable} characterize the data quite well in
spite of the formally unacceptable fit; a more complicated model
(such as an additional thermal plasma component) does not improve the fit.  

Figure~\ref{fig:globalspectra}a illustrates the global spectral fit for
all components of the main 30~Dor nebula as outlined in
Figure~\ref{fig:introimage}, including point sources, the SNR N157B,
its pulsar, and its cometary nebula.  The spectrum is fit with a single
absorbing column via the {\it wabs} model \citep{Morrison83}, which
assumes solar abundances for the absorbing material, and a spectral
model consisting of two components:  a variable-abundance thermal
plasma and a power law.  The average absorption is a factor of 2
smaller than the {\em ROSAT} result \citep{Norci95}, although the
plasma temperature is similar.  \citet{Norci95} do not state what
abundances they assume for the absorbing material and \citet{Mignani05}
note that abundances of $0.4Z_{\odot}$ give $N_H$ values a factor of
two higher than those for solar abundances when fitting the spectrum of
PSR~J0537$-$6910, so perhaps different assumptions regarding the
metallicity of the intervening material account for this discrepancy in
the average absorption.  The ACIS soft-band intrinsic
(absorption-corrected) luminosity for the main 30~Dor nebula is a
factor of 2.5--5 smaller than the {\em ROSAT} 0.1--2.4~keV estimate
\citep{Norci95}. 

Figure~\ref{fig:globalspectra}b covers the same region of X-ray
emission as Figure~\ref{fig:globalspectra}a minus the N157B SNR (a
composite spectrum of the SNR is shown in \S\ref{sec:n157bspectra}).
Since N157B is quite bright and contains substantial non-thermal
emission that would not necessarily be present in a more distant GEHR
unless it contained a recent supernova, removing it from the overall
spectrum of 30~Dor might yield a more typical GEHR spectrum.  The {\it
wabs*(vapec + powerlaw)} fit is actually quite similar to that in
Figure~\ref{fig:globalspectra}a, with nearly the same $N_H$, $kT$, and
plasma abundances as that fit, but a steeper power law slope.  This
shows that, although the non-thermal N157B pulsar and cometary nebula
are bright compared to other point sources in the field, the overall
spectral shape of the GEHR is still dominated by a soft thermal plasma
with fairly simple spatially-averaged spectral properties---no
prominent lines are present and there is no obvious need for multiple
plasma components or absorbing columns to fit the composite spectrum.
The only indication of the presence of the pulsar is at high energies,
where it flattens the power law component of the fit.

The same extraction region is used for Figure~\ref{fig:globalspectra}c
as was used in Figure~\ref{fig:globalspectra}b, but now the 158 point
sources that this region contains have been removed from the spectrum.
The composite spectrum of those point sources is shown in
Figure~\ref{fig:globalspectra}d.  With only the diffuse emission
component remaining in Figure~\ref{fig:globalspectra}c, the power law
is no longer needed in the spectral model; rather the emission can be
represented by a single, slightly hotter thermal plasma with enhanced
abundances.  The composite point source spectrum in
Figure~\ref{fig:globalspectra}d requires a hotter thermal plasma and a
flatter power law; it constitutes less than 10\% of the full-band
emission in this region but dominates the hard-band (2--8~keV) emission
by a factor of 3 over the diffuse component.

These fits suggest that unresolved GEHRs in more distant galaxies might
appear as soft, moderately luminous ($L_X \sim 10^{36}$~ergs~s$^{-1}$)
X-ray sources, quite distinct from harder and often brighter ($L_X \sim
10^{37}$--$10^{40}$~ergs~s$^{-1}$) X-ray binaries.  Distinguishing
GEHRs from individual SNRs, though, is probably only possible for young
SNRs, which might present hotter thermal plasmas or occasionally
non-thermal emission and may have prominent line features.  Even though
the N157B SNR is quite young, it does not show prominent lines and,
when averaged with the emission from the rest of 30~Dor, it is not
bright enough to harden the composite spectrum.  If its non-thermal
central components were absent, it would not be spectrally distinct
from the other diffuse structures in 30~Dor.  This is illustrated in
the spectral fitting presented in \S\ref{sec:superbubbles} and
\S\ref{sec:N157B} below.

Despite its complex spatial morphology, 30~Dor's integrated thermal
emission has simple spectral properties:  it is well-fit by a
single-temperature plasma with $T \simeq 4$~MK, roughly solar line
strengths, and a single absorbing column density.  We
will see in \S\ref{sec:superspectra} that a considerable variety of
plasma temperatures and absorbing columns are present on small spatial
scales.  In the ACIS soft band (0.5--2~keV), the X-ray luminosity comes
primarily from the diffuse structures, which are brighter by an order
of magnitude than the point sources.  Above 2~keV, though, the diffuse
emission disappears and the integrated emission is dominated by the
point sources.  Paper~II demonstrates that this point source emission
is itself dominated by a few bright sources; one third of the counts
that compose the spectrum in Figure~\ref{fig:globalspectra}d come from
a single source \citep[probably a colliding-wind binary,][]{PPL02}.

%==========================================================================
%==========================================================================
\section{THE X-RAY SUPERBUBBLES
\label{sec:superbubbles}}

\subsection{Morphology
\label{sec:supermorph}}

As outlined in \S\ref{sec:globalmorph}, the five plasma-filled
superbubbles first defined by \citet{Wang91a} from {\em Einstein}
observations of 30~Dor appear to be much more complicated structures in
these ACIS images, exhibiting knots, whisps, loops, and voids with a
range of surface brightnesses and X-ray colors.  In some cases the
X-ray morphology may be affected by the clumpy molecular clouds in the
region \citep{Johansson98}, the shreds of the material from which R136
formed.  This neutral material may shadow soft X-rays; comparing CO
cloud locations from \citet{Johansson98} to our smoothed images, this
effect may account for some of the X-ray voids close to R136, although
probably not the larger, more distant voids seen in Shells 2 and 3.

Some X-ray regions show distinctive sharp edges, such as the long
north-central bubble comprising Shell 5 and the long arc stretching
across the southeast edge of Shell 2.  Other regions such as the large
loop cut off by the northwest edge of the ACIS-I array have softer
edges that seem to blend into faint, larger-scale diffuse emission.
The two bubbles of Shells 4 and 5 running almost due north and south of
R136 appear quite red in Figures~\ref{fig:csmoothimage} and
\ref{fig:diffuseintro}, perhaps implying that they are less absorbed
than other parts of the field or are filled with cooler gas.  They are
also distinctly center-filled, while other superbubbles appear as loops
with central voids.  The ACIS-S structures 30~Dor~C and the faint loop
that arcs to the northeast of the Honeycomb SNR are good examples of
such loops, but the main 30~Dor complex shows them as well.

\begin{figure}
\centering
  \includegraphics[width=0.99\textwidth]{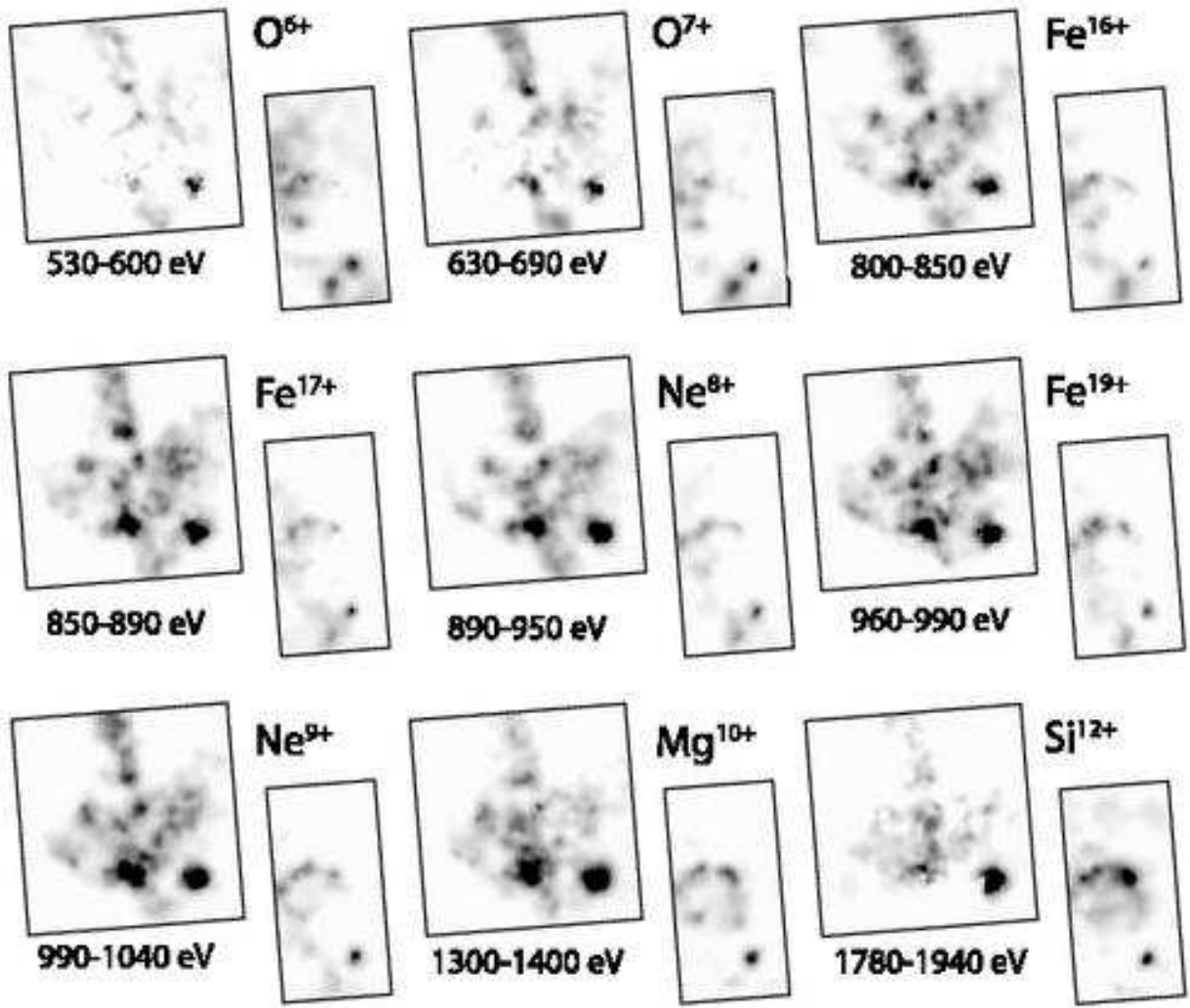}
\caption{ Narrow-band smoothed images of the 30~Dor complex in energies
corresponding to ions commonly seen in SNRs:  O$^{6+}$ (530--600~eV), O$^{7+}$ (630--690~eV), Fe$^{16+}$ (800--850~eV), Fe$^{17+}$ (850--890~eV), Ne$^{8+}$ (890--950~eV), Fe$^{19+}$ (960--990~eV), Ne$^{9+}$ (990--1040~eV),
Mg$^{10+}$ (1300--1400~eV), and Si$^{12+}$ (1780--1940~eV).  
\label{fig:lineims}} 
\end{figure}

In the hot thermal plasma models that we use for spectral fitting of
these data (see \S\ref{sec:globalspectra}), most of the emission
between 0.5 and 2~keV comes from a large number of blended spectral
lines, with only a small remainder caused by thermal bremsstrahlung.
In order to understand the morphology of 30~Dor's superbubbles and its
energy dependence, Figure~\ref{fig:lineims}\footnote{These images
(point sources removed) were made with the {\it adaptive\_density\_2d}
procedure in {\it ACIS Extract} using the smoothing scales from a
broadband (350--2000~eV) image, so that all narrow-band images have the
same smoothing scales.  To account for the fact that each narrow-band
image was made from a different number of events, each image was
normalized by its median intensity, then all were scaled the same in
the display.} shows narrow-band smoothed images of the full ACIS field
centered on some of the most prominent spectral lines in the thermal
plasma models, specifically those often seen in SNRs, following {\it
Chandra} studies of Magellanic Cloud diffuse nebulae by \citet{Behar01}
and \citet{Naze02}.  The continuum emission dominates the
highest-energy image ($\sim$65\%) but contributes only $\sim$10--30\%
to the other narrow-band images.

While small-scale changes between narrow-band images should be viewed
with caution due to limited photon statistics, it is clear that all of
these images hint at the presence of complex fine spatial structure and
that the morphology of large-scale structures changes substantially
with energy.  A good example of this is 30~Dor~C, which seems to emerge
out of faint, unorganized swaths of soft emission to coalesce into the
well-defined hard loop that we see in broad-band images.  The large,
soft superbubble at the south-center of the ACIS-I array (Shell 4) and
the loop containing the Honeycomb SNR are prominent in the softer
images, but fade above 950~eV.  Conversely, the sharp southeast ridge
seen on CCD I1 (edge of Shell 2) becomes more prominent at higher
energies and the void to its northwest appears to be filling in, with
no void left in the 1780--1940~eV image.  Smaller-scale knots and
whisps near the center and west-center of 30~Dor also come and go with
energy, with some prominent in the softer panels and others bright at
higher energies.

\begin{figure}
\centering
  \includegraphics[width=0.9\textwidth]{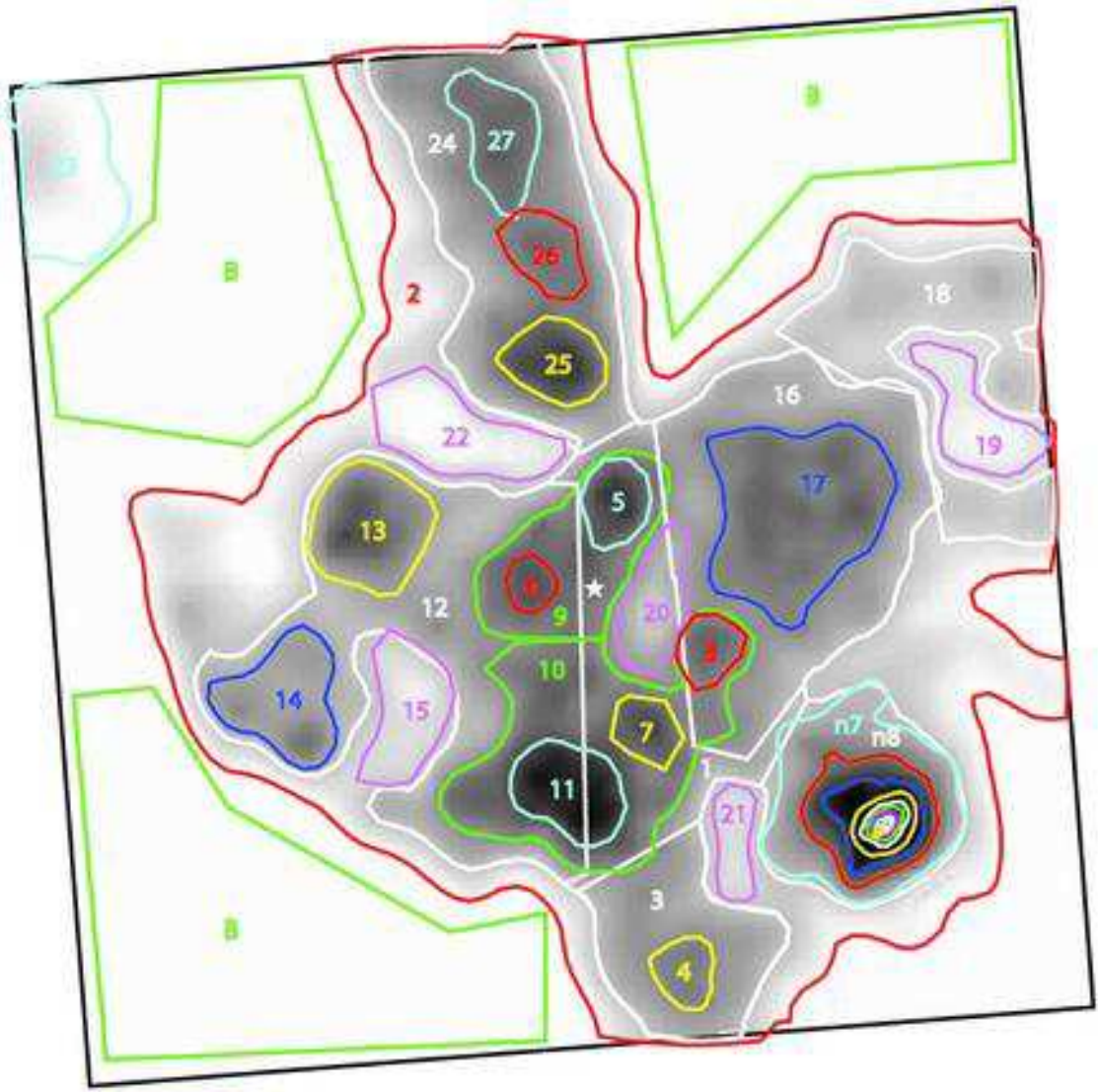} 

\caption{ Spectral extraction regions for diffuse structures in 30~Dor,
overlaid on an adaptively smoothed broadband (350--2000~eV) image of
the {\em Chandra} ACIS-I data, where point sources have been removed to
study the diffuse spectra.  The small white star in the middle of the
figure shows the approximate location of the R136 cluster.
\label{fig:diffregions}}
\end{figure}

These images show that substantial spectral variation exists in the
30~Dor superbubbles but do not by themselves reveal whether the
variations are produced by spatial changes in absorption columns,
plasma temperatures, abundances, or other phenomena.  To investigate
this question, we divide the diffuse emission on the ACIS-I array as
shown in Figure~\ref{fig:diffregions}.  These regions were chosen
primarily on the basis of apparent surface brightness of the diffuse
emission and overall morphology of the 30~Dor complex, so they form a
phenomenological rather than a physical parameterization of the
X-ray emission.  We emphasize that they were not chosen by considering
the morphology of 30~Dor in any other waveband; this will become relevant
later in this paper.  

The three large green polygons labeled ``B'' are the regions used to
obtain the background spectrum used in all spectral fitting of diffuse
regions described in this paper.  From {\em XMM} data of this region
\citep{Dennerl01} and {\em ROSAT} maps of the LMC
\citep[e.g.][]{Points01,Sasaki02}, we know that these regions contain
faint diffuse emission, but this may be the most appropriate background
to use, as the whole field may well be pervaded by such faint diffuse
emission.

The regions in Figure~\ref{fig:diffregions} are best understood by
considering them hierarchically by size, starting with the outermost,
largest regions and working inward and to smaller sizes.  The large red
contour outlines the region used earlier for the global spectral
properties of 30~Dor.  The six white regions define large-scale diffuse
structures, including the N157B SNR.  Region 1 encompasses the
brightest emission from the whole nebula, including regions 3, 12, 16,
24, n8, and the central polygon that these regions abut.  Region 2 is
the perimeter between region 1 and the outer red contour, also
excluding region 18.  The five purple contours define ``voids,'' or
areas of low apparent surface brightness.  Two central green regions (9
and 10) outline relatively bright central structures.  Contained within
these large-scale regions are smaller regions defining locally bright
areas, in blue, yellow, cyan, and red.  A cyan contour in the far
northeastern corner (region 23) defines an area of faint diffuse
emission known from {\em ROSAT} observations \citep{Points01}.  A
series of annular regions (n0--n7) is used to explore the spectral
properties of the N157B SNR; \S\ref{sec:N157B} gives more detail.

After choosing these extraction regions based on the X-ray apparent
surface brightness in our {\em Chandra} images, we compared them to the
$\sim 100$~ks {\em XMM} first light image in \citet{Dennerl01}, which
contains the western half of the main 30~Dor nebula.  This image
closely matches the {\em Chandra} data, showing very similar diffuse
X-ray structures, and justifies our choice of extraction regions even
for faint features such as the outer loop in Shell 3 that we delineate
as region 18, soft features such as regions 3 and 4 that fill Shell 4,
and the X-ray voids in regions 19, 20, and 21.

%-------------------------------------------------------------------------
\subsection{Spectra
\label{sec:superspectra}}

The spectral variety in these diffuse regions is shown in
Table~\ref{tbl:diffspectable}.  Column 1 gives the identifier of the
regions shown in Figure \ref{fig:diffregions}.  Column 2 gives the net
full-band (0.5--8~keV) counts in each diffuse region.  Spectral fit
parameters, allowing for both {\em vapec} thermal plasma and power law
fits, are given in Columns 3--7.  Column 8 gives the reduced $\chi^2$
value for the fit.  All elemental abundances were allowed to vary;
Columns 9-12 show abundances relative to solar that exceeded the
nominal value $0.3Z_{\odot}$.  These abundances were not always
well-constrained in the fits; in cases where the fit was improved with
higher abundances but actual values were not well-determined, they are
listed as ``hi'' in the table.  Derived X-ray luminosities are given in
Columns 13-16; since the diffuse emission is quite soft in most
extended regions (apart from the center of N157B), we report the
intrinsic (absorption-corrected) soft-band luminosity as well as the
usual intrinsic full-band luminosity.  The area subtended by each
diffuse region is given in Column 17 and the table is completed by a
list of intrinsic soft-band surface brightnesses in Column 18.

Using methods similar to those developed by Jeremy Sanders for study of
extragalactic plasmas \citep[e.g.][]{Sanders05}, we have used thermal
plasma fits and the regions defined in Figure~\ref{fig:diffregions} to
produce maps of $N_H$, $kT$, and intrinsic X-ray surface brightness in
30~Dor, as shown in Figure~\ref{fig:nhktmaps}.  For uniformity, the
spectral fits used to make these maps were single-temperature thermal
plasmas, so the results are not necessarily the same as those presented
in Table~\ref{tbl:diffspectable}, where power law components were added
for some regions.  Characteristics of the largest regions are recorded
first and are overlaid by those of smaller regions.  This approach has
the misleading effect of leaving apparent ``rings'' around void
regions (e.g.\ regions 15 and 19) where the underlying large-scale fit
is not masked by an overlying smaller region.

Although crude, these maps are useful for understanding the physical
state of the plasmas that make up the 30~Dor diffuse emission.  For
example, region 4 is not intrinsically more luminous than the
surrounding region 3, it is just less absorbed and, since it has the
same temperature as region 3, its apparent surface brightness is
larger.  The same explanation applies to region 8.  Region 11 is
particularly bright, even though it has the same plasma temperature as
its surrounding region 10, because it is slightly less absorbed plus it
has intrinsically higher surface brightness.  Regions 25 and 26 suffer
more absorption than the surrounding region 24 and have cooler plasma
temperatures, but they stand out because they have higher intrinsic
surface brightness.  Regions 18 and 19 suffer similar high obscuration
and have the same cool plasma; region 19 appears as a void because it
is intrinsically fainter than its surrounding region 18, which is only
visible to us in regular images because it has relatively high
intrinsic surface brightness.  Region 22 has similar obscuration and
plasma temperature, but appears as a void because it lacks this
enhanced intrinsic brightness.

Particularly unusual is region 15, which has high obscuration and low
intrinsic surface brightness so it appears to us as a void, but it has
the highest plasma temperature in the nebula.  Region 5 is also
interesting, exhibiting a low temperature and relatively high
obscuration, but it is very bright in regular images due to its
intrinsic brightness, one of the largest in the nebula.  Of even higher
intrinsic brightness, though, is region 7, yet it is much less
noticeable because of its large obscuration.  The intrinsically faint
and soft emission in region 23, probably associated with more eastern
structures rather than with 30~Dor itself, is visible only because of
its minimal obscuration.

In order to get enough counts for reliable spectral fitting, the
regions used here average over many tens of square parsecs.  From the
complex energy variations seen in Figures~\ref{fig:csmoothimage},
\ref{fig:diffuseintro}, and \ref{fig:lineims}, it is quite possible
that each region averages over many distinct plasma components, with
different pressures, densities, temperatures, absorbing columns, and
possibly even different abundances.  Study of such features and their
energetics will be possible when longer ACIS exposures of the region
are made.

%==========================================================================
%==========================================================================
\section{THE SUPERNOVA REMNANT N157B AND PSR~J0537$-$6910
\label{sec:N157B}}

%==========================================================================
\subsection{Morphology
\label{sec:n157bmorph}}

The {\em Chandra}/HRC \citep{Wang01} and ACIS images of the composite
SNR N157B are incremental improvements to the {\em ROSAT} data, which
also showed a comet-shaped nebula embedded in a diffuse remnant,
suggesting that the neutron star received a kick in the supernova blast
and is plowing through the surrounding ISM \citep{Wang98a}.  Smoothed,
soft-band and hard-band ACIS images of N157B are shown in
Figure~\ref{fig:N157B-regs}.  The soft-band image
(Figure~\ref{fig:N157B-regs}a) shows the point source extraction
regions outlined in red; the hard-band image
(Figure~\ref{fig:N157B-regs}b) additionally shows the concentric
diffuse extraction regions n7--n0 as defined in
Figure~\ref{fig:diffregions} and  Table~\ref{tbl:diffspectable}.  The
extraction region for the pulsar is shown in white.

The ACIS soft-band smoothed image shows diffuse emission from N157B
extending somewhat farther than that shown in the HRC data
\citep[][Figure 1]{Wang01}, giving dimensions for the X-ray SNR of
$\sim 3\arcmin.3 \times 3\arcmin.5$, or roughly 50~pc in diameter.  The
soft X-ray emission might be partially shadowed along the southern edge
of the SNR by the dark cloud described in \citet{Chu92}.  The hard
X-ray emission in Figure~\ref{fig:N157B-regs}b, which would not be
shadowed because it penetrates such material, does not appear to
extend into this region, though.  Thus it does not appear that the
SNR extends behind the southern dark cloud.   
  
Figure~\ref{fig:imageN157B} shows ACIS images of the embedded cometary
nebula and pulsar.  The SNR is imaged $6\arcmin.9$ off-axis, where the
PSF is $\sim 5\arcsec \times 10\arcsec$ in size (shown in white in
Figure~\ref{fig:imageN157B}a).  The central source is still quite
centrally peaked and sharp enough that photon pile-up affects its
spectrum.  Since it is immersed in its pulsar wind nebula and may have
further corruption from the surrounding cometary nebula and SNR, our
estimate of the pulsar's power law slope ($\Gamma = 2.0$) should be
considered qualitative.  \citet{Mignani05} estimate $\Gamma = 1.8$ for
the pulsar using an ACIS subarray observation that minimizes photon
pile-up for the pulsar and images it on-axis, so the small PSF limits
the contamination from diffuse structures around the pulsar.

Figure~\ref{fig:imageN157B}b shows the maximum likelihood
reconstruction of Figure~\ref{fig:imageN157B}a, using the PSF of the
pulsar for the reconstruction of the whole field. We recover an image
very similar to the on-axis {\em Chandra}/HRC images in
\citet{Wang01}:  the pulsar is a circular point source centered on its
pulsar wind nebula and a bright, $\sim 4\arcsec \times 7\arcsec$ region
of emission is oriented perpendicular to the larger, trailing cometary
nebula.  Although this image is not useful for spectral analysis, it is
reassuring that even a simple reconstruction algorithm can recover
information far off-axis, for sufficiently bright sources.

%==========================================================================
\subsection{Spectra
\label{sec:n157bspectra}}

Figure~\ref{fig:n157bspectra}a shows the spectral fit for the entire
N157B SNR (labeled in Figure~\ref{fig:introimage} and seen as region n8
in Figure~\ref{fig:diffregions}), including its pulsar (suffering from
moderate photon pile-up with $\sim 0.9$ counts per frame), its pulsar
wind nebula, and the surrounding cometary nebula.  Fit parameters are
given on the last line of Table~\ref{tbl:globalspectable}.  Using the
same spectral model as for Figure~\ref{fig:globalspectra}a and b, we
obtain the same $N_H$ but a hotter thermal plasma.  Some elements
require abundances $>0.3Z_{\odot}$ but are not well-constrained.

Our thermal plasma fit results are quite similar to the {\em ROSAT}
results \citep{Wang98a}, but the power law component is substantially
flatter than the results given by other X-ray observations of N157B
(see \S\ref{sec:n157bbkgd}).  Since N157B is a young SNR, we also
characterized its spectrum with a variable-abundance non-equilibrium
ionization model, using the {\it XSPEC} model {\it vnei + powerlaw}.
The fit was essentially the same as the thermal plasma fit described
above, with $kT \sim 0.9$~keV and $\Gamma = 2.2$, but the fit
parameters were not as well-constrained as those for the thermal plasma fit.

Removing all the point sources from this region leaves the diffuse
emission associated with the SNR (Figure~\ref{fig:n157bspectra}b).  Its
spectral fit parameters are listed under region n8 in
Table~\ref{tbl:diffspectable}.  The cometary nebula associated with the
pulsar contributes the power law component of the spectrum.  The
thermal emission averaged over the whole SNR is quite soft and possible
emission lines of Ne and Mg are seen.  While our fit results are
consistent with the {\em XMM} results for the thermal component and
overabundances of Ne and Mg are consistent with {\em XMM's} discovery
of emission lines \citep{Dennerl01}, our power law slope is slightly
flatter than the $\Gamma = 2.8$ result from {\em XMM}
\citep{Dennerl01}, even though our fit excluded the pulsar.
\citet{Dennerl01} note that, if N157B is similar to the Crab Nebula, we
would expect a difference between the core power law slope and that for
the entire nebula of $\sim 0.5$ due to synchrotron losses.  This is
seen in our {\em Chandra} data, confirming the analogy with the Crab.

The {\em XMM} value for the absorption (assuming an abundance of
$0.5Z_{\odot}$) was $N_H = 1.9 \times 10^{22}$~cm$^{-2}$, more than a
factor of 6 higher than our result; the highest absorption we see for
any part of N157B is at its core, where our $N_H$ is still a factor of
3 lower than the {\em XMM} value.  As discussed above, our assumption
of solar abundance for the absorption can account for a factor of $\sim
2$ discrepancy in the measured column, but the cause of the remaining
difference in $N_H$ estimates is unknown.  One possibility might be
calibration uncertainties in the early {\em XMM} results or 
cross-calibration issues between the two observatories.  

To study the SNR on finer spatial scales, spectral fits to the annular
regions n0--n7 defined in Figure~\ref{fig:N157B-regs} were performed
and are given in Table~\ref{tbl:diffspectable}.  Our spectral fitting
confirms the results of \citet{Dennerl01} and \citet{Wang98a} that
N157B is a composite nebula, with a thermal plasma showing hints of
line emission in the outer regions giving over to synchrotron emission
in the bright core.  The spectral model in these fits allowed for both
a thermal plasma and a power law component, reverting to a single
component when an adequate fit could be obtained with just the thermal
plasma or the power law model.

We have included two point source extraction regions in this table and
show their spectra in Figure~\ref{fig:n157bspectra}:  p1 contains the
pulsar (CXOU~J053747.41$-$691019.8), p2 (CXOU~J053745.61$-$691011.1) is the
pointlike component of the cometary nebula found by {\em wavdetect}.
These sources are included because their large off-axis PSFs make it
difficult to isolate the point source emission from the surrounding
bright diffuse emission, so these point sources include diffuse
spectral components.  In fact it is not clear in the ACIS subarray
observation of N157B that p2 is a point source at all; it is most
likely another example of concentrated diffuse emission that is
consistent with the ACIS PSF at this large off-axis angle.  The
spectral fits in Table~\ref{tbl:diffspectable} are not identical to
those in Paper II because the former were performed by hand, while the
latter were performed automatically in {\em ACIS Extract}.  The results
are consistent to within errors. 

Comparing the spectrum of p1 (core, Figure~\ref{fig:n157bspectra}c) to
p2 (cometary nebula bright point, Figure~\ref{fig:n157bspectra}d) and
the surrounding bright cometary nebula (n0, the smallest blue annulus
in Figure~\ref{fig:N157B-regs}b), we see that all three regions are
adequately fit by a simple power law, but its slope steepens rapidly
away from the pulsar.  As we progress through the larger concentric
annular regions n1--n7, an additional thermal component is needed for
an adequate fit and the power law component apparently becomes very
steep.  In the outermost annular region (n7), the power law component
is no longer necessary.  The spectral fits of these annular regions are
problematic due to small number counts; there can be interplay between
the thermal plasma temperature, the power law slope, and the absorption
that can lead to incongruous results, as we see for region n6.  Here
the power law slope is quite flat, but its normalization is low.  The
general trend of these annular spectra, though, seems to indicate
electron cooling through the steepening power law slope, a transition
from non-thermal to thermal spectra with distance from the pulsar, and
higher abundances in the SNR, possibly hinting at the presence of
spectral lines.

%==========================================================================
%==========================================================================
\section{DIFFUSE STRUCTURES IN THE CONTEXT OF 30~DOR KINEMATICS
\label{sec:otherdiffuse}}

From {\em Einstein} and {\em ROSAT} data, it has long been known that
many of the X-ray concentrations in the 30~Dor superbubbles are
spatially associated with high-velocity optical emission line clouds
\citep[][hereafter CK94]{Chu94}.  These kinematic data are invaluable
for sorting out the many overlapping X-ray features seen in the {\em
Chandra} data as well.  For example, the large arc that appears to be
associated with the Honeycomb SNR is actually not kinematically related
to it, but kinematic data indicate that the Honeycomb itself is the
result of a cavity supernova explosion.  The part of it seen in X-rays
is due to the collision of this cavity SNR with an intervening porous
gas sheet that could be associated with its slow-moving giant shell
\citep{Chu95b,Redman99}.  
  
We place our bright X-ray features in the context of CK94's echelle
study in Table~\ref{tbl:kinematics}.  Several prominent X-ray features
(identified solely on the basis of their apparent X-ray surface
brightness in a smoothed ACIS image, see \S\ref{sec:supermorph}) show
high velocities in the kinematic data:  our West Ring is adjacent to
CK94's NW Loop; our region 5 is CK94's R139W; our region 6 is CK94's
R136E.  This correspondence between high-velocity features and diffuse
X-ray emission was noticed and explained by CK94 as X-rays produced in
the shocks between high-velocity material and the surrounding
slow-moving shells.  The {\em Chandra} data allow us to explore this
general idea in more detail, with spatial resolution  better matched to
that of CK94's echelle data.

%==========================================================================
\subsection{X-rays from High-velocity Features
\label{sec:highvel}}

The {\em Chandra} data show a more complex spatial distribution of
X-ray plasmas than that described by \citet{Wang99}, who inferred from
{\em ASCA} data that the diffuse emission was hot in the core of the
nebula and cooler in its outer regions.  This complex distribution more
closely matches the kinematic portrait of the nebula built by CK94,
though.  Many of the bright X-ray regions that match high-velocity
features listed in Table~\ref{tbl:kinematics} show cooler plasma
temperatures than their surroundings (Figure~\ref{fig:nhktmaps}b).  We
interpret this as evidence that these regions are denser due to the
high-velocity shocks, hence they are able to cool faster than the 
surrounding superbubbles.

The high velocities measured near this X-ray-emitting material are more
consistent with CK94's interpretation of them as supernova shocks
rather than as regions of high mass-loading, as proposed by
\citet{Wang99}, although our data do not exclude mass loading as a
secondary mechanism for increasing X-ray luminosity in some regions.
We would expect this mechanism to be most important in regions where
substantial neutral material is known to exist, thus mass loading may
contribute to the bright diffuse X-rays seen near the center of the
nebula, which contains the remains of the GMC.

Our bright, soft region 5 is coaligned with CK94's fast shell R139W,
which shows blueshifted emission with velocities up to
150~km~s$^{-1}$.  The region of bright X-ray emission is $\sim 15$~pc
in diameter and fills a prominent, well-known hole in the H$\alpha$
emission that also appears in the {\em Spitzer} data (see
\S\ref{sec:hotcold}).  Nearby is another bright, soft X-ray patch
(region 6) also $\sim 15$~pc in diameter that is well-matched to CK94's
fast shell R136E.  This region exhibits both redshifted and blueshifted
emission with speeds in excess of 130~km~s$^{-1}$.  The X-ray emission
is centered on the WR star R145, but CK94 note that it is unlikely to
be caused solely by the winds from that one star because the energy
requirements to support the X-ray emission and expansion velocities
seen would require at least 5 average WR stars.  Both X-ray regions 5
and 6 are cooler and intrinsically brighter than their surrounding
region 9 (see Figure~\ref{fig:nhktmaps}).  They may well be parts of
the bubble blown by R136 that have been brightened by off-center SNe
\citep{Chu90}.

We find that not every high-velocity feature shows bright X-ray
emission.  A good example is region 15, which coincides with a 20~pc
shell expanding at 110~km~s$^{-1}$ described by CK94.
Figure~\ref{fig:nhktmaps} shows that this region is intrinsically X-ray
faint and exhibits the hottest plasma in the 30~Dor nebula.  Perhaps
this is the site of a recent cavity supernova that occurred far enough
from the edge of Shell 2 that it did not produce an X-ray-bright shock
but instead deposited its energy into heating the surrounding plasma
\citep{Chu90}.  A more detailed velocity study of this region is
warranted to elucidate the cause of this unusual X-ray feature.

%==========================================================================
\subsection{Is the West Ring a Cavity Supernova Remnant?
\label{sec:westring}}

A prominent feature in all X-ray images featured in this paper is a
nearly-complete, clumpy ring structure in the middle of Shell 3, as
described in \S\ref{sec:globalmorph}.  We called this structure the
``West Ring'' and find that it sits $\sim 1\arcmin$ west of CK94's
high-velocity ``NW Loop,'' which shows expansion velocities of
200~km~s$^{-1}$.  CK94 note that such high velocities are never found
in wind-blown bubbles.  In Figure~\ref{fig:nhktmaps}, the West Ring is
contained in region 17, which is indistinguishable from its surrounding
region 16 in obscuration and temperature but has slightly higher
intrinsic surface brightness.  

The West Ring could be a SNR:  \citet{Meaburn88} performed an echelle
study of this region and concluded that the large expansion velocities
make a supernova origin the most likely explanation.  It resembles the
``shell'' SNRs catalogued in \citet{Williams99}, both in size and
structure.  Such a ring structure is predicted by \citet{Velazquez03}
when a SNR hits its cluster's stellar winds.  Recent echelle
spectroscopy concentrating on this region also finds knots and shells
consistent with a supernova interpretation \citep{Redman03}.

This structure is also very similar to the central X-ray ring
\citep{Leahy85} in the over-sized Galactic SNR HB3 (G132.7+1.3),
considered to be a cavity SNR \citep{Routledge91}.  The West Ring is
about $2\arcmin.5$ in diameter, or $\sim 36$~pc.  The X-ray ring in HB3
is about $35\arcmin$ in diameter, which at $D = 2.3$~kpc corresponds to
$\sim 23$~pc.  HB3's X-ray ring is also very clumpy and not quite
complete \citep{Landecker87}; it is strikingly similar in appearance to
the West Ring.  From Table~\ref{tbl:diffspectable}, diffuse region 17
has 0.5--8~keV intrinsic luminosity $L_{X,corr} = 4.7 \times
10^{35}$~ergs~s$^{-1}$ and $kT = 0.5$~keV.  The {\em Einstein}
(0.2--4~keV) X-ray luminosity of HB3 totals $1.6 \times
10^{35}$~ergs~s$^{-1}$ and shows a hot central region ($kT \sim 1$~keV)
evolving to a cooler limb ($kT \sim 0.3$~keV) \citep{Leahy85}.  Our
spectral fit to region 17 shows solar abundances for O and Ne; a recent
analysis of a short {\em XMM} observation of HB3 shows the possibility
of enhanced abundances of O, Ne, and Mg in its X-ray ring
\citep{Lazendic05}.  Just as in the West Ring, HB3 lacks strong X-ray
emission lines.

Examining the visual and radio images from \citet{Dickel94}, it is
clear that there is no prominent visual or radio continuum source at
the location of the West Ring.  As noted by \citet{Chu00} in their
study of the large LMC X-ray ring RX~J050736-6847.8, though, SNRs in
low-density media are not expected to show prominent visual or radio
features.  Given the similarity of the West Ring to HB3 and other LMC
SNRs, we suspect that it is in fact a cavity SNR, perhaps produced by a
star in the Hodge~301 cluster.

Although more detailed comparisons of the data from CK94 and other
kinematic studies to the diffuse X-ray structures revealed by {\em
Chandra} are beyond the scope of this paper, CK94 noted that such
comparisons likely hold the key to understanding the complex ISM in
30~Dor.  We will search for the parsec-scale high-velocity knots seen
by CK94 and \citet{Meaburn84} in the longer {\em Chandra} observation.
We suspect that this dataset will reveal small-scale diffuse X-ray
features that will merit new kinematic studies as well.

%==========================================================================
%==========================================================================
\section{DISCUSSION
\label{sec:discussion}}

%==========================================================================
\subsection{The Relationship between Hot, Warm, and Cool Interstellar 
Material in 30~Dor
\label{sec:hotcold}}

The morphology of the hot X-ray emitting plasma is extremely
complicated and bears little resemblance to theoretical calculations of
individual SNRs or superbubbles, which generally predict
quasi-spherical structures.  Insight into the origins of these
structures emerges from comparison with the distribution of
interstellar material traced by H$\alpha$ emission from ionized gas and
IR emission from dust.  The comparison with H$\alpha$ has been made
since the earliest X-ray images of 30~Dor were obtained by {\em
Einstein} \citep[e.g.][]{Chu90,Walborn91,Wang91a}.

Figure~\ref{fig:Halpha} shows a high-resolution H$\alpha$ image from
the MCELS project \citep{Smith00} in red, with adaptively-smoothed
soft-band ACIS images (point sources removed) in green and blue.  This
image is reminiscent of earlier combinations of H$\alpha$ and X-ray
data that showed that diffuse soft X-ray emission was anticorrelated
with H$\alpha$ emission, often filling the cavities outlined by ionized
gas \citep[e.g.][]{Wang99}.  These high-resolution images allow us to
refine that picture somewhat, giving more detailed information on the
relative locations of harder and softer X-rays and H$\alpha$ emission.

Comparing Figure~\ref{fig:Halpha} to a similar image based on {\em
ROSAT}/HRI data in \citet[][Figure 3]{Wang99}, both images show that
the northern Shell 5 and southern Shell 2 are very clearly outlined by
large-scale H$\alpha$ structures.  X-rays from the northern part of SNR
N157B coincide with a dark cavity in the H$\alpha$ emission caused by a
dust lane \citep{Chu92}.  Conversely, the outer loop of Shell 3,
labeled as X-ray emitting region 18 in Figure~\ref{fig:diffregions},
coincides with a similar loop in H$\alpha$; in Figure~\ref{fig:Halpha}
we can see that the northern part of this loop is softer than the
southern part.  Both images show that a prominent H$\alpha$ dark region
in Shell 2 (southeast of R136) contains faint X-ray emission, while the
bright H$\alpha$ filaments that outline it bifurcate our X-ray void
labeled region 15.  Figure~\ref{fig:Halpha} shows that Shell 2 X-rays
are relatively hard and Figure~\ref{fig:nhktmaps} reveals that this is
due to a combination of intrinsic spectral hardness and heavy
obscuration in this part of 30~Dor.  The West Ring is now more clearly
seen to be a clumpy ring of X-ray emission, fainter in its center, with
some soft X-ray emission coincident with the Hodge 301 cluster on the
northeast edge of the ring.  Small-scale clumps (such as our regions 7
and 8) are now distinct from the smoother underlying X-ray emission.
With {\em Chandra's} high on-axis spatial resolution, we have been able
to excise the point source X-ray emission from Figure~\ref{fig:Halpha},
so it is clearer that bright diffuse regions 5 and 6 are distinct from
the R136 cluster and from the nearby WR stars R139, R140, and R145.

The 8$\mu$m {\em Spitzer} data (Figure~\ref{fig:spitzer}) highlighting
warm dust also add insight into the diffuse X-ray structures.  Hot
X-ray plasma fills the interiors of superbubbles that are outlined by
warm dust and emission from PAHs (Brandl et al.\ 2006, in
preparation).  The X-ray morphology and possible confinement are more
fully appreciated when anchored by these IR data.  Heated dust provides
an envelope to the base of the cylindrical X-ray Shell 5.  X-ray
emission could be suppressed in this region or it could be present but
absorbed.  An IR-bright V-shaped ridge of emission (reminiscent of the
Carina Nebula) separating Shell 3 from Shell 5 is devoid of observable
soft X-ray plasma, while the bright X-ray spot dominating Shell 1 is
nearly devoid of IR emission.  The thermal plasma that constitutes the
outer regions of the N157B SNR fills a large cavity in the warm dust.
One of the brightest clumps of diffuse X-rays (our region 5) fills a
distinct hole in both the H$\alpha$ and IR emission.

To give a more complete view of the wide range of emission from 30~Dor,
Figure~\ref{fig:3bands} combines the 8$\mu$m data tracing PAH emission
and warm dust (red) from Figure~\ref{fig:spitzer}, the H$\alpha$ data
tracing ionized $10^4$~K gas (green) from Figure~\ref{fig:Halpha}, and
the 1120--2320~eV X-ray data tracing $10^7$~K plasma (blue) from
Figure~\ref{fig:csmoothimage}.  Note that here the X-ray image comes
from the ACIS data processed with {\em csmooth}, with the point sources
left in place, not the adaptively-smoothed images used in
Figures~\ref{fig:Halpha} and \ref{fig:spitzer} where the point sources
were removed. 

Figure~\ref{fig:3bands} shows the full ACIS-I field of view and is scaled
to show the full extent of faint diffuse X-ray emission across the
field.  The zoomed image in Figure~\ref{fig:3bandszoom} is scaled to
show just the brighter patches of X-ray emission in the center of the
field and removes the color saturation in the core.  Although the {\em
Chandra} data are not deep enough to match the spatial resolution seen
in the H$\alpha$ and {\em Spitzer} data, it is clear that the 30~Dor
complex cannot be understood by visual and IR studies alone, no matter
how high the quality of those datasets.  The high-energy emission is a
near-perfect complement to the longer-wavelength emission, filling
cavities in the complex that are outlined by H{\sc II} regions.  These
are in turn outlined by warm dust, because ultraviolet radiation in the
H{\sc II} regions destroys PAHs.  

X-rays from the northern parts of SNR N157B fill a prominent hole in
the 8$\mu$m emission.  The eastern side of the Shell 5 X-ray
emission, which appeared unconfined by warm dust in
Figure~\ref{fig:spitzer}, is clearly defined by a large region of
H$\alpha$ emission, while its western side shows a narrower H{\sc II}
region bordered by dust.  The southwestern side of the 30~Dor complex
shows more warm dust than the northern or eastern sides.

There is a notable absence of X-ray emission in the southeastern corner
of the image, where the stellar cluster SL~639
\citep{Shapley63,Melnick87,Bica99} is clearly seen in the H$\alpha$ and
8$\mu$m data.  Some ionized gas is present around the cluster along
with substantial amounts of heated dust.  This cluster may be too young
to have produced a substantial wind-blown bubble or any supernovae, or
its ionizing stars may not generate enough wind power to blow a
substantial bubble.  Conversely, if it is older than $\sim$10~Myr it
may have dispersed its hot gas.  It will be an interesting site to
search for faint diffuse X-rays in longer observations.

The famous central arcs north and west of R136 are bright at 8$\mu$m as
well as in H$\alpha$, illustrating the transition layers from cold
molecular material to heated dust to ionized gas that characterize much
of the 30~Dor complex.  Figure~\ref{fig:3bandszoom} shows that the
bright X-ray emission lies in the cavities interior to these ionization
fronts, but the saturated center of Figure~\ref{fig:3bands} suggests
that some diffuse X-rays are seen superposed on the H$\alpha$ and IR
emission.  This is true in other parts of the image as well and implies
that the soft X-rays come from regions that lie in front of the denser
IR-emitting material since they avoided being absorbed.  This is a
reminder that the complex ISM in 30~Dor may be absorbing similar soft
X-rays along other lines of sight.  Thus the regions exhibiting hot
plasma may be connected via tunnels or fissures that are not visible to
us in these images.  The regions where we do not see soft X-ray
emission are not necessarily lacking in hot plasma, especially if it is
clear from the {\em Spitzer} data that substantial absorbing material
lies along the line of sight.

These data thus augment and support the ideas developed over the last
twenty years for the evolution of the 30~Dor complex.  Powerful stellar
winds from extremely massive stars are carving holes in an extensive
GMC and filling those holes with hot plasma.  Additional hot plasma is
added as those stars explode inside the cavities they created.  In this
process, cold gas is pushed aside and confines the hot plasma into
shapes otherwise difficult to understand.  Since this plasma appears to
emit only soft X-rays for most of its lifetime, some of it may be
absorbed by intervening molecular material.  Tunnels and fissures could
exist in the GMC that allow hot plasma to flow between apparently
unconnected regions, although none of the hot plasma seen by {\em
Chandra} appears completely unconfined (there are no empty bubbles that
appear to have vented their X-ray plasma into the surrounding ISM).
The H$\alpha$ emission traces the ionization fronts, H{\sc II} region
``sheets'' that mark the transition between the million-degree plasma
and the neutral material.  This transition layer is heated, evaporated,
and is accelerating away from the cloud interface.  Evolved stars are
distributed across the entire 30~Dor nebula, providing the fuel for the
SNRs that occasionally brighten the superbubbles in X-rays and the
enhanced density that allows the plasma to cool.  Neutral clumps may
also enhance the X-ray emission, especially in regions close to the
remains of the GMC.

%==========================================================================
\subsection{Summary of Findings}

Diffuse emission dominates the morphology of 30~Dor in soft X-rays,
caused by hot plasma filling the large superbubbles created by
generations of star formation and subsequent SNe in this region.  This
is dramatically illustrated by combining the {\em Chandra} data with
H$\alpha$ and {\em Spitzer} images
(Figures~\ref{fig:Halpha}--\ref{fig:3bandszoom}); the $10^6$--$10^7$~K
X-ray plasmas are enveloped and probably confined in most cases by the
cooler gas and warm dust that define the classic picture of 30~Dor.
This example illustrates the need for high-quality X-ray observations
of star-forming regions; new facets of both stellar and diffuse
components are revealed by high-energy data.

The combined spatial and spectral resolution afforded by ACIS shows
that these superbubbles are not uniformly filled with a
single-temperature gas; great variety is seen, on a range of spatial
scales, in absorption, plasma temperature, and intrinsic surface
brightness (Figure~\ref{fig:nhktmaps}).  Some bubbles are
center-filled, perhaps revealing interactions with cold gas left in
shell interiors \citep{Arthur96,Wang99}, while others are
edge-brightened with distinct central voids not caused by obscuring
foreground material.  The faintest regions (voids and the periphery of
the main 30~Dor nebula) are a factor of 20 fainter in intrinsic surface
brightness than the brightest regions, which have surface brightnesses
$>1 \times 10^{33}$~ergs~s$^{-1}$~pc$^{-2}$.  Our diffuse regions range
in size from $1\arcmin$ ($\sim 14.5$~pc) for small surface brightness
enhancements to $>7\arcmin$ for the large shells.  For comparison with
other GEHRs, we have provided integrated spectra and fits for the main
30~Dor nebula and for the N157B SNR (Table~\ref{tbl:globalspectable}).

Spectral fits to the diffuse emission show moderate absorption ($N_H =
1$--$6 \times 10^{21}$~cm$^{-2}$) and soft thermal plasmas with $kT =
0.3$--0.8~keV (3--9~MK).  Many diffuse regions exhibit elevated
abundances (above the nominal value of $0.3Z_{\odot}$), perhaps
indicative of line emission, but none show prominent emission lines.
Although not a definitive demonstration of a supernova origin for the
X-ray emission, these spectral fit results are consistent with that
interpretation.  SNR N157B, also possibly the result of an explosion
within a pre-existing cavity, also lacks strong emission lines, as does
the Galactic example of a possible cavity SNR discussed here, HB3.

The brightest point source in the field is PSR~J0537$-$6910 in the SNR
N157B, imaged almost $7\arcmin$ off-axis.  Although the large PSF there
prevents us from performing accurate spectral analysis of the pulsar
and its pulsar wind nebula, these data clearly show spectral changes as
a function of distance from the pulsar; the pure non-thermal spectrum
steepens and combines with a thermal component, finally giving over to
a pure thermal spectrum in the outer regions of the SNR
(\S\ref{sec:N157B}).

Several interesting smaller-scale structures emerge with the
high-resolution {\em Chandra} observations.  One feature (the ``West
Ring'') may be the relic signature of a past cavity supernova that
exploded inside a pre-existing wind-blown bubble generated by the
Hodge~301 cluster.  As first shown by CK94, bright patches of X-ray
emission often coincide with high-velocity features known from
H$\alpha$ echelle studies, indicating that X-rays are produced in
shocks in 30~Dor's ISM (\S\ref{sec:otherdiffuse}).

The MCELS H$\alpha$ image, the {\em Spitzer} image, and the recent
mosaic of {\em HST} data \citep{Walborn02} are all dominated by
highly-structured arcs and shells on 1--10~pc scales; in the central
region, these reveal the interfaces between the central cavity created
by R136 and the surrounding molecular clouds \citep{Scowen98} and
demonstrate how R136 is shredding its natal environment
(\S\ref{sec:hotcold}).  Here and throughout the 30~Dor nebula, an
appreciation of the X-ray emission is necessary to further our
understanding of the processes working to shape this complex.  As the
images in \S\ref{sec:hotcold} show, it is imperative that the diffuse
X-ray emission become part of the ``classic'' picture of 30~Dor.

New stellar clusters are now forming in the dense knots that remain in
30~Dor's GMC; their collapse was probably triggered by R136
\citep{Walborn02}.  These clusters are not yet resolved in X-rays,
demonstrating that there is much more work to be done to achieve a
complete picture of the X-ray emission from 30~Dor.

%==========================================================================
\subsection{30~Dor and Other Massive Star-forming Regions
\label{sec:othermsfrs}}

30~Dor's diffuse X-ray emission is, overall, substantially different
than that seen in Galactic MSFRs that are too young to have produced
SNe.  M~17 shows strong diffuse soft X-rays probably due to wind-wind
and/or wind-cloud collisions \citep{Townsley03}, but the primary
component of this emission is hotter (kT = 0.6~keV) than many 30~Dor
regions and has low surface brightness
($10^{31.9}$~ergs~s$^{-1}$~pc$^{-2}$).  These quantities are most
similar to the regions we call ``voids'' in 30~Dor (regions 15, 19, 20,
21, and 22 in Figure~\ref{fig:diffregions}).  The diffuse X-ray
emission reported for the massive Galactic clusters NGC~3603
\citep{Moffat02} and Arches \citep{Yusef-Zadeh02} has higher surface
brightness more comparable to the brighter regions in 30~Dor
\citep[$10^{32.6}$ and $10^{33.1}$~ergs~s$^{-1}$~pc$^{-2}$
respectively, calculated from Table 4 of][]{Townsley03} but the plasmas
are much hotter (kT = 3.1~keV and 5.7~keV respectively).  Either a very
different mechanism is generating hard diffuse emission in these
clusters \citep[e.g. wind collisions,][]{Canto00}, young SNe dominate
the diffuse emission, or substantial unresolved point source emission
is corrupting the measurements.

\citet{Chu93} proposed that the soft diffuse X-ray emission seen in the
Carina star-forming complex \citep{Seward82} could be due to a cavity
SNR inside a superbubble blown by Carina's many massive stellar
clusters; thus the Carina complex should be a Galactic analog to one of
the superbubbles seen in 30~Dor.  The entire Carina complex shows
integrated X-ray emission with $kT \sim 0.8$~keV and surface brightness
$\geq 10^{32.2}$~ergs~s$^{-1}$~pc$^{-2}$ \citep{Seward82,Townsley03}.
{\em Chandra} and {\em XMM} resolve this X-ray emission into thousands
of harder but comparatively faint stellar sources
\citep[e.g.][]{Albacete03,Evans03} and bright diffuse emission
pervading the complex yet not centered on the O or WR stars or the
stellar clusters.  No obvious SNR is present, but the complex is old
enough to have produced SNe, as evidenced by the presence of evolved
stars.  This bright, soft diffuse emission is quite comparable in
surface brightness and extent to the smaller regions sampled in
30~Dor.  Using new {\em Chandra} data centered on the Trumpler~14 OB
cluster in Carina, we show that the brightest regions of diffuse
emission are well-separated from the massive stars in the Carina
complex and show filamentary morphology, both consistent with a cavity
supernova origin \citep{TownsleyIAUS227}.  Thus we suspect that Carina
serves as a good microscope for understanding the processes powering
the X-ray emission from 30~Dor.

Given the morphological complexity that 30~Dor displays at all
wavelengths, it is surprisingly similar to other GEHRs, most notably
NGC~604 in M33, the second-largest GEHR in the Local Group.  Although
not dominated by a massive central star cluster, the large-scale loops
and voids of NGC~604 are also characteristic of superbubbles and are
filled with hot gas emitting soft X-rays \citep[][Brandl et al.\ 2006,
in preparation]{Maiz04}.  The second-largest GEHR in the LMC, N11, also
exhibits a central massive stellar cluster surrounded by a superbubble
and containing bright, soft diffuse X-rays \citep{MacLow98,Naze04}.
Although more distant GEHRs are not resolvable with current technology,
we should expect all GEHRs to be made up of a complex mix of pointlike
and diffuse X-ray components with the hard-band X-ray luminosity
dominated by massive stars and the soft-band X-ray luminosity dominated
by the effects of recent cavity SNe expoding near the edges of
superbubbles, as we see in 30~Dor.

%==========================================================================
\subsection{Concluding Comments}

We have analyzed an early {\em Chandra}/ACIS observation of 30~Doradus,
concentrating here on the diffuse X-ray structures and in Paper II on
the point sources.  This study documents a wide variety of diffuse
X-ray-emitting sources:  a complex hierarchy of diffuse structures,
from small-scale knots and wisps to huge superbubbles; a composite SNR
including its pulsar, pulsar wind nebula, and cometary tail; an X-ray
ring possibly due to a cavity SNR; bright X-ray patches associated with
high-velocity H$\alpha$ structures.  A coherent understanding of the
structures is beginning to emerge from a multiwavelength comparison of
the X-ray, H$\alpha$, and mid-IR maps.  We demonstrate a variety of
data analysis tools for study of ACIS fields with mixtures of pointlike
sources and diffuse structures.  All software used in this study is
publicly available.

\begin{figure}
\centering
  \includegraphics[width=0.90\textwidth]{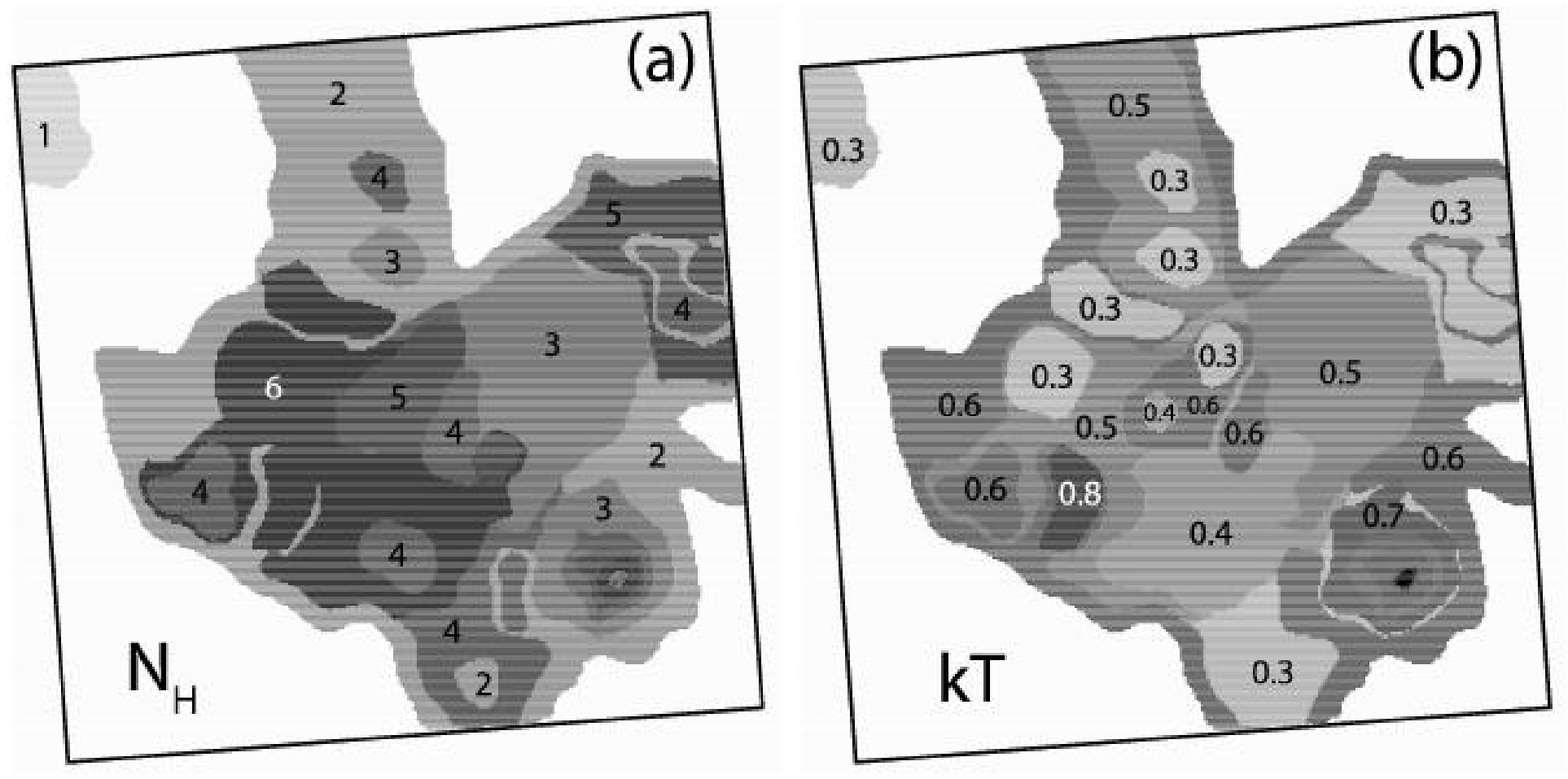} 
  \includegraphics[width=0.6\textwidth]{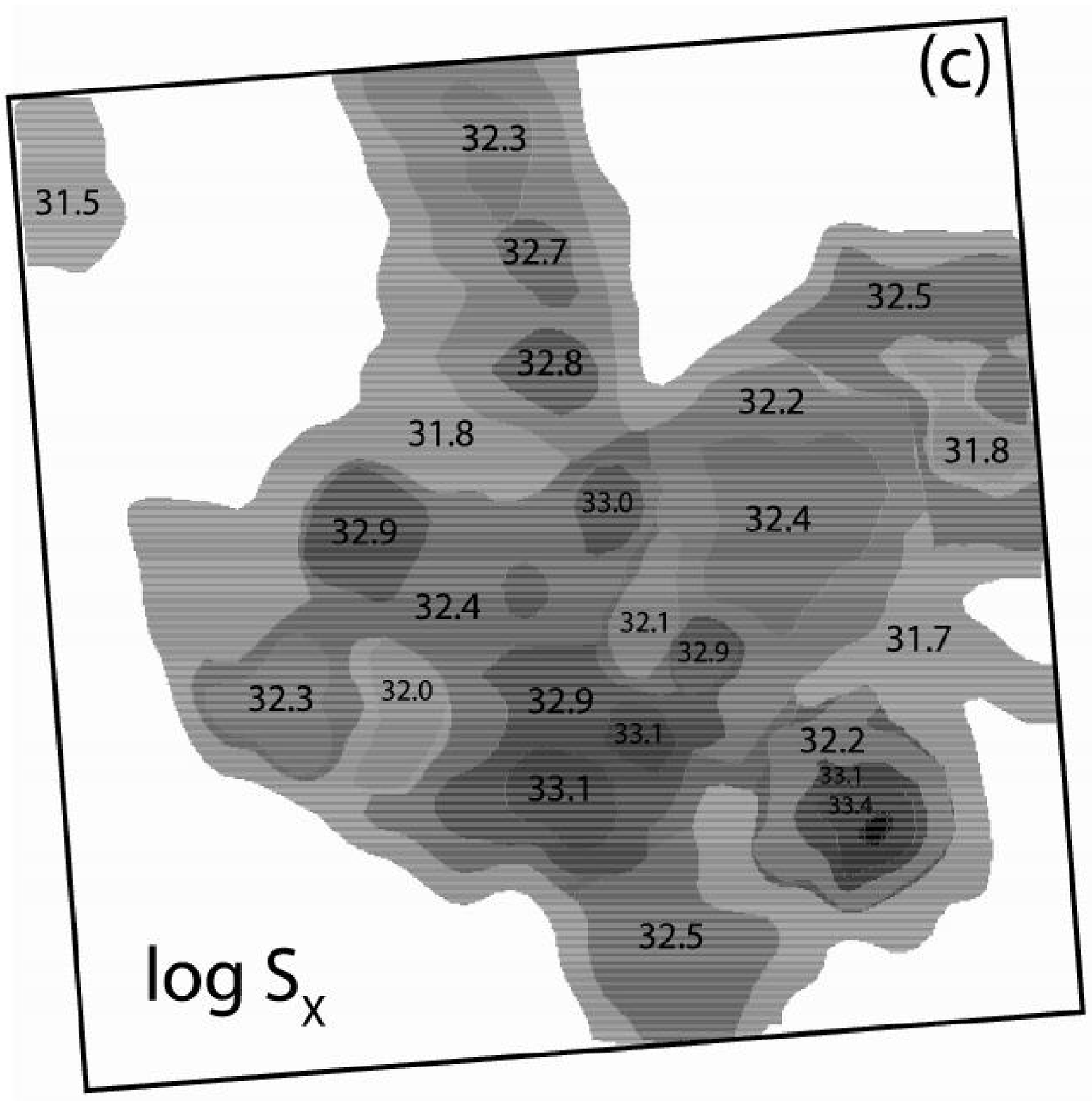}
\caption{ Maps of spectral fit parameters for diffuse regions defined in 
Figure~\ref{fig:diffregions}.
(a) Absorption $N_H$ in units of $10^{21}$~cm$^{-2}$. 
(b) Plasma temperature expressed as kT (keV).
(c) Log of the absorption-corrected soft-band (0.5--2~keV) surface brightness
$S_X$ in units of ergs~s$^{-1}$~pc$^{-2}$.
Spectral fits for each diffuse region consisted of a single absorbed thermal plasma with variable abundances.  Apparent ``rings'' around voids are
an artifact of the map-making; see \S\ref{sec:superspectra} for details.
\label{fig:nhktmaps}}
\end{figure}

\begin{figure}
\centering
  \includegraphics[width=0.9\textwidth]{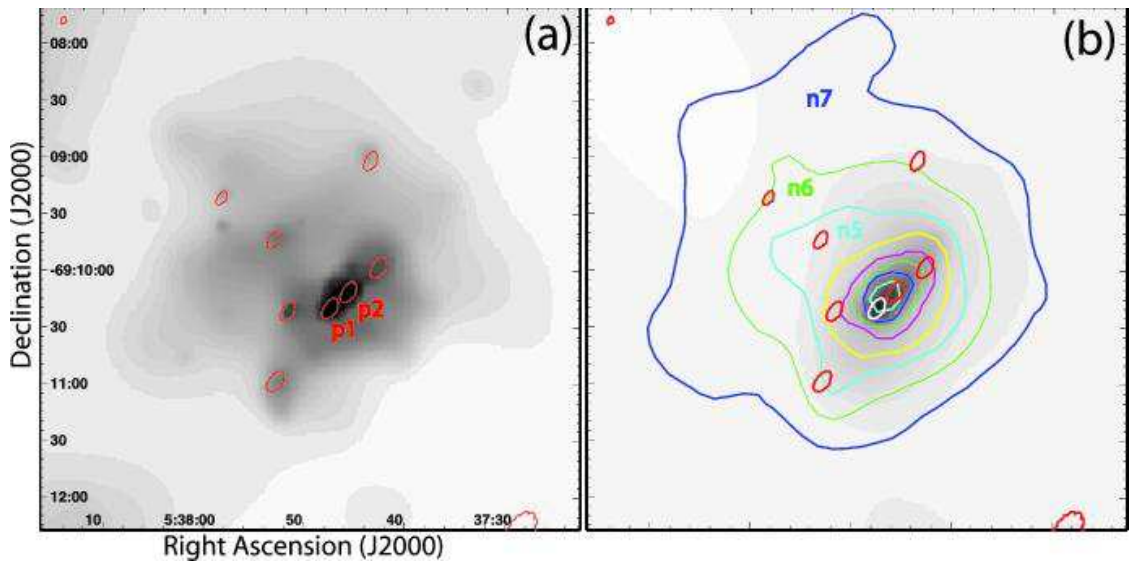}
\caption{The SNR N157B.
(a) Soft-band {\em csmooth} image with pointlike sources shown in red using 
polygons that define their 90\% PSF extraction regions.
(b) Diffuse spectral extraction regions (based on full-band smoothed
images) overlaid on a hard-band {\em csmooth} image, same size as (a).   
\label{fig:N157B-regs}} 
\end{figure}

\begin{figure}
\centering
  \includegraphics[width=0.9\textwidth]{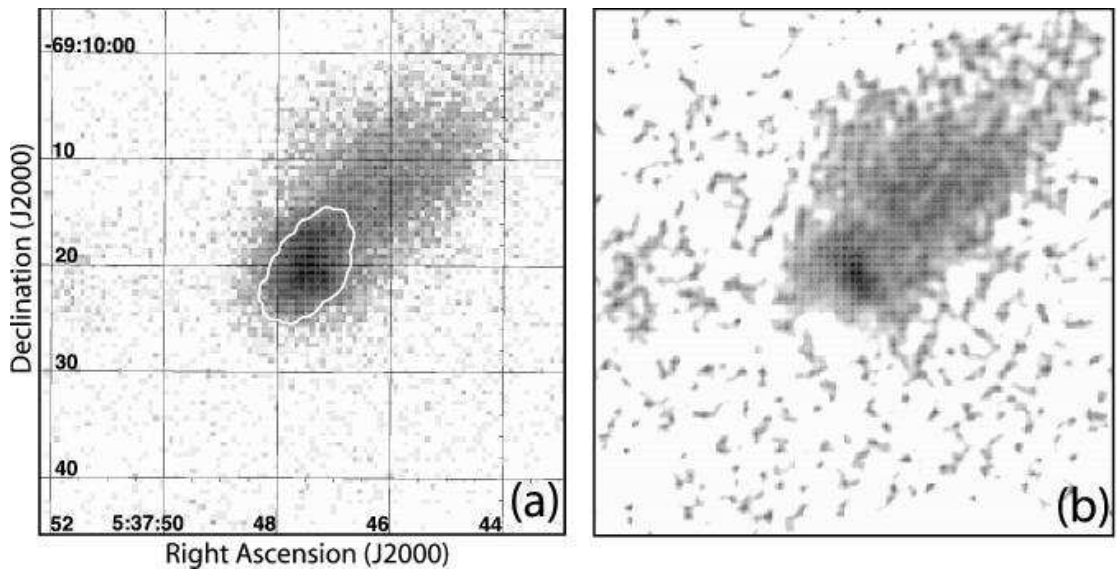}
\caption{The composite SNR N157B, $6.9\arcmin$ off-axis.  
(a) ACIS data, full-band, binned into $0.\arcsec5$ pixels.  The 90\%
contour of the 1.5~keV PSF at the location of the pulsar is shown in
white.
(b) Maximum likelihood reconstruction of (a), 64 iterations.
\label{fig:imageN157B}} 
\end{figure}

\begin{figure}
\centering
\includegraphics[width=0.4\textwidth]{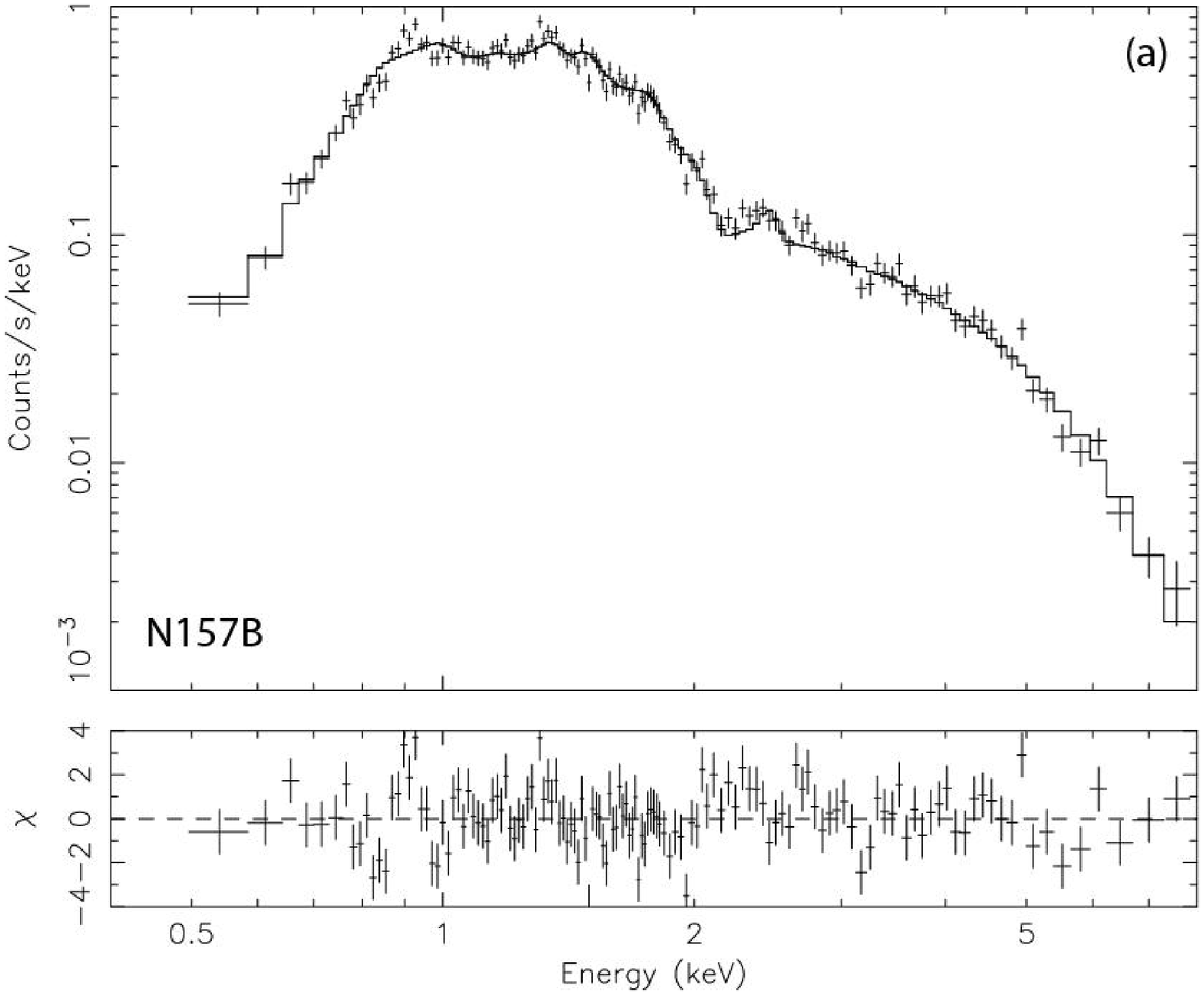}
\hspace*{0.15in}
\includegraphics[width=0.4\textwidth]{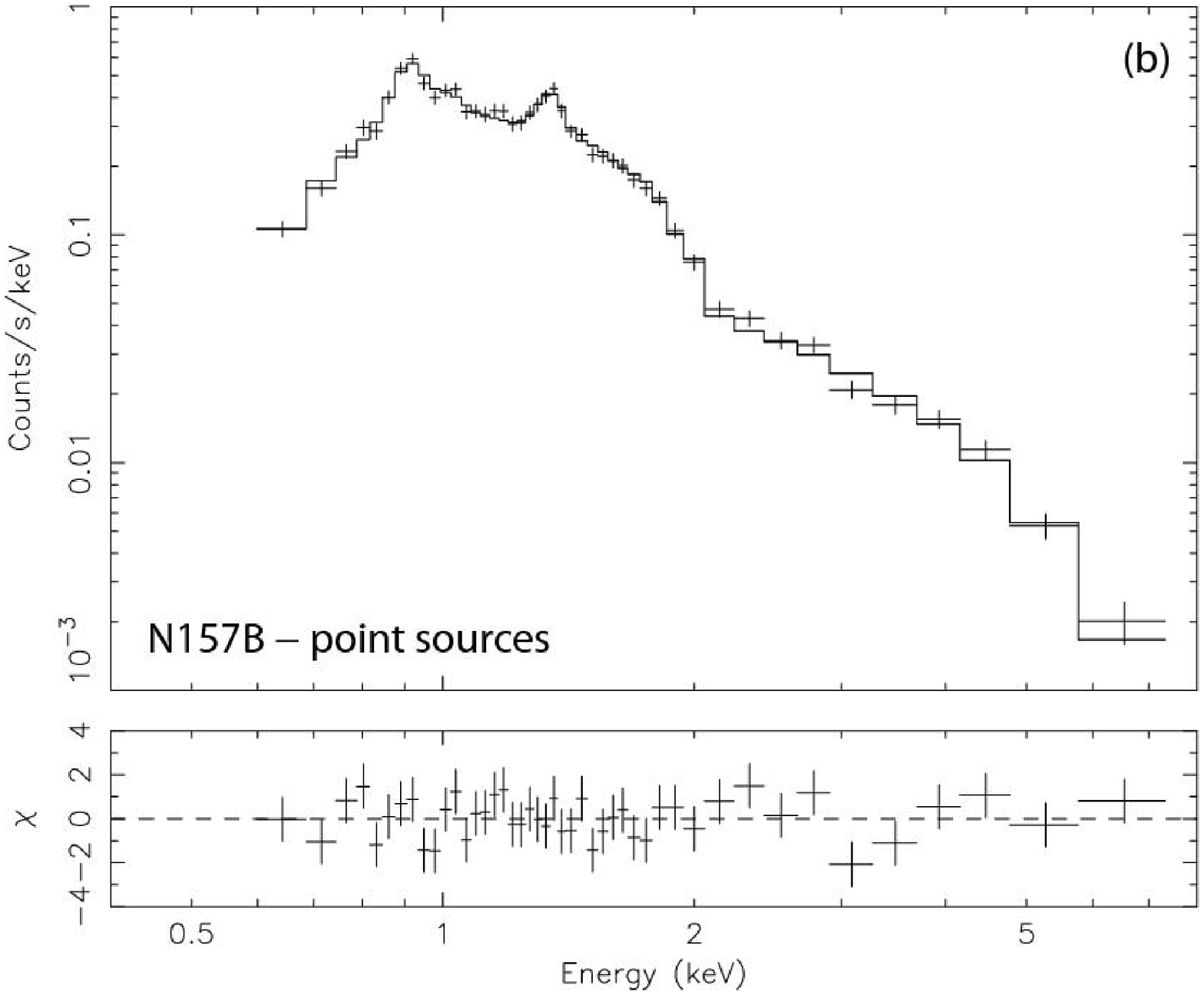}\\
\bigskip
\includegraphics[width=0.4\textwidth]{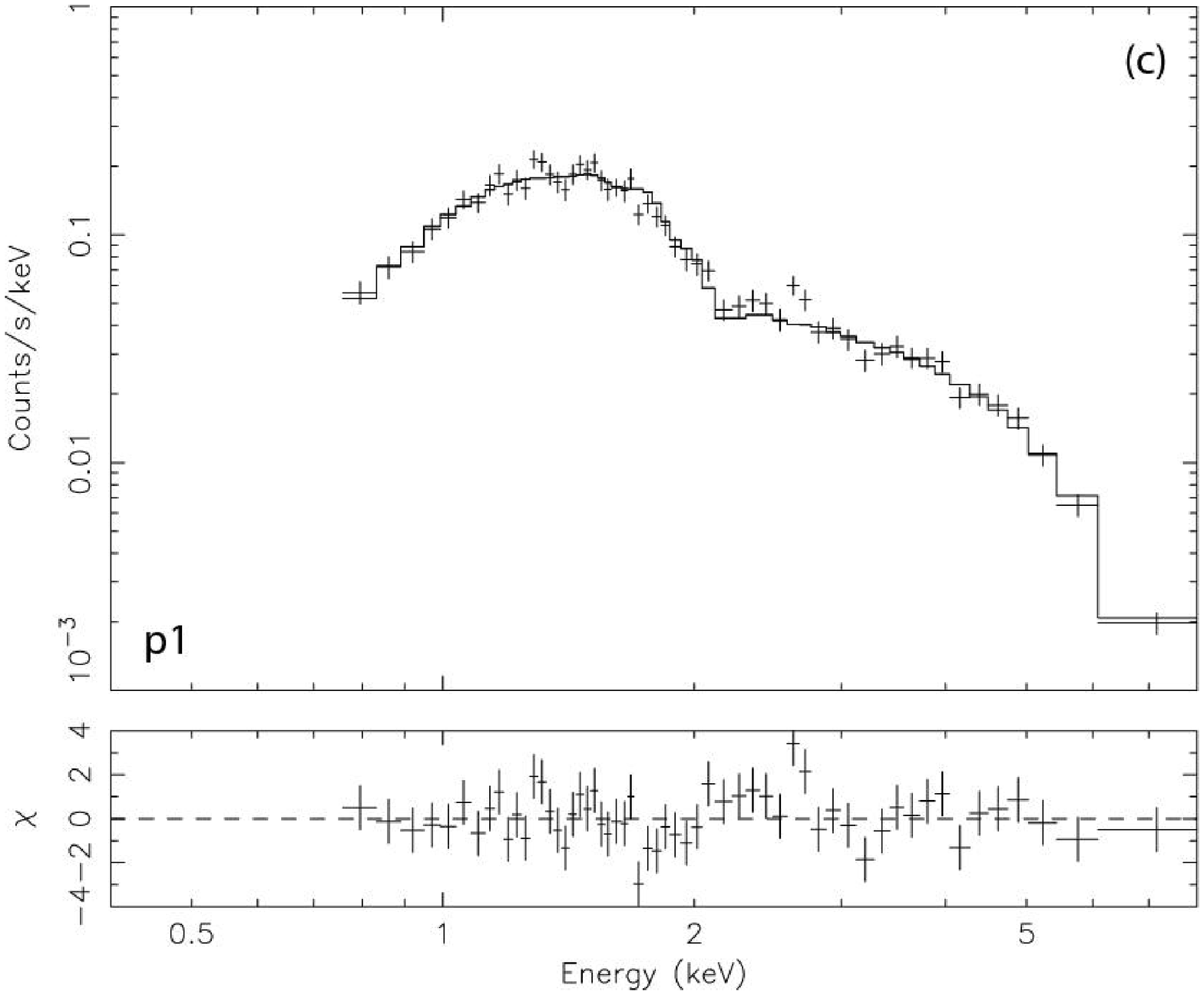}
\hspace*{0.15in}
\includegraphics[width=0.4\textwidth]{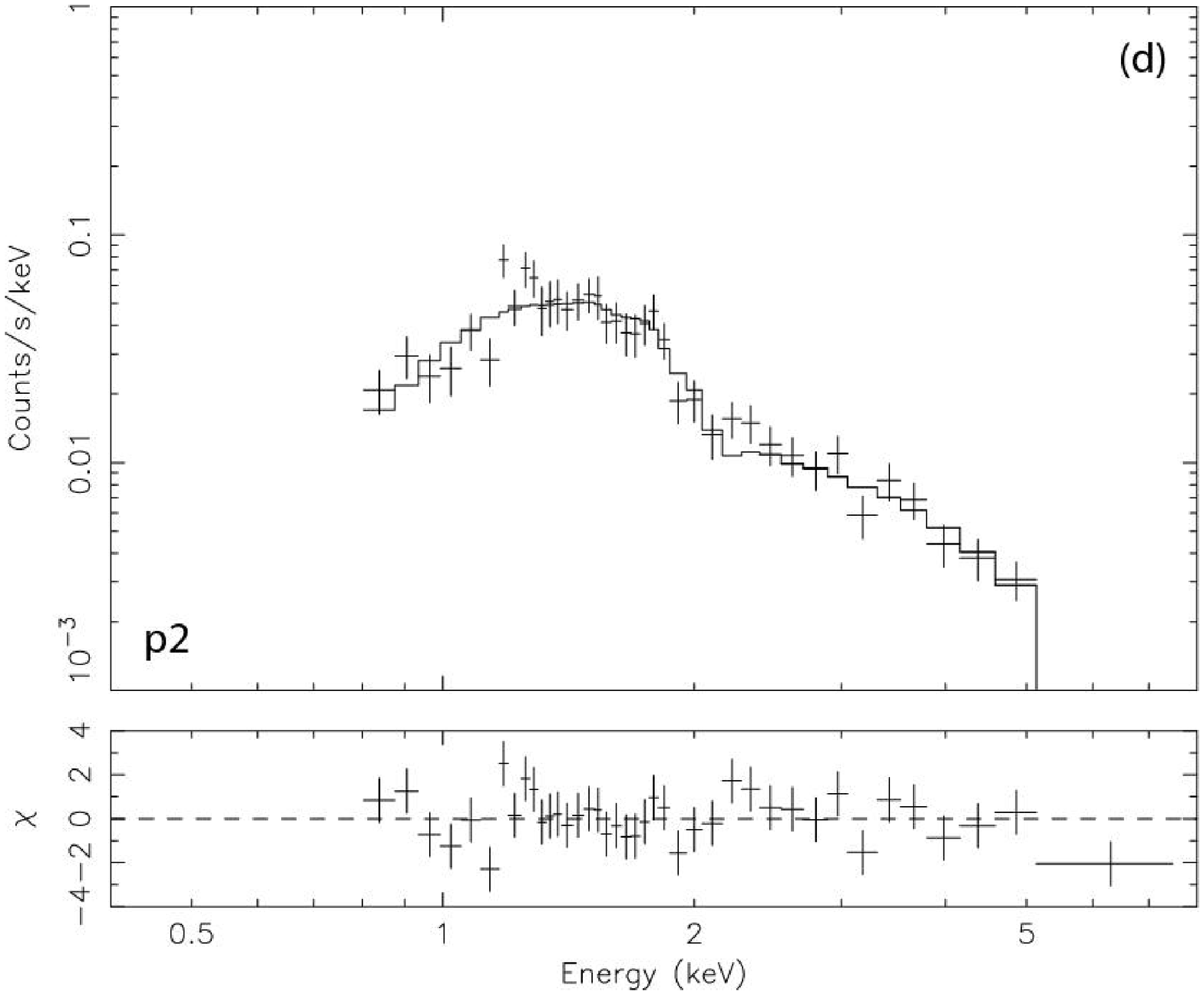}
\caption{Spectra of the X-ray emission in N157B. 
(a) The composite spectrum of the entire N157B SNR (as outlined in
Figure~\ref{fig:introimage}), including all point sources.
(b) Diffuse emission in the same region (excluding all point sources).
(c) The slightly piled-up spectrum of the pulsar (region p1).
(d) Region p2, the pointlike structure in the N157B cometary nebula.
\label{fig:n157bspectra}} 
\end{figure}

\begin{figure}
\centering
  \includegraphics[width=1.0\textwidth]{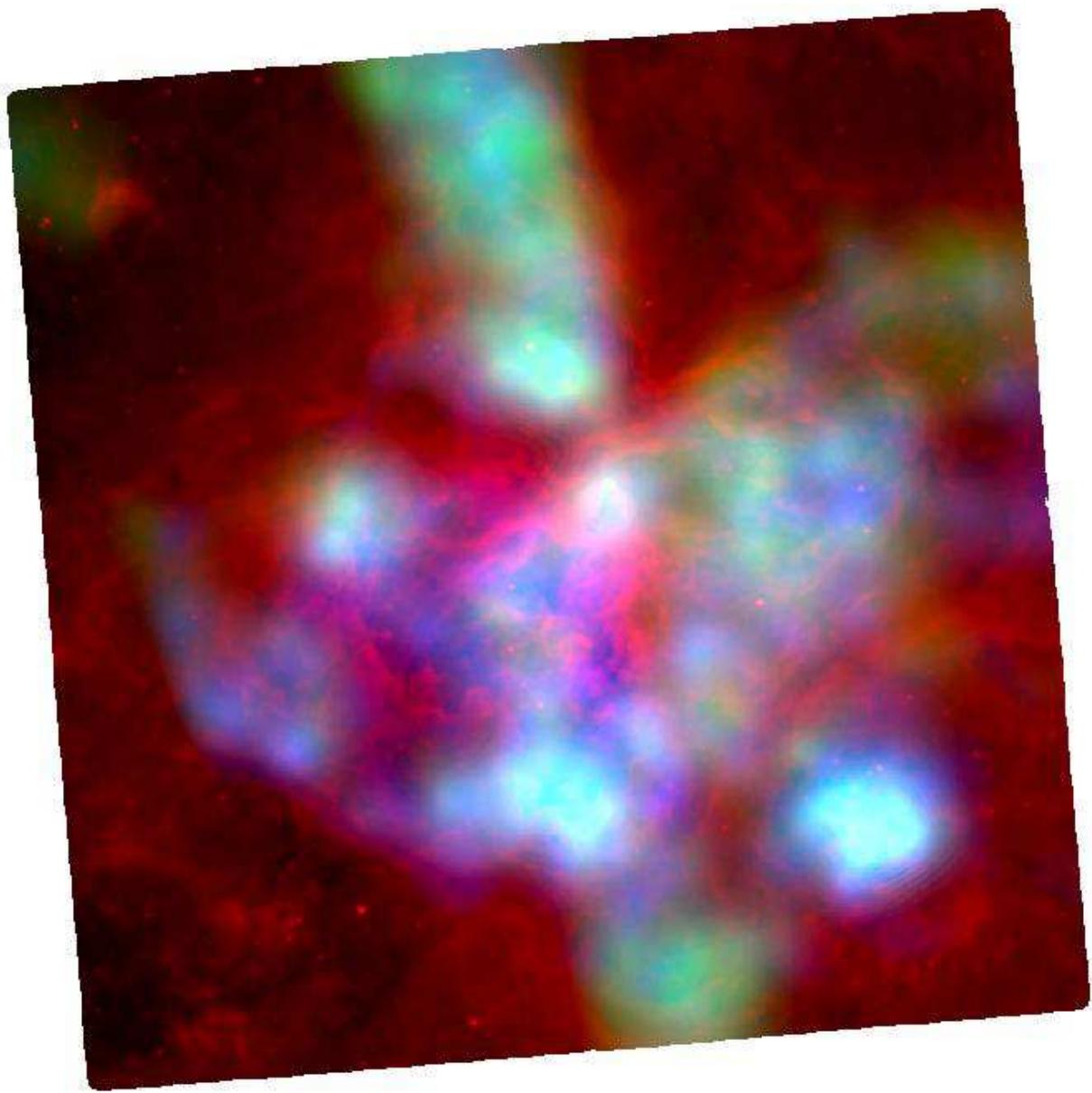}
\caption{X-ray diffuse structures in the H{\sc II} region context: red = MCELS H$\alpha$, green = ACIS 350--900~eV, blue = ACIS 900--2300~eV.       
\label{fig:Halpha}} 
\end{figure}

\begin{figure}
\centering
  \includegraphics[width=1.0\textwidth]{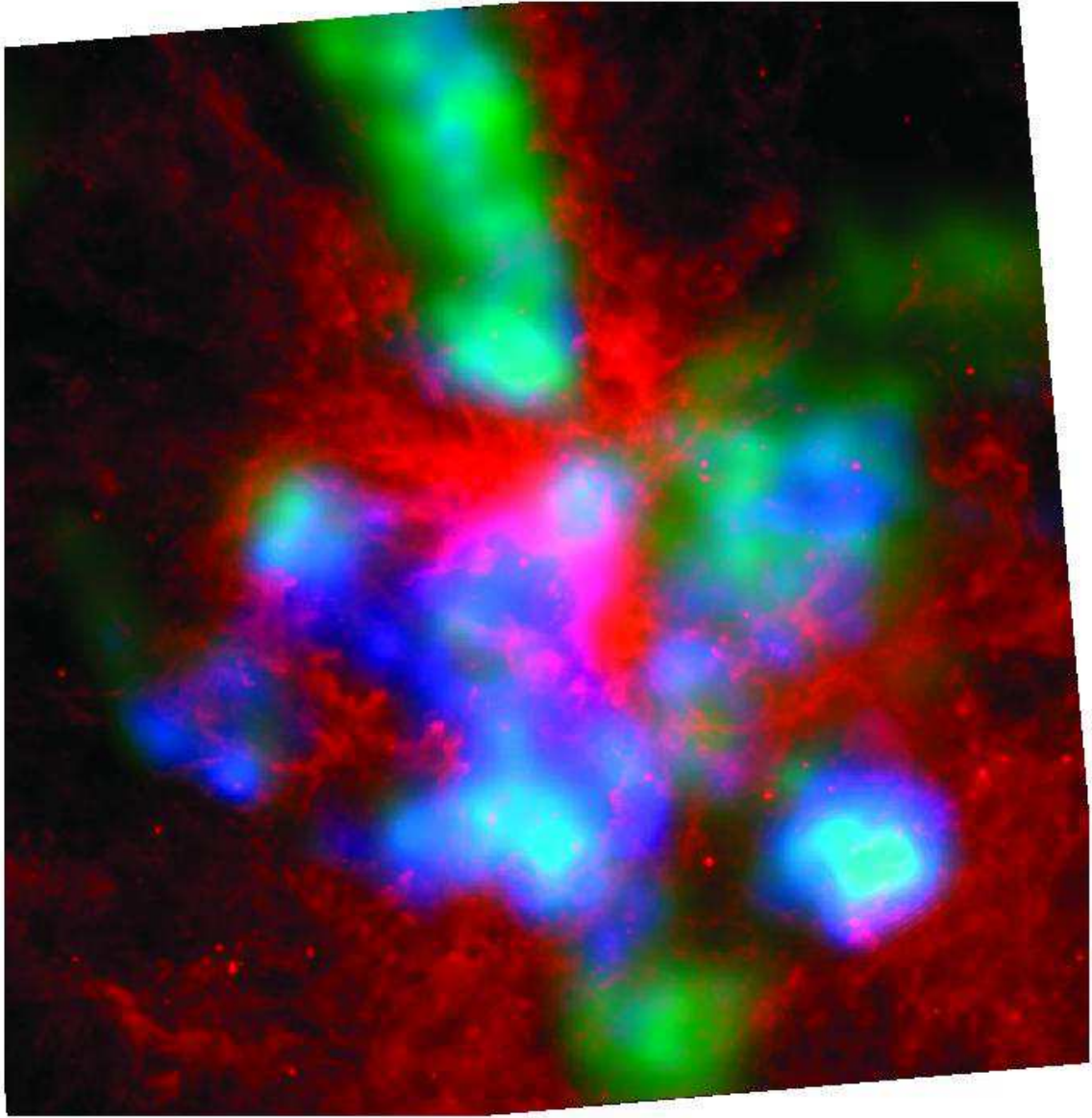}
\caption{A composite of the new mid-infrared (6.5--9.4$\mu$m) {\em
Spitzer}/IRAC image of 30~Dor (in red) with the 350--900~eV and
900--2300~eV adaptively smoothed {\em Chandra}/ACIS images (in green and
blue).  This {\em Spitzer} image was the subject of a January 2004 NASA
press release (BRB, PI).
\label{fig:spitzer}} 
\end{figure}

\begin{figure}
\centering
  \includegraphics[width=1.0\textwidth]{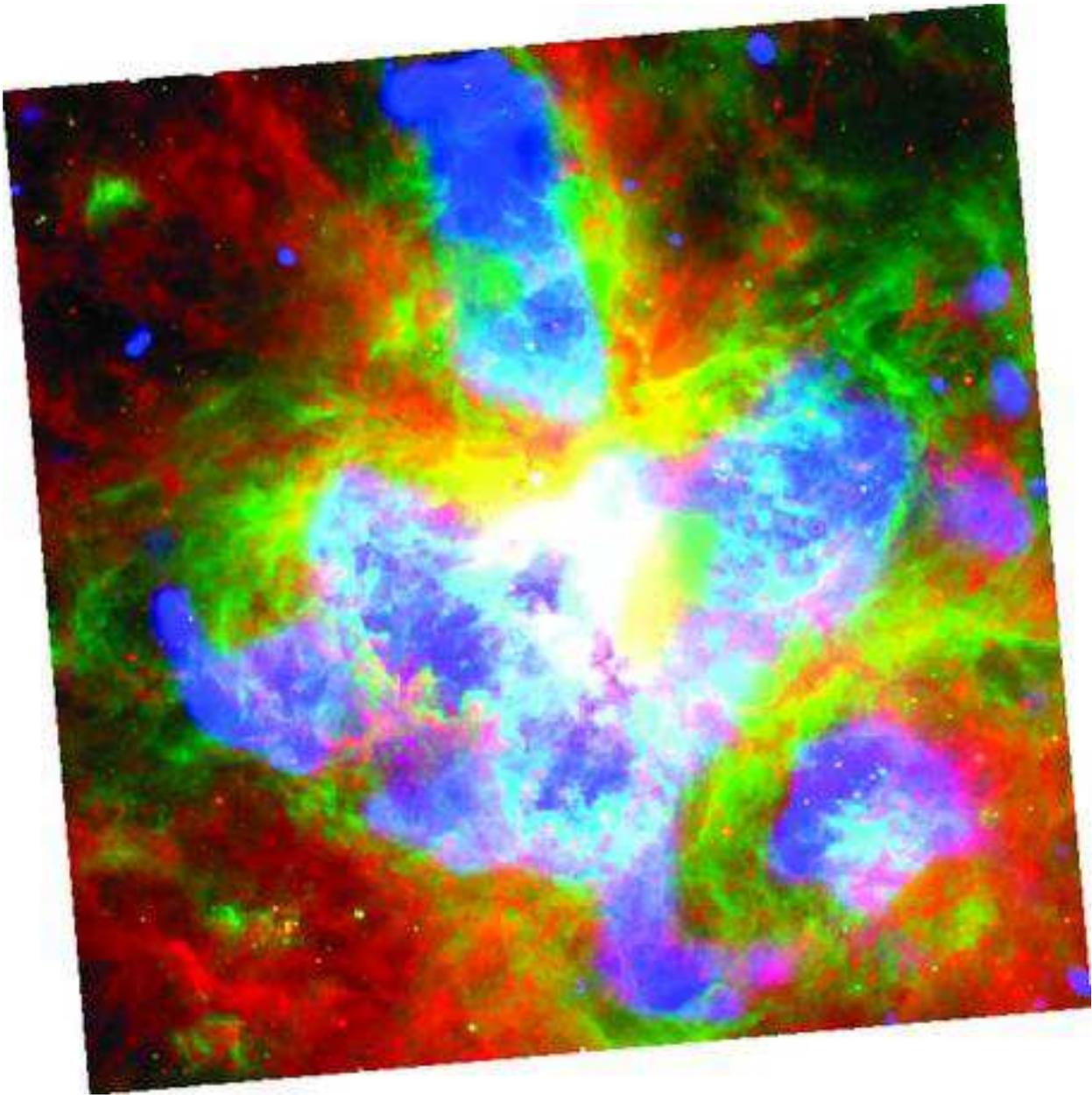}
\caption{ X-ray diffuse structures in the broader context: red = {\em
Spitzer}/IRAC 6.5--9.4$\mu$m, green = MCELS H$\alpha$, blue = {\em
Chandra}/ACIS 900--2300~eV.  X-ray emission is scaled to show large-scale
structures.      
\label{fig:3bands}} 
\end{figure}

\begin{figure}
\centering
  \includegraphics[width=1.0\textwidth]{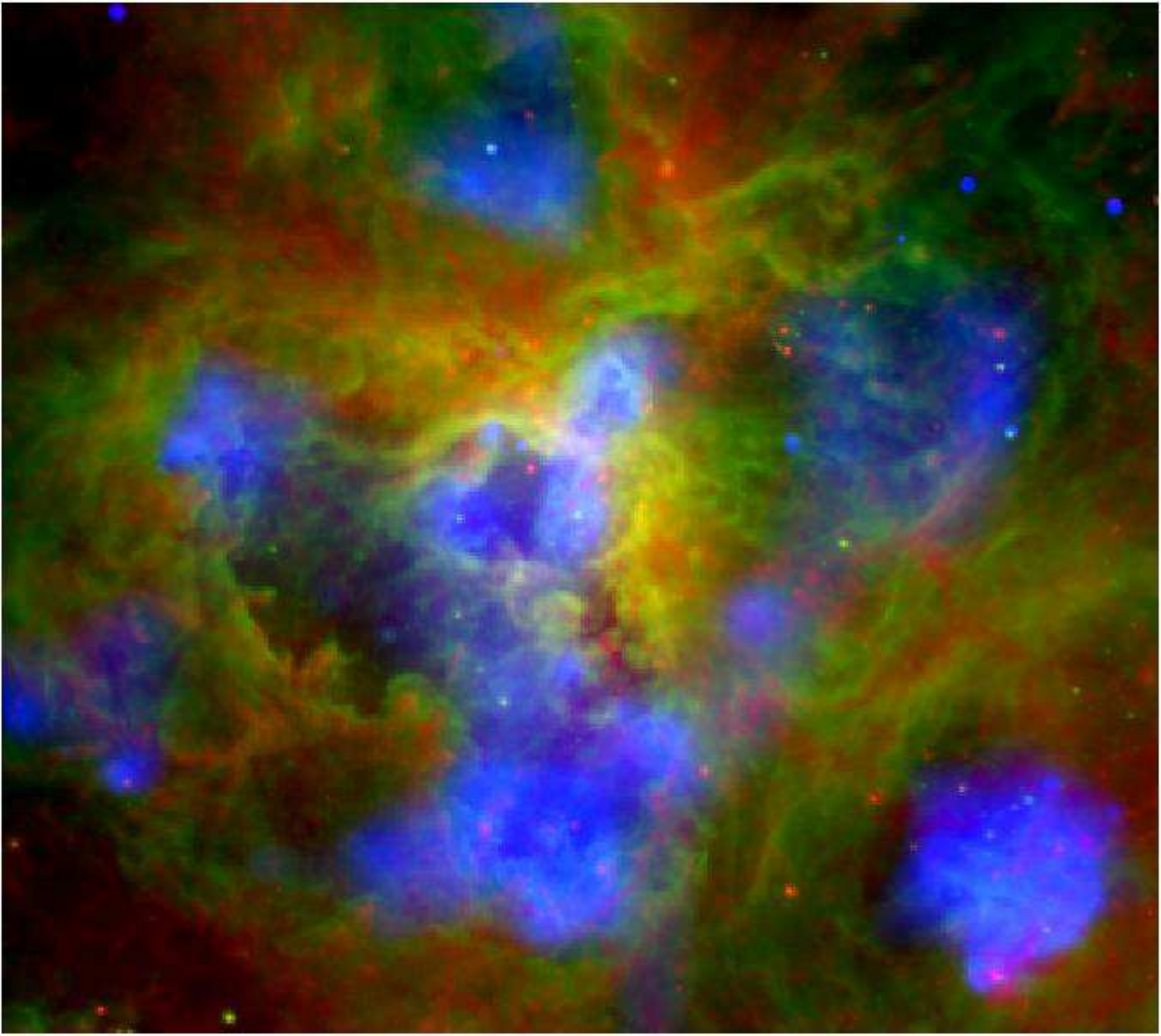}
\caption{ The central $13\arcmin.9 \times 12\arcmin.5$ of
Figure~\ref{fig:3bands}, centered on R136 and scaled to show just the
brighter diffuse X-ray structures.       
\label{fig:3bandszoom}} 
\end{figure}

From this work, we conclude the following:
\begin{itemize}
\item  30~Dor's integrated diffuse emission is well-fit by a
single-temperature plasma with $T \simeq 7$~MK, roughly solar line
strengths, and a single absorbing column density of $N_H \simeq 3
\times 10^{21}$~cm$^{-2}$.  The soft-band luminosity is dominated by the
diffuse structures, but these disappear above 2~keV and the integrated
emission is dominated by just a few bright point sources
(\S\ref{sec:globalspectra} and Paper~II).
\item  ACIS spectra of 30~Dor's diffuse emission regions often suggest
chemical enrichment, another argument in favor of a SNR origin for the
X-ray emission (Table~\ref{tbl:diffspectable}).
\item  Annular spectra around the SNR N157B, centered on
PSR~J0537$-$6910, indicate electron cooling through a steepening power
law slope, a transition from non-thermal to thermal spectra with
distance from the pulsar, and chemical enrichment in the SNR
(\S\ref{sec:N157B}).
\item  A powerful method for understanding the complex spectral and
spatial information contained in high-resolution X-ray studies of
massive star-forming complexes like 30~Dor is to study maps of column
density, plasma temperature, and intrinsic surface brightness derived
from X-ray spectral fitting (Figure~\ref{fig:nhktmaps}).
\item  The main 30~Dor nebula exhibits several X-ray ``voids'' not
caused by obscuration, most notably the hot region 15
(\S\ref{sec:superspectra}).  These resemble the wind collision plasma
seen in nearby H{\sc II} regions like M~17 (\S\ref{sec:othermsfrs}).
\item  Comparing the diffuse X-ray emission with kinematic studies of
the warm gas in H{\sc II} region complexes is essential to disentangle
the confusing morphological information.  While the fast shocks found
in 30~Dor by CK94 often correlate well with regions of high X-ray
surface brightness, the correlation is not complete; some fast shells
lack bright X-rays while some X-ray features are not known to exhibit
high velocities (\S\ref{sec:highvel}).
\item Some shells do not currently contain massive clusters to supply
wind-generated X-rays.  The source of diffuse X-rays may be centered
SNe that heated the interior of these shells without producing fast
shocks against the shell walls \citep{Chu90}.
\item  The X-ray structure that we call the West Ring (our region 17)
is adjacent to CK94's NW Loop and has X-ray properties very similar to
the Galactic cavity SNR HB3.  We propose that it is a cavity SNR that
exploded inside 30~Dor's Shell 3 and that its projenitor most likely
came from the massive cluster Hodge~301 (\S\ref{sec:westring}).
\item  Bright X-ray patches often have cooler plasma temperatures than
their surrounding shells, possibly indicating that the increased
densities associated with fast shocks allow the associated X-ray plasma
to cool more efficiently than the plasma associated with the large,
low-density shells.
\end{itemize}

Our upcoming 100~ks ACIS-I observation of 30~Dor will provide the
high-quality X-ray dataset that is needed to place the X-ray emission
in context with state-of-the-art observations in other wavebands.  We
expect this observation to reveal more of the high-mass stellar
population in 30~Dor and to give even more detailed information on the
complex morphology of the wind-blown bubbles and superbubbles, both
with higher spatial resolution imaging and with spectroscopy on finer
spatial scales.  True progress in understanding, though, will require
cooperation.  By combining the power of today's Great Observatories and
high-quality ground-based data, we are confident that unique insight
awaits.

%==========================================================================
\acknowledgments

Support for this work was provided to Gordon Garmire, the ACIS
Principal Investigator, by the National Aeronautics and Space
Administration (NASA) through NASA Contract NAS8-38252 and {\em
Chandra} Contract SV4-74018 issued by the {\em Chandra X-ray
Observatory} Center, which is operated by the Smithsonian Astrophysical
Observatory for and on behalf of NASA under contract NAS8-03060.  LKT
appreciates technical contributions by Jeremy Sanders and Konstantin
Getman and helpful discussions with Mike Eracleous and Bruce
Elmegreen.  We thank the Magellanic Cloud Emission Line Survey (MCELS)
team for the use of their H$\alpha$ image of 30~Doradus.  We also
thank our anonymous referee for investing the time to review this
paper and for several helpful suggestions.

This work is based in part on observations made with the {\em Spitzer
Space Telescope}, which is operated by the Jet Propulsion Laboratory,
California Institute of Technology under NASA Contract 1407.  This
research made use of data products from the Two Micron All Sky Survey,
which is a joint project of the University of Massachusetts and the
Infrared Processing and Analysis Center/California Institute of
Technology, funded by NASA and the National Science Foundation.  This
research also made use of the SIMBAD database and VizieR catalogue
access tool, operated at CDS, Strasbourg, France.  We are grateful for
the invaluable tools of NASA's Astrophysics Data System.

%==========================================================================

\onecolumn
%=============================================================================
%----------------------------------FIGURES------------------------------------
\clearpage

%-----------------------------------------------------------------------------

%-----------------------------------------------------------------------------
\newpage

%  ***NOTE TO AJ:  THIS FIGURE SHOULD BE IN COLOR ONLY IN THE ELECTRONIC
%  EDITION.  Greyscale version is supplied as f2_grey_bit.ps.***
%-----------------------------------------------------------------------------

%-----------------------------------------------------------------------------
\newpage

%-----------------------------------------------------------------------------

%-----------------------------------------------------------------------------
\newpage

%-----------------------------------------------------------------------------

%-----------------------------------------------------------------------------
\newpage

%-----------------------------------------------------------------------------

%-----------------------------------------------------------------------------
\newpage

%-----------------------------------------------------------------------------

%-----------------------------------------------------------------------------
\newpage

%-------------------------------------------------------------------------

%-----------------------------------------------------------------------------
\newpage

%-------------------------------------------------------------------------

%-----------------------------------------------------------------------------
\newpage

%-----------------------------------------------------------------------------

%-----------------------------------------------------------------------------
\newpage

%-----------------------------------------------------------------------------

%-----------------------------------------------------------------------------
\newpage

%-----------------------------------------------------------------------------

%-----------------------------------------------------------------------------
\newpage

%-----------------------------------------------------------------------------
 
%-----------------------------------------------------------------------------
\newpage

%-----------------------------------------------------------------------------

%-----------------------------------------------------------------------------
\newpage

%-----------------------------------------------------------------------------

%-----------------------------------------------------------------------------
\newpage

%-----------------------------------------------------------------------------

%=============================================================================
%---------------------------------TABLES------------------------------------

\begin{deluxetable}{lcrccc}
\centering 
\rotate
\tabletypesize{\normalsize} \tablewidth{0pt}
\tablecolumns{6}

\tablecaption{ Log of {\it Chandra} Observations 
 \label{tbl:obslog}}

\tablehead{
\colhead{Obs ID} & 
\colhead{Start Time (UT)} & 
\colhead{Exposure} & 
\colhead{Frame} & 
\colhead{Active} & 
\colhead{Mode\tablenotemark{a}} \\
\colhead{} & 
\colhead{1999 Sep 21} & 
\colhead{Time (s)} & 
\colhead{Time (s)} &
\colhead{CCDs} & 
\colhead{} 
}

\startdata
         22 & 19:48 &  1049 & 3.2 & 012367 & F \\
 62520 long & 20:40 & 20613 & 3.3 & 012378 & VF \\
62520 short & 20:40 &   208 & 0.3 & 012378 & VF \\
\enddata

\tablenotetext{a}{ The observing mode:  F=Faint, VF=Very Faint.}

\tablecomments{  Exposure times are the net usable times after various
filtering steps are applied in the data reduction process.   Nominal
pointing (in decimal degrees) for all observations was RA (J2000) =
84.68730356, Dec (J2000) = $-$69.0954973, with a roll angle of
$85^{\circ}$.  These quantities are obtained from the satellite aspect
solution before astrometric correction is applied.}

\end{deluxetable}
%-----------------------------------------------------------------------------

%-----------------------------------------------------------------------------
%\clearpage

\begin{deluxetable}{crclcccclccccccccccc}
\centering \rotate 
\tabletypesize{\tiny} \tablewidth{0pt}
\tablecolumns{20}

\tablecaption{X-ray Spectroscopy of 30~Dor Global Emission
\label{tbl:globalspectable}}

\tablehead{
\colhead{Emission} & \colhead{Net} &
  &
\multicolumn{6}{c}{Spectral Fit Parameters\tablenotemark{a}} &
  &
\multicolumn{4}{c}{Abundances $>0.3Z_{\odot}$} &
  &
\multicolumn{5}{c}{X-ray Luminosities\tablenotemark{b}} \\ 
\cline{4-9} \cline{11-14} \cline{16-20}

\colhead{Component} & \colhead{Counts} &
  &
\colhead{$\log N_H$} & \colhead{$kT$} & \colhead{$\log EM$} & \colhead{$\Gamma$} & \colhead{$\log N_{\Gamma}$} &  \colhead{$\chi^2$/dof} &
  &
\colhead{O} & \colhead{Ne} & \colhead{Mg} & \colhead{S} &
  &
\colhead{$\log L_s$} & \colhead{$\log L_h$} & \colhead{$\log L_{h,c}$} & \colhead{$\log L_t$} & \colhead{$\log L_{t,c}$}   \\

\colhead{} & \colhead{} &
  &
\colhead{(cm$^{-2}$)} & \colhead{(keV)} & \colhead{(cm$^{-3}$)} & \colhead{} & \colhead{} & \colhead{} &
  &
\colhead{} & \colhead{} & \colhead{} & \colhead{} &
  &
\multicolumn{5}{c}{(ergs s$^{-1}$)} \\

\colhead{(1)} & \colhead{(2)} & 
  &
\colhead{(3)} & \colhead{(4)} & \colhead{(5)} & \colhead{(6)} &  \colhead{(7)} & \colhead{(8)} &
  &
\colhead{(9)} & \colhead{(10)} & \colhead{(11)} & \colhead{(12)} &  
  &
\colhead{(13)} & \colhead{(14)} &  \colhead{(15)} & \colhead{(16)} &
\colhead{(17)}  
}

\startdata

all & 55767 && 21.6 & 0.35 & 60.03 & 2.3 & $-$2.3 & 256/157 && 0.6 & 0.8 & 0.6 & \nodata && 36.51 & 36.29 & 36.31 & 36.71 & 37.15 \\
all $-$ N157B & 35634 && 21.6 & 0.37 & 59.93 & 2.8 & $-$2.8 & 149/95 && 0.6 & 0.8 & 0.6 & 1.5 && 36.35 & 35.61 & 35.64 & 36.42 & 36.95 \\
diffuse & 32306 && 21.4 & 0.58 & 59.58 & \nodata & \nodata & 146/85 && 1.2 & 1.1 & 0.7 & \nodata && 36.32 & 35.09 & 35.12 & 36.34 & 36.67 \\
pt srcs & 2954 && 21.5 & 1.0 & 58.19 & 1.8 & $-$3.3 & 106/103 && \nodata & \nodata & \nodata & \nodata && 35.31 & 35.63 & 35.65 & 35.80 & 35.95 \\
N157B & 20216 && 21.6 & 0.71 & 58.88 & 2.1 & $-$2.6 & 243/130 && 0.8 & \nodata & 1.1 & 2.1 && 35.98 & 36.17 & 36.20 & 36.39 & 36.62 \\

\enddata

\tablenotetext{a}{ All thermal plasma fits were {\it wabs*(apec +
powerlaw)} or {\it wabs*(vapec + powerlaw)} in {\it XSPEC} and assumed
abundances of $0.3Z_{\odot}$ for all elements unless otherwise noted.
$\Gamma$ is the power law photon index, $N_{\Gamma}$ is the power law
normalization (in units of photons~cm$^{-2}$~s$^{-1}$~keV$^{-1}$ at
1~keV), ``dof'' = degrees of freedom, and $\chi^2$/dof represents
the goodness of fit.}

\tablenotetext{b}{ X-ray luminosities:  s = soft band (0.5--2 keV); h =
hard band (2--8 keV); t = total band (0.5--8 keV).
Absorption-corrected luminosities are subscripted with a $c$.}

\tablecomments{See \S\ref{sec:globalspectra} for descriptions of the
columns.  Uncertainties (90\% confidence intervals) on $\log N_H$ are
$\leq \pm 0.07$.  Uncertainties on $kT$ are $\leq \pm 0.04$ except for
the composite point source fit, where $kT = 1.0^{+0.23}_{-0.14}$.
Uncertainties on the power law slope $\Gamma$ are $\leq \pm 0.3$.
Uncertainties on $\log EM$, $\log N_{\Gamma}$, and abundances are
typically $\pm 0.1$ or smaller except for the sulfur abundance, which
is not as well constrained.  A distance of 50.0~kpc was assumed
throughout. }

\end{deluxetable}
%-----------------------------------------------------------------------------

%-----------------------------------------------------------------------------
\clearpage

\thispagestyle{empty}

\begin{deluxetable}{crccccccccccccccccccrc}
\centering \rotate 
\tabletypesize{\tiny} \tablewidth{0pt}
\tablecolumns{22}

\tablecaption{X-ray Spectroscopy of Diffuse Regions
\label{tbl:diffspectable}}

\tablehead{
\colhead{Diff} & \colhead{Net} &
  &
\multicolumn{6}{c}{Spectral Fit Parameters\tablenotemark{a}} &
  &
\multicolumn{4}{c}{Abundances $>0.3Z_{\odot}$} &
  &
\multicolumn{4}{c}{X-ray Luminosities\tablenotemark{b}} &
  &
\colhead{Diffuse} & \colhead{$\log S_X$\tablenotemark{c}} \\ 
\cline{4-9} \cline{11-14} \cline{16-19}

\colhead{Reg} & \colhead{Counts} &
  &
\colhead{$\log N_H$} & \colhead{$kT$} & \colhead{$\log EM$} & \colhead{$\Gamma$} & \colhead{$\log N_{\Gamma}$} &  \colhead{$\chi^2$/dof} &
  &
\colhead{O} & \colhead{Ne} & \colhead{Mg} & \colhead{Other} &
  &
\colhead{$\log L_s$} & \colhead{$\log L_{s,c}$} & \colhead{$\log L_t$} & \colhead{$\log L_{t,c}$} &
  &
\colhead{Area} & \colhead{(ergs~s$^{-1}$}  \\

\colhead{\#} & \colhead{} &
  &
\colhead{(cm$^{-2}$)} & \colhead{(keV)} & \colhead{(cm$^{-3}$)} & \colhead{} & \colhead{} & \colhead{} &
  &
\colhead{} & \colhead{} & \colhead{} & \colhead{hi} &
  &
\multicolumn{4}{c}{(ergs~s$^{-1}$)} &
  &
\colhead{(arcmin$^2$)}  & \colhead{pc$^{-2}$)} \\

\colhead{(1)} & \colhead{(2)} & 
  &
\colhead{(3)} & \colhead{(4)} & \colhead{(5)} & \colhead{(6)} &  \colhead{(7)} & \colhead{(8)} &
  &
\colhead{(9)} & \colhead{(10)} & \colhead{(11)} & \colhead{(12)} &  
  &
\colhead{(13)} & \colhead{(14)} &  \colhead{(15)} & \colhead{(16)} &
  &
\colhead{(17)} & \colhead{(18)} 
}

\startdata
  
 1 &37299 && 21.6 &  0.3 & 60.0 & 2.8 & $-$2.6 & 163/102 && 0.6 & 0.8 & 0.6 & \nodata && 36.4 & 37.0 & 36.4 & 37.0 && 101.37 & 32.7 \\
 2 & 4039 && 21.4 &  0.6 & 58.7 & \nodata & \nodata & 59/51 && 1.2 & 0.5 & 0.6 & \nodata && 35.4 & 35.8 & 35.4 & 35.8 &&  56.38 & 31.7 \\
 3 & 1355 && 21.6 &  0.3 & 58.8 & \nodata & \nodata & 60/40 && 0.5 & 0.7 & 0.7 & \nodata && 35.0 & 35.7 & 35.0 & 35.7 &&   6.91 & 32.5 \\
 4 &  360 && 21.3 &  0.3 & 58.0 & \nodata & \nodata & 11/13 && 0.5 & 0.9 & 1.2 & \nodata && 34.4 & 34.8 & 34.4 & 34.9 &&   1.00 & 32.5 \\
 5 &  840 && 21.7 &  0.3 & 58.8 & \nodata & \nodata & 28/31 && \nodata & \nodata & \nodata & \nodata && 34.7 & 35.5 & 34.7 & 35.5 &&   1.40 & 33.1 \\
 6 &  355 && 21.7 &  0.4 & 57.9 & \nodata & \nodata & 24/25 && hi & hi & hi & Al      && 34.3 & 34.9 & 34.3 & 34.9 &&   0.63 & 32.8 \\
 7 &  608 && 21.8 &  0.4 & 58.3 & \nodata & \nodata & 39/41 && 1.8 & 1.4 & \nodata & \nodata && 34.5 & 35.5 & 34.6 & 35.5 &&   1.06 & 33.1 \\
 8 &  458 && 21.7 &  0.4 & 58.1 & \nodata & \nodata & 29/36 && 1.2 & 1.3 & 0.8 & \nodata && 34.4 & 35.2 & 34.4 & 35.2 &&   1.07 & 32.9 \\
 9 & 1460 && 21.7 &  0.6 & 58.5 & \nodata & \nodata & 44/39 && 1.3 & \nodata & \nodata & \nodata     && 34.9 & 35.5 & 35.0 & 35.6 &&   4.49 & 32.6 \\
10 & 3454 && 21.8 &  0.4 & 59.3 & \nodata & \nodata & 56/43 && 0.5 & 0.7 & 0.6 & \nodata && 35.3 & 36.2 & 35.3 & 36.2 &&   8.93 & 32.9 \\
11 & 2338 && 21.7 &  0.4 & 58.8 & \nodata & \nodata & 44/34 && 1.0 & 1.3 & 0.9 & \nodata && 35.2 & 35.8 & 35.2 & 35.8 &&   2.54 & 33.1 \\
12 & 2194 && 21.8 &  0.5 & 58.9 & \nodata & \nodata & 35/36 && \nodata & 0.9 & 0.7 & \nodata && 35.1 & 35.8 & 35.2 & 35.8 &&  12.49 & 32.4 \\
13 & 1223 && 21.8 &  0.3 & 59.1 & \nodata & \nodata & 20/34 && \nodata & \nodata & \nodata & \nodata && 34.9 & 35.8 & 34.9 & 35.8 &&   3.65 & 32.9 \\
14 &  854 && 21.5 &  0.6 & 58.2 & \nodata & \nodata & 34/40 && \nodata & \nodata & \nodata & \nodata && 34.7 & 35.2 & 34.8 & 35.2 &&   3.26 & 32.3 \\
15 &  222 && 21.8 &  0.8 & 57.8 & \nodata & \nodata & 26/32 && \nodata & \nodata & \nodata & \nodata && 34.1 & 34.7 & 34.2 & 34.8 &&   2.73 & 32.0 \\
16 & 2598 && 21.4 &  0.5 & 58.7 & \nodata & \nodata & 70/38 && 0.7 & 0.8 & 0.5 & \nodata && 35.2 & 35.6 & 35.2 & 35.6 &&  12.48 & 32.2 \\
17 & 2939 && 21.4 &  0.5 & 57.6 & \nodata & \nodata & 28/21 && 1.1 & 1.1 & \nodata & \nodata && 35.3 & 35.7 & 35.3 & 35.7 &&   7.97 & 32.4 \\
18 & 1092 && 21.7 &  0.3 & 58.7 & \nodata & \nodata & 41/23 && 1.3 & 1.1 & \nodata & \nodata && 34.9 & 35.8 & 34.9 & 35.8 &&  10.00 & 32.5 \\
19 &  122 && 21.6 &  0.3 & 57.8 & \nodata & \nodata & 19/13 && \nodata & \nodata & \nodata & \nodata && 33.9 & 34.6 & 33.9 & 34.6 &&   2.42 & 31.8 \\
20 &  309 && 21.6 &  0.6 & 57.8 & \nodata & \nodata & 36/35 && \nodata & \nodata & \nodata & \nodata && 34.2 & 34.7 & 34.2 & 34.7 &&   2.07 & 32.1 \\
21 &   86 &&{\it 21.5}&0.6&57.2 & \nodata & \nodata & 11/13 && \nodata & \nodata & \nodata & \nodata && 33.7 & 34.2 & 33.7 & 34.2 &&   1.16 & 31.8 \\
22 &  158 && 21.7 &  0.3 & 57.8 & \nodata & \nodata & 20/18 && \nodata & hi & hi & Al && 34.0 & 34.7 & 34.0 & 34.7 &&   3.62 & 31.8 \\
23 &  228 && 21.1 &  0.3 & 57.8 & \nodata & \nodata & 30/35 && \nodata & \nodata & \nodata & \nodata && 34.3 & 34.5 & 34.3 & 34.5 &&   4.59 & 31.5 \\
24 & 2692 && 21.2 &  0.5 & 58.4 & \nodata & \nodata & 59/39 && 1.1 & 0.8 & 0.6 & \nodata && 35.2 & 35.5 & 35.3 & 35.5 &&  12.70 & 32.1 \\
25 & 1050 && 21.5 &  0.3 & 58.6 & \nodata & \nodata & 41/37 && 0.6 & \nodata & \nodata & Si=0.8 && 34.8 & 35.5 & 34.9 & 35.5 &&   1.95 & 32.8 \\
26 &  493 && 21.6 &  0.3 & 58.5 & \nodata & \nodata & 21/20 && \nodata & \nodata & \nodata & \nodata && 34.5 & 35.2 & 34.5 & 35.2 &&   1.54 & 32.7 \\
27 &  880 && 21.2 &  0.5 & 58.0 & \nodata & \nodata & 29/35 && 1.3 & 0.9 & \nodata & \nodata && 34.8 & 35.0 & 34.8 & 35.1 && 2.38 & 32.3 \\
\\
p1 & 6129 && 21.8 &\nodata &\nodata & 2.0 & $-$2.9 & 67/56 && \nodata & \nodata & \nodata & \nodata && 35.4 & 35.9 & 36.1 & 36.2 && 0.02 & 35.3 \\
p2 & 1515 && 21.8 &\nodata &\nodata & 2.3 & $-$3.3 & 38/36 && \nodata & \nodata & \nodata & \nodata && 34.9 & 35.5 & 35.4 & 35.7 && 0.02 & 34.8 \\
\\
n0 & 1754 && 21.8 &\nodata &\nodata & 2.5 & $-$3.3 & 28/47 && \nodata & \nodata & \nodata & \nodata && 34.9 & 35.5 & 35.4 & 35.7 && 0.05 & 34.5 \\
n1 & 1869 && 21.8 &  0.6 & 57.2 & 2.4 & $-$3.4 & 45/42 && \nodata & hi & hi & Al, Si, Ar && 34.9 & 35.5 & 35.4 & 35.7 &&   0.09 & 34.2 \\
n2 &  849 && 21.7 &  0.9 & 57.5 & 2.7 & $-$3.8 & 33/29 && \nodata & \nodata & \nodata &S, Ca && 34.6 & 35.2 & 35.0 & 35.3 && 0.11 & 33.8 \\
n3 &  649 && 21.5 &  0.7 & 57.7 & 2.6 & $-$4.2 & 34/23 && \nodata & \nodata & \nodata & Al && 34.6 & 35.0 & 34.7 & 35.0 &&   0.19 & 33.4 \\
n4 & 1130 &&{\it 21.6}&0.3&56.3 & 3.4 & $-$3.7 & 33/37 && \nodata & hi & hi & Al, Si, S && 34.8 & 35.4 & 34.9 & 35.4 &&   0.45 & 33.4 \\
n5 & 1264 && 21.7 &  0.3 & 58.5 & 3.5 & $-$3.7 & 47/27 && 0.7 & hi & hi & \nodata && 34.9 & 35.7 & 35.0 & 35.7 && 1.01 & 33.4 \\
n6 & 1032 && 21.8 &  0.3 & 58.8 & 2.2 & $-$4.3 & 30/26 && 0.7 & 0.8 & 0.6 & \nodata && 34.8 & 35.7 & 34.9 & 35.7 && 1.89 & 33.1 \\
n7 & 1022 && 21.5 &  0.7 & 58.2 & \nodata & \nodata & 54/31 && 1.4 & \nodata & 0.6 & \nodata && 34.8 & 35.2 & 34.8 & 35.2 && 5.03 & 32.2 \\
\\
n8 & 9755 && 21.5 &  0.3 & 58.4 & 2.6 & $-$2.8 & 36/38 && \nodata & hi & hi & \nodata && 35.7 & 36.1 & 36.0 & 36.3 && 9.78 & 32.8 \\

\enddata

\tablenotetext{a}{ All thermal plasma fits for regions 1--27 were {\it
wabs*(apec)} or {\it wabs*(vapec)} in {\it XSPEC} and assumed
$0.3Z_{\odot}$ abundances for all elements unless otherwise noted.
Fits to region 1 and N157B (p1--n8) additionally included a power law
component.  $\Gamma$ is the power law photon index, $N_{\Gamma}$ is the
power law normalization, ``dof'' = degrees of freedom, and $\chi^2$/dof
represents the goodness of fit.}

\tablenotetext{b}{ X-ray luminosities:  s = soft band (0.5--2 keV); h =
hard band (2--8 keV); t = total band (0.5--8 keV).
Absorption-corrected luminosities are subscripted with a $c$.}

\tablenotetext{c}{ Soft band surface brightness, corrected for absorption.}

\tablecomments{See \S\ref{sec:superspectra} for descriptions of the
columns.  Quantities {\it in italics} were frozen in the spectral fit;
see the text for details.  A distance of 50.0~kpc was assumed
throughout; at that distance, $1\arcmin^2 \simeq 211$~pc$^2$.  Uncertainties
are omitted due to space constraints in the table; they depend on the
number of counts used in the fit of course but are typically similar to
the uncertainties described in the Table~\ref{tbl:globalspectable} notes. }

\end{deluxetable}

%-----------------------------------------------------------------------------
\newpage

\begin{deluxetable}{lcll}
\centering 
\rotate
\tabletypesize{\normalsize} \tablewidth{0pt}
\tablecolumns{4}

\tablecaption{ ACIS Spectral Extraction Regions and H$\alpha$ Kinematics 
 \label{tbl:kinematics}}

\tablehead{
\colhead{ACIS Region} & 
\colhead{CK94 Echellogram} & 
\colhead{CK94 Notes} &
\colhead{Expansion velocity (km~s$^{-1}$)} 
}

\startdata
 5		& Fig.\ 7b	& R139W fast shell	& $-$120	\\
 6		& Fig.\ 7a	& R136E fast shell	& $-$130	\\
 5 left edge, 9	& Fig.\ 6d	& R136's massive slow shell	& $\pm$40 \\
 7, 10, 11	& Fig.\ 3	& Shell 1 network	& 70 and higher	\\
 8 & Fig.\ 7g	& $\sim 1\arcmin$ extended high-velocity feature &  90--120 \\
11		& Fig.\ 3	& Shell 1 south rim	& 70 and higher	\\
13 left	of center & Fig.\ 6b	& $2.4\arcmin$ slow shell & 85	\\
13 right half	& Fig.\ 7h	& $30\arcsec$ high-velocity feature & $>-$100  \\
15 left half	& \nodata	& 20~pc fast shell between Shells 1 \& 2 & 110 \\
16--19		& Fig.\ 8	& Shell 3 network	& 20--200  \\
17 (West Ring)	& Fig.\ 7c \& d	& NW Loop fast shell	& $-$200	\\
24, 26, 27	& Fig.\ 7f	& Shell 5 network	& 60 and higher	\\
25 top		& Fig.\ 6c	& slow shell		& 70	\\
\enddata

\tablecomments{  CK94 = \citet{Chu94}; echellogram labels refer to figure
numbers from that paper.  }

\end{deluxetable}
%-----------------------------------------------------------------------------


\newpage

\begin{thebibliography}{}

\bibitem[Albacete Colombo et al.(2003)]{Albacete03} Albacete 
Colombo, J.~F., M{\' e}ndez, M., \& Morrell, N.~I.\ 2003, \mnras, 346, 704 
 
\bibitem[Arnaud(1996)]{Arnaud96} Arnaud, K.~A.\ 1996, in ASP
Conf.~Ser.~101, Astronomical Data Analysis Software and Systems V, ed.\
G.~H.~Jacoby \& J.~Barnes (San Francisco: ASP), 17

\bibitem[Arthur \& Henney(1996)]{Arthur96} Arthur, S.~J., \& 
Henney, W.~J.\ 1996, \apj, 457, 752 
 
\bibitem[Bamba et al.(2004)]{Bamba04} Bamba, A., Ueno, M., 
Nakajima, H., \& Koyama, K.\ 2004, \apj, 602, 257 

\bibitem[Behar et al.(2001)]{Behar01} Behar, E., Rasmussen, 
A.~P., Griffiths, R.~G., Dennerl, K., Audard, M., Aschenbach, B., \& 
Brinkman, A.~C.\ 2001, \aap, 365, L242 

\bibitem[Bica et al.(1999)]{Bica99} Bica, E.~L.~D., Schmitt, 
H.~R., Dutra, C.~M., \& Oliveira, H.~L.\ 1999, \aj, 117, 238 
 
\bibitem[Brandl(2005)]{Brandl05} Brandl, B.~R.\ 2005, in ASSL 
Vol.~329, Starbursts: From 30 Doradus to Lyman Break Galaxies, ed.\
R.~de Grijs \& R.~M.~Gonz{\' a}lez Delgado (Dordrecht: Springer), 49 
 
\bibitem[Broos et al.(2000)]{Broos00} Broos, P., et al.\ 2000, User's Guide for the TARA Package (University Park: Pennsylvania State Univ.) 

\bibitem[Broos et al.(2002)]{Broos02} Broos, P.~S., Townsley, L.~K., Getman, K.~V., \& Bauer, F.~E.\ 2002, ACIS Extract, An ACIS Point Source Extraction Package (University Park: Pennsylvania State Univ.)

\bibitem[Cant\'{o}, Raga, \& Rodr\'{\i}guez(2000)]{Canto00} Cant\'{o}, J.,
Raga, A.~C., \& Rodr\'{\i}guez, L.~F.\ 2000, \apj, 536, 896

\bibitem[Chen et al.(2005)]{Chen05} Chen, C.-H.~R., Chu, Y., 
\& Johnson, K.~E.\ 2005, \apj, 619, 779 
 
\bibitem[Chu \& Mac Low(1990)]{Chu90} Chu, Y.~\& Mac Low, M.\ 1990,
\apj, 365, 510

\bibitem[Chu et al.(1992)]{Chu92} Chu, Y., Kennicutt, R.~C., 
Schommer, R.~A., \& Laff, J.\ 1992, \aj, 103, 1545 

\bibitem[Chu et al.(1993)]{Chu93} Chu, Y., Low, M.~M., 
Garcia-Segura, G., Wakker, B., \& Kennicutt, R.~C.\ 1993, \apj, 414, 213 
 
\bibitem[Chu \& Kennicutt(1994)]{Chu94} Chu, Y., \& 
Kennicutt, R.~C.\ 1994, \apj, 425, 720 

\bibitem[Chu et al.(1995a)]{Chu95} Chu, Y., Chang, H., Su, Y., 
\& Mac Low, M.\ 1995a, \apj, 450, 157 

\bibitem[Chu et al.(1995b)]{Chu95b} Chu, Y., Dickel, J.~R., 
Staveley-Smith, L., Osterberg, J., \& Smith, R.~C.\ 1995b, \aj, 109, 1729 

\bibitem[Chu et al.(2000)]{Chu00} Chu, Y., Kim, S., Points, 
S.~D., Petre, R., \& Snowden, S.~L.\ 2000, \aj, 119, 2242 

\bibitem[Chu et al.(2004)]{Chu04} Chu, Y., Gruendl, R.~A., 
Chen, C.-H.~R., Lazendic, J.~S., \& Dickel, J.~R.\ 2004, \apj, 615, 727 

\bibitem[Clark et al.(1982)]{Clark82} Clark, D.~H., Tuohy, 
I.~R., Dopita, M.~A., Mathewson, D.~S., Long, K.~S., Szymkowiak, A.~E., \& 
Culhane, J.~L.\ 1982, \apj, 255, 440 

\bibitem[Crawford et al.(2005)]{Crawford05} Crawford, F., 
McLaughlin, M., Johnston, S., Romani, R., \& Sorrelgreen, E.\ 2005, in 35th 
COSPAR Scientific Assembly, 854 

\bibitem[Cusumano et al.(1998)]{Cusumano98} Cusumano, G., 
Maccarone, M.~C., Mineo, T., Sacco, B., Massaro, E., Bandiera, R., \& 
Salvati, M.\ 1998, \aap, 333, L55 

\bibitem[Dennerl et al.(2001)]{Dennerl01} Dennerl, K.~et al.\ 2001,
\aap, 365, L202

\bibitem[Dickel et al.(1994)]{Dickel94} Dickel, J.~R., Milne, 
D.~K., Kennicutt, R.~C., Chu, Y., \& Schommer, R.~A.\ 1994, \aj, 107, 1067 

\bibitem[Dunne et al.(2001)]{Dunne01} Dunne, B.~C., Points, 
S.~D., \& Chu, Y.\ 2001, \apjs, 136, 119 

\bibitem[Evans et al.(2003)]{Evans03} Evans, N.~R., Seward, 
F.~D., Krauss, M.~I., Isobe, T., Nichols, J., Schlegel, E.~M., \& Wolk, 
S.~J.\ 2003, \apj, 589, 509 
 
\bibitem[Feast et al.(1960)]{Feast60} Feast, M.~W., Thackeray, 
A.~D., \& Wesselink, A.~J.\ 1960, \mnras, 121, 337 
 
\bibitem[Freeman et al.(2002)]{Freeman02} Freeman,
P.~E., Kashyap, V., Rosner, R., \& Lamb, D.~Q.\ 2002, \apjs, 138, 185

\bibitem[Getman et al.(2005)]{Getman05} Getman, K.~V., et al.\ 
2005, \apjs, 160, 319

\bibitem[Grebel \& Chu(2000)]{Grebel00} Grebel, E.~K., \& Chu, 
Y.\ 2000, \aj, 119, 787 

\bibitem[Haberl \& Pietsch(1999)]{Haberl99} Haberl, F., \& 
Pietsch, W.\ 1999, \aaps, 139, 277 

\bibitem[Haberl et al.(2001)]{Haberl01} Haberl, F., Dennerl, K., 
Filipovi{\' c}, M.~D., Aschenbach, B., Pietsch, W., \& Tr{\" u}mper, J.\ 
2001, \aap, 365, L208 

\bibitem[Higdon et al.(1998)]{Higdon98} Higdon, J.~C., 
Lingenfelter, R.~E., \& Ramaty, R.\ 1998, \apjl, 509, L33 

\bibitem[Johansson et al.(1998)]{Johansson98} Johansson, L.~E.~B., 
et al.\ 1998, \aap, 331, 857 

\bibitem[Landecker et al.(1987)]{Landecker87} Landecker, T.~L., 
Dewdney, P.~E., Vaneldik, J.~F., \& Routledge, D.\ 1987, \aj, 94, 111 

\bibitem[Lazendic et al.(2003)]{Lazendic03} Lazendic, J.~S., 
Dickel, J.~R., \& Jones, P.~A.\ 2003, \apj, 596, 287 

\bibitem[Lazendic \& Slane(2005)]{Lazendic05} Lazendic, J., \& 
Slane, P.\ 2005, \apj, submitted (astro-ph/0505498)
 
\bibitem[Leahy et al.(1985)]{Leahy85} Leahy, D.~A., Venkatesan, 
D., Long, K.~S., \& Naranan, S.\ 1985, \apj, 294, 183 

\bibitem[Long \& Helfand(1979)]{Long79} Long, K.~S., \& 
Helfand, D.~J.\ 1979, \apjl, 234, L77 

\bibitem[Long et al.(1981)]{Long81} Long, K.~S., Helfand, 
D.~J., \& Grabelsky, D.~A.\ 1981, \apj, 248, 925 

\bibitem[Mac Low et al.(1998)]{MacLow98} Mac Low, M., Chang, 
T.~H., Chu, Y., Points, S.~D., Smith, R.~C., \& Wakker, B.~P.\ 1998, \apj, 
493, 260 
 
\bibitem[Ma{\'{\i}}z-Apell{\' a}niz et al.(2004)]{Maiz04} 
Ma{\'{\i}}z-Apell{\' a}niz, J., P{\' e}rez, E., \& Mas-Hesse, J.~M.\ 2004, 
\aj, 128, 1196 

\bibitem[Mark et al.(1969)]{Mark69} Mark, H., Price, R., 
Rodrigues, R., Seward, F.~D., \& Swift, C.~D.\ 1969, \apjl, 155, L143 

\bibitem[Marshall et al.(1998)]{Marshall98} Marshall, F.~E., 
Gotthelf, E.~V., Zhang, W., Middleditch, J., \& Wang, Q.~D.\ 1998, \apjl, 
499, L179 

\bibitem[Meaburn(1984)]{Meaburn84} Meaburn, J.\ 1984, \mnras, 211, 521 

\bibitem[Meaburn(1988)]{Meaburn88} Meaburn, J.\ 1988, \mnras, 235, 375 
 
\bibitem[Melnick(1987)]{Melnick87} Melnick, J.\ 1987, in IAU Symp.~121,
Observational Evidence of Activity in Galaxies, ed.\ E.~E.~Khachikian,
K.~J.~Fricke \& J.~Melnick (Dordrecht: Kluwer), 545 
 
\bibitem[Meurer et al.(1995)]{Meurer95} Meurer, G.~R., Heckman, T.~M.,
Leitherer, C., Kinney, A., Robert, C., \& Garnett, D.~R.\ 1995, \aj,
110, 2665

\bibitem[Mignani et al.(2005)]{Mignani05} Mignani, R.~P., Pulone, 
L., Iannicola, G., Pavlov, G.~G., Townsley, L., \& Kargaltsev, O.~Y.\ 2005, 
\aap, 431, 659 

\bibitem[Moffat et al.(2002)]{Moffat02} Moffat, A.~F.~J.~et al.\ 2002,
\apj, 573, 191

\bibitem[Mori et al.(2001)]{Mori01} Mori, K., Tsunemi, H., Miyata, E.,
Baluta, C.~J., Burrows, D.~N., Garmire, G.~P., \& Chartas, G.\ 2001, in
ASP Conf.\ Ser.~251, New Century of X-ray Astronomy, ed.\ H.\ Inoue \&
H.\ Kunieda (San Francisco: ASP), 576 

\bibitem[Morrison \& McCammon(1983)]{Morrison83} Morrison, R.~\&
McCammon, D.\ 1983, \apj, 270, 119

\bibitem[Naz{\' e} et al.(2002)]{Naze02} Naz{\' e}, Y., 
Hartwell, J.~M., Stevens, I.~R., Corcoran, M.~F., Chu, Y.-H., 
Koenigsberger, G., Moffat, A.~F.~J., \& Niemela, V.~S.\ 2002, \apj, 580, 
225 

\bibitem[Naz{\' e} et al.(2004)]{Naze04} Naz{\' e}, Y., 
Antokhin, I.~I., Rauw, G., Chu, Y.-H., Gosset, E., \& Vreux, J.-M.\ 2004, 
\aap, 418, 841 
 
\bibitem[Norci \& {\" O}gelman(1995)]{Norci95} Norci, L., \& 
{\" O}gelman, H.\ 1995, \aap, 302, 879 

\bibitem[Parizot et al.(2004)]{Parizot04} Parizot, E., Marcowith, A.,
van der Swaluw, E., Bykov, A.~M., \& Tatischeff, V.\ 2004, \aap, 424,
747

\bibitem[Park et al.(2005)]{Park05} Park, S., Zhekov, S.~A., 
Burrows, D.~N., Garmire, G.~P., \& McCray, D.\ 2005, Advances in Space 
Research, 35, 991 

\bibitem[Points et al.(2001)]{Points01} Points, S.~D., Chu, 
Y.-H., Snowden, S.~L., \& Smith, R.~C.\ 2001, \apjs, 136, 99 

\bibitem[Portegies Zwart et al.(2002)]{PPL02} Portegies 
Zwart, S.~F., Pooley, D., \& Lewin, W.~H.~G.\ 2002, \apj, 574, 762 

\bibitem[Redman et al.(1999)]{Redman99} Redman, M.~P., 
Al-Mostafa, Z.~A., Meaburn, J., Bryce, M., \& Dyson, J.~E.\ 1999, \aap, 
345, 943 
 
\bibitem[Redman et al.(2003)]{Redman03} Redman, M.~P., 
Al-Mostafa, Z.~A., Meaburn, J., \& Bryce, M.\ 2003, \mnras, 344, 741 
 
\bibitem[Routledge et al.(1991)]{Routledge91} Routledge, D., 
Dewdney, P.~E., Landecker, T.~L., \& Vaneldik, J.~F.\ 1991, \aap, 247, 529 

\bibitem[Sanders et al.(2005)]{Sanders05} Sanders, J.~S., Fabian, 
A.~C., \& Taylor, G.~B.\ 2005, \mnras, 356, 1022 

\bibitem[Sasaki et al.(2000)]{Sasaki00} Sasaki, M., Haberl, F., 
\& Pietsch, W.\ 2000, \aaps, 143, 391 

\bibitem[Sasaki et al.(2002)]{Sasaki02} Sasaki, M., Haberl, F., 
\& Pietsch, W.\ 2002, \aap, 392, 103 

\bibitem[Scowen et al.(1998)]{Scowen98} Scowen, P.~A., et al.\ 
1998, \aj, 116, 163 

\bibitem[Seward \& Chlebowski(1982)]{Seward82} Seward, F.~D.~\&
Chlebowski, T.\ 1982, \apj, 256, 530

\bibitem[Shapley \& Lindsay(1963)]{Shapley63} Shapley, H., \& 
Lindsay, E.~M.\ 1963, Irish Astron.~J., 6, 74 
 
\bibitem[B.\ Smith et al.(2005)]{Smith05} Smith, B.~J., Struck, C., 
\& Nowak, M.~A.\ 2005, \aj, 129, 1350 
 
\bibitem[C.\ Smith et al.(2000)]{Smith00} Smith, C., Leiton, R., \&
Pizarro, S.\ 2000, in ASP Conf.~Ser.~221, Stars, Gas and Dust in
Galaxies:  Exploring the Links, ed.\ D.~Alloin, K.~Olsen, \& G.~Galaz
(San Francisco: ASP), 83 

\bibitem[D.\ Smith \& Wang(2004)]{Smith04} Smith, D.~A., \& Wang, 
Q.~D.\ 2004, \apj, 611, 881 

\bibitem[R.\ Smith et al.(2001)]{Smith01} Smith, R.~K., Brickhouse, N.~S.,
Liedahl, D.~A., \& Raymond, J.~C.\ 2001, \apjl, 556, L91

\bibitem[Snowden \& Petre(1994)]{Snowden94} Snowden, S.~L., \& 
Petre, R.\ 1994, \apjl, 436, L123 

\bibitem[Stevens \& Hartwell(2003)]{Stevens03} Stevens, I.~R., \& 
Hartwell, J.~M.\ 2003, \mnras, 339, 280 

\bibitem[Strickland et al.(2002)]{Strickland02} Strickland, D.~K.,
Heckman, T.~M., Weaver, K.~A., Hoopes, C.~G., \& Dahlem, M.\ 2002,
\apj, 568, 689

\bibitem[Townsley et al.(1999)]{Townsley99} Townsley, L., Feigelson,
E., Burrows, D., Chu, Y.-H., Garmire, G., Broos, P., Tsuboi, Y., \&
Griffiths, R.\ 1999, \baas, 31, 1453 

\bibitem[Townsley et al.(2000a)]{Townsley00} Townsley, L.~K., 
Broos, P.~S., Garmire, G.~P., \& Nousek, J.~A.\ 2000, \apjl, 534, L139 

\bibitem[Townsley et al.(2000b)]{Townsley00b} Townsley, L., et al.\ 
2000, \baas, 32, 1594 

\bibitem[Townsley et al.(2002a)]{Townsley02} Townsley, L., et al.\ 
2002, APS Meeting Abstracts, 17061 

\bibitem[Townsley et al.(2002b)]{Townsley02a} Townsley, L.~K., Broos,
P.~S., Nousek, J.~A., \& Garmire, G.~P.\ 2002, Nuclear Instr.\ \&
Methods, 486, 751

\bibitem[Townsley et al.(2002c)]{Townsley02b} Townsley, L.~K., Broos,
P.~S., Chartas, G., Moskalenko, E., Nousek, J.~A., \& Pavlov, G.~G.\
2002, Nuclear Instr.\ \& Methods, 486, 716

\bibitem[Townsley et al.(2003)]{Townsley03} Townsley, L.~K., 
Feigelson, E.~D., Montmerle, T., Broos, P.~S., Chu, Y., \& Garmire, G.~P.\ 
2003, \apj, 593, 874 

\bibitem[Townsley et al.(2005a)]{TownsleyIAUS227} Townsley, L.~K., 
Broos, P.~S., Feigelson, E.~D., \& Garmire, G.~P.\ 2005, in IAU Symp.~227,
Massive Star Birth -- A Crossroads of Astrophysics, ed.\ R.~Cesaroni, E.~Churchwell, M.~Felli, \& C.~M.~Walmsley (astro-ph/0506418)

\bibitem[Townsley et al.(2005b)]{Townsley05b} Townsley, L.~K., Broos, P.~S.,
Feigelson, E.~D., Garmire, G.~P., \& Getman, K.~V.\ 2005, \aj, submitted

\bibitem[Tsunemi et al.(2001)]{Tsunemi01} Tsunemi, H., Mori, K., 
Miyata, E., Baluta, C., Burrows, D.~N., Garmire, G.~P., \& Chartas, G.\ 
2001, \apj, 554, 496 

\bibitem[Vel{\' a}zquez et al.(2003)]{Velazquez03} Vel{\' a}zquez, 
P.~F., K{\" o}nigsberger, G., \& Raga, A.~C.\ 2003, \apj, 584, 284 

\bibitem[Walborn(1991)]{Walborn91} Walborn, N.~R.\ 1991, in IAU 
Symp.~148, The Magellanic Clouds, ed.\ R.~Haynes \& D.~Milne (Dordrecht: Kluwer), 145 
 
\bibitem[Walborn et al.(2002)]{Walborn02} Walborn, N.~R., 
Ma{\'{\i}}z-Apell{\' a}niz, J., \& Barb{\' a}, R.~H.\ 2002, \aj, 124, 1601 

\bibitem[Wang et al.(1991)]{Wang91c} Wang, Q., Hamilton, T., 
Helfand, D.~J., \& Wu, X.\ 1991, \apj, 374, 475 

\bibitem[Wang \& Helfand(1991a)]{Wang91a} Wang, Q., \& Helfand, 
D.~J.\ 1991a, \apj, 370, 541 

\bibitem[Wang \& Helfand(1991b)]{Wang91b} Wang, Q., \& Helfand, 
D.~J.\ 1991b, \apj, 373, 497 

\bibitem[Wang(1995)]{Wang95} Wang, Q.~D.\ 1995, \apj, 453, 783 

\bibitem[Wang \& Gotthelf(1998a)]{Wang98a} Wang, Q.~D., \& 
Gotthelf, E.~V.\ 1998, \apj, 494, 623 

\bibitem[Wang \& Gotthelf(1998b)]{Wang98b} Wang, Q.~D., \& 
Gotthelf, E.~V.\ 1998, \apjl, 509, L109 

\bibitem[Wang(1999)]{Wang99} Wang, Q.~D.\ 1999, \apjl, 510, L139 

\bibitem[Wang et al.(2001)]{Wang01} Wang, Q.~D., Gotthelf, 
E.~V., Chu, Y.-H., \& Dickel, J.~R.\ 2001, \apj, 559, 275 

\bibitem[Williams et al.(1999)]{Williams99} Williams, R.~M., Chu, 
Y., Dickel, J.~R., Petre, R., Smith, R.~C., \& Tavarez, M.\ 1999, \apjs, 
123, 467 

\bibitem[Yusef-Zadeh et al.(2002)]{Yusef-Zadeh02} Yusef-Zadeh, F., Law,
C., Wardle, M., Wang, Q.~D., Fruscione, A., Lang, C.~C., \& Cotera,
A.\ 2002, \apj, 570, 665

\end{thebibliography}
\end{document}